\documentclass{ws-ijmpa}
\usepackage{here}
\usepackage[super]{cite}
\usepackage{xcolor}
\usepackage[verbose,hypertexnames=false]{hyperref}
\hypersetup{colorlinks=false,allbordercolors=blue,pdfborderstyle={/S/U/W 1}}

\begin{document}

\markboth{Andr\'e Sopczak}{High-mass new scalars at the LHC, with H in the final state}

\title{High-mass new scalars at the LHC, with H in the final state}

\author{Andr\'e Sopczak}

\address{Institute of Experimental and Applied Physics, Czech Technical University in Prague, \\Husova 240/5, 11000 Prague 1, Czech Republic; \\
andre.sopczak@cern.ch
}

\maketitle

\begin{abstract}
The search for new particles, conducted at the LHC by the ATLAS and CMS collaborations, covers several production and decay modes, leading to a large variety of final states that could be observed in both detectors. This review focuses on the production of new heavy scalars that have a Higgs boson in the final state. Three cases are covered: resonant production of a heavy scalar $X$ decaying into a Higgs boson and a lighter scalar $S$, heavy neutral Higgs boson decaying into another neutral Higgs boson and a $Z$ boson, and a heavy charged Higgs boson production with subsequent decay into another neutral Higgs boson and a $W$ boson. The reviewed searches are based on the complete LHC Run-2 dataset and partial Run-3 dataset.
\end{abstract}


\section{Introduction}
A new high-mass scalar produced at the Large Hadron Collider (LHC) could decay into a Higgs boson, which subsequently decays into Standard Model (SM) particles. Such processes have been searched for in several production and decay modes. This review is structured according to the published results by the experimental collaborations ATLAS and CMS.

\subsection{ATLAS and CMS experiments
at the Large Hadron Collider}
\begin{figure}[H]
\centering
\includegraphics[width=0.49\linewidth]
{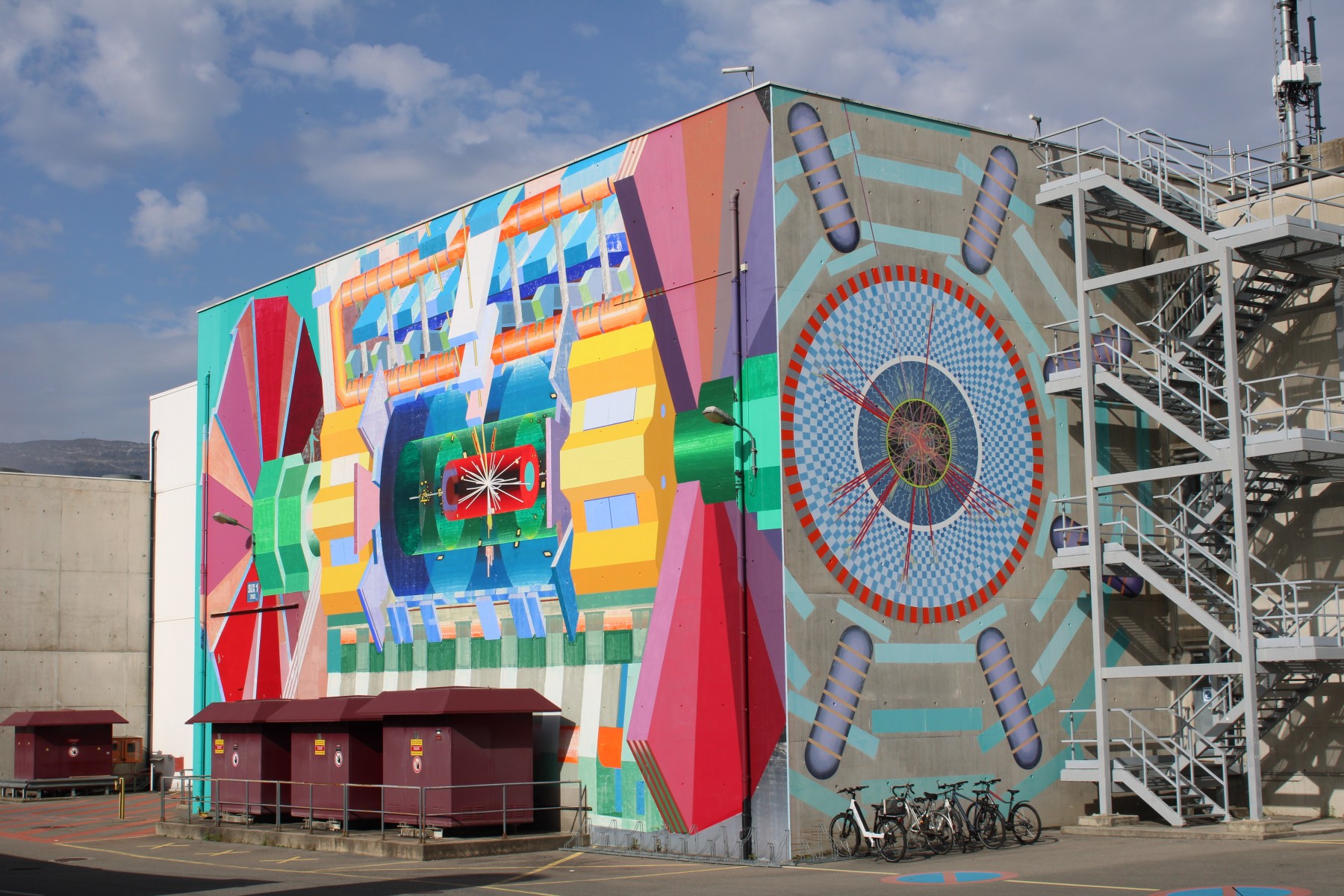}
\includegraphics[width=0.49\linewidth]
{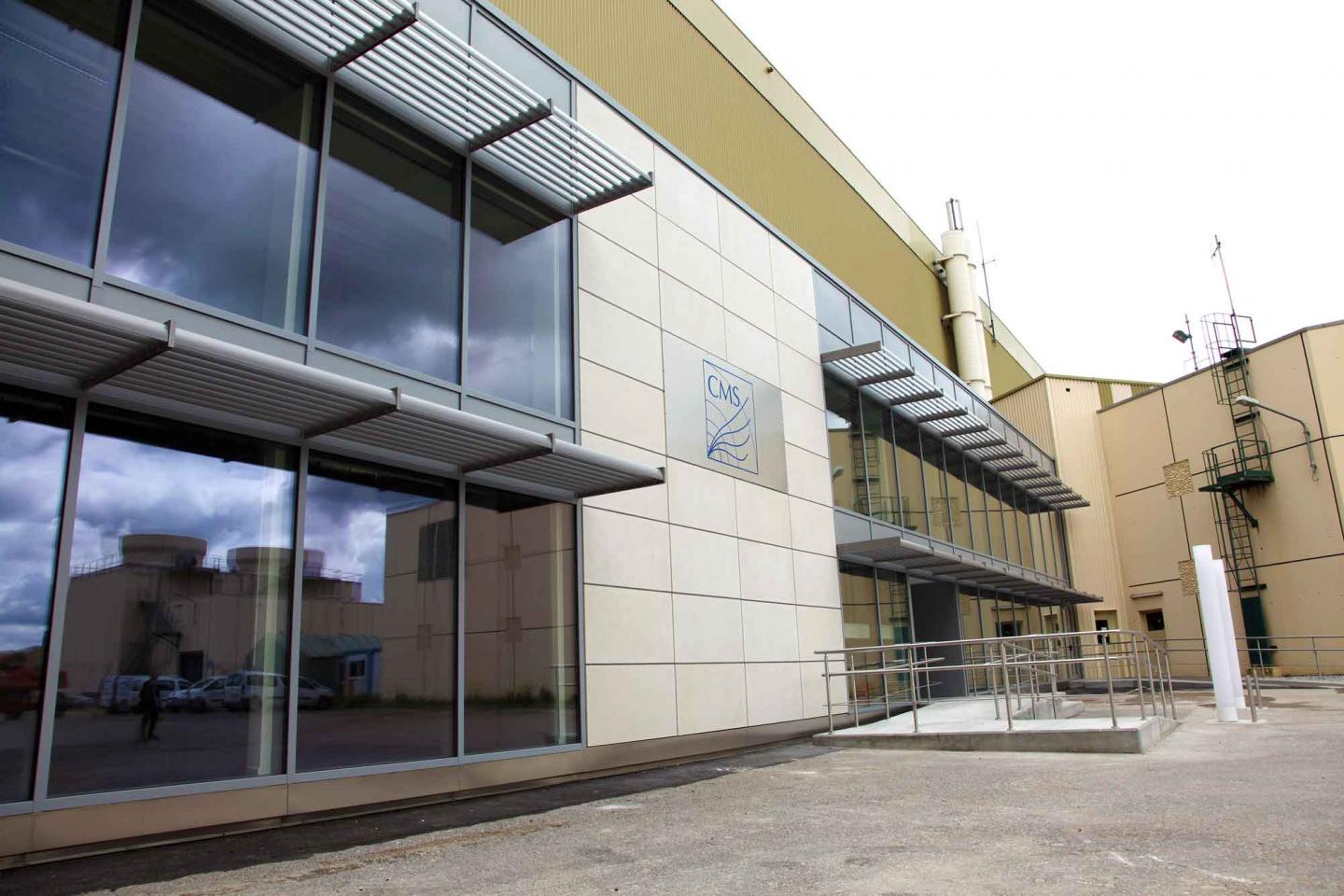}
\caption{ATLAS (left) and CMS (right) experimental buildings. {ATLAS (left) and CMS (right) experimental buildings. The ATLAS detector is located at Point~1 of the LHC ring on the CERN site near Geneva, Switzerland, while the CMS detector is located at Point~5 in Cessy, France. The two detectors are situated on diametrically opposite sides of the 26.7~km Large Hadron Collider (LHC) ring.}
\copyright~CERN/ATLAS Collaboration (left); \copyright~CERN/CMS Collaboration (right).}
\label{fig:ATLASCMS}
\end{figure}

\begin{figure}[H]
\vspace{-10mm}
\centering
\includegraphics[width=\linewidth]{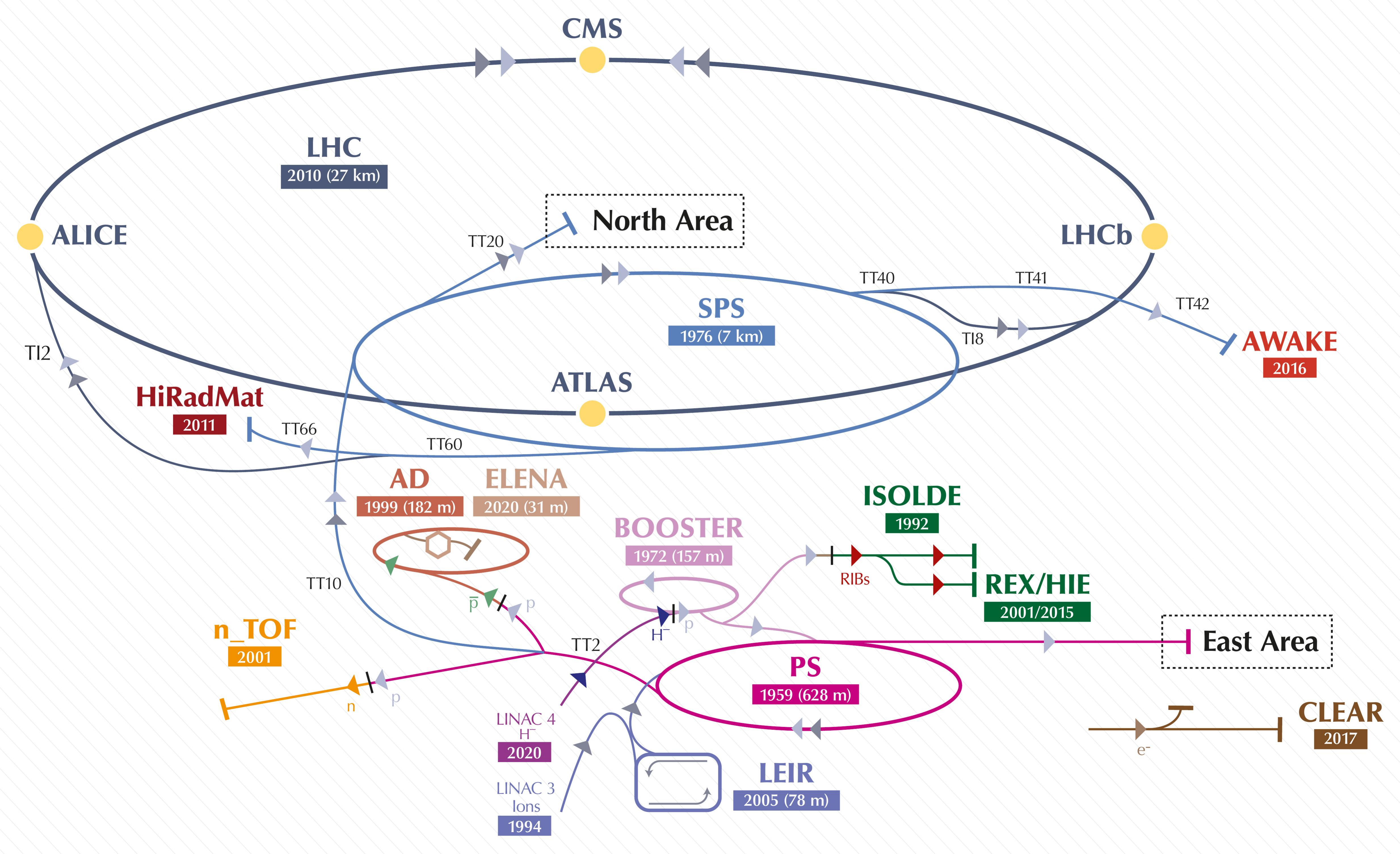}
\vspace{-7mm}
\caption{Schematic view of the LHC accelerator complex and the location of the four main experiments. The LHC injection chain includes the LINAC4 linear accelerator, the Proton Synchrotron Booster (BOOSTER), the Proton Synchrotron (PS) and the Super Proton Synchrotron (SPS), which delivers protons to the LHC ring.
\copyright~CERN
}
\label{fig:LHC}
\end{figure}

The data from proton-proton collisions were recorded by the ATLAS and CMS experiments~\cite{ATLAS_detector,CMS_detector}. Their surface buildings are shown in Fig.~\ref{fig:ATLASCMS}. They are located at opposite positions in the LHC accelerator complex (Fig.~\ref{fig:LHC}){. The LHC is the world's largest and highest-energy particle accelerator, designed to collide beams of protons at center-of-mass energies up to 14\,TeV, thereby providing access to physics at the TeV scale. The LHC injection chain, comprising the LINAC4 linear accelerator, the proton synchrotron booster, the proton synchrotron and the super proton synchrotron, delivers proton beams to the LHC ring, where they are accelerated to their final collision energy}.

\subsection{Dataset}
\begin{figure}[H]
\vspace{-3mm}
\centering
\includegraphics[width=0.49\linewidth]
{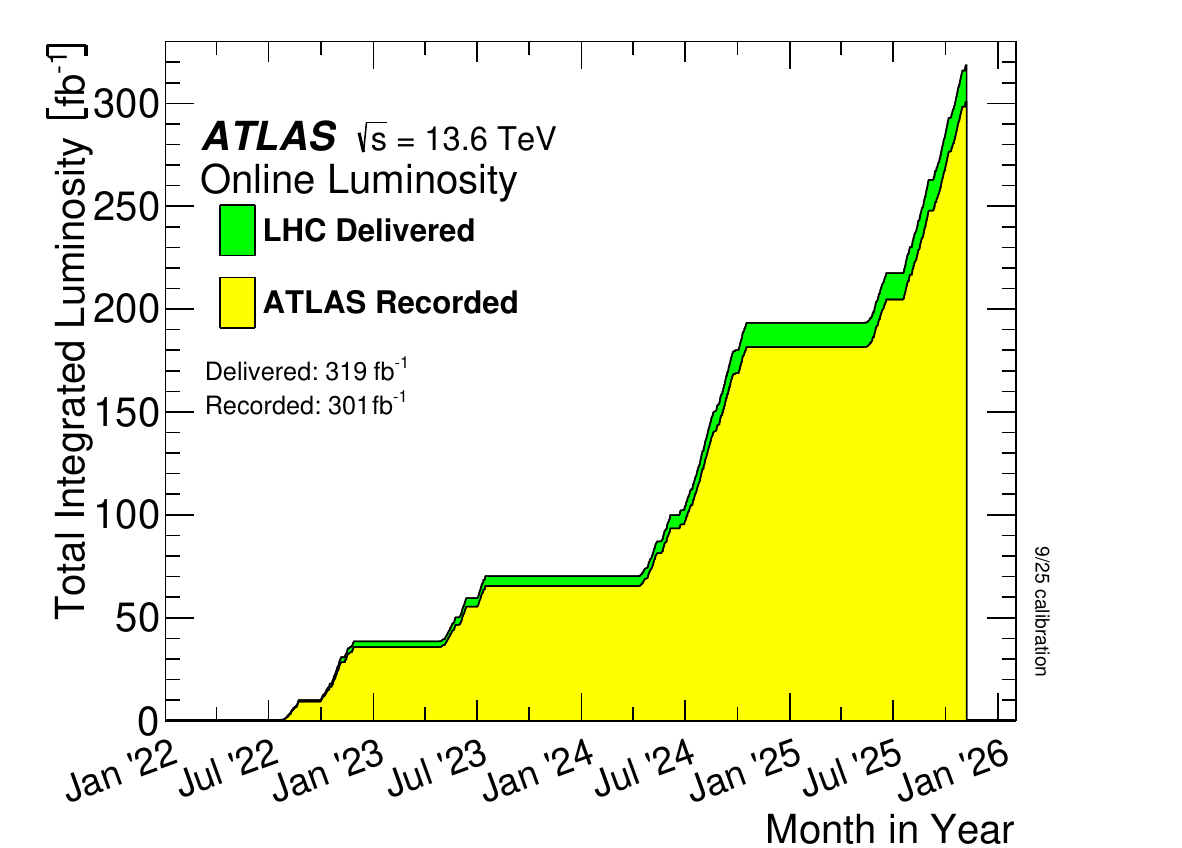}
\includegraphics[width=0.49\linewidth]
{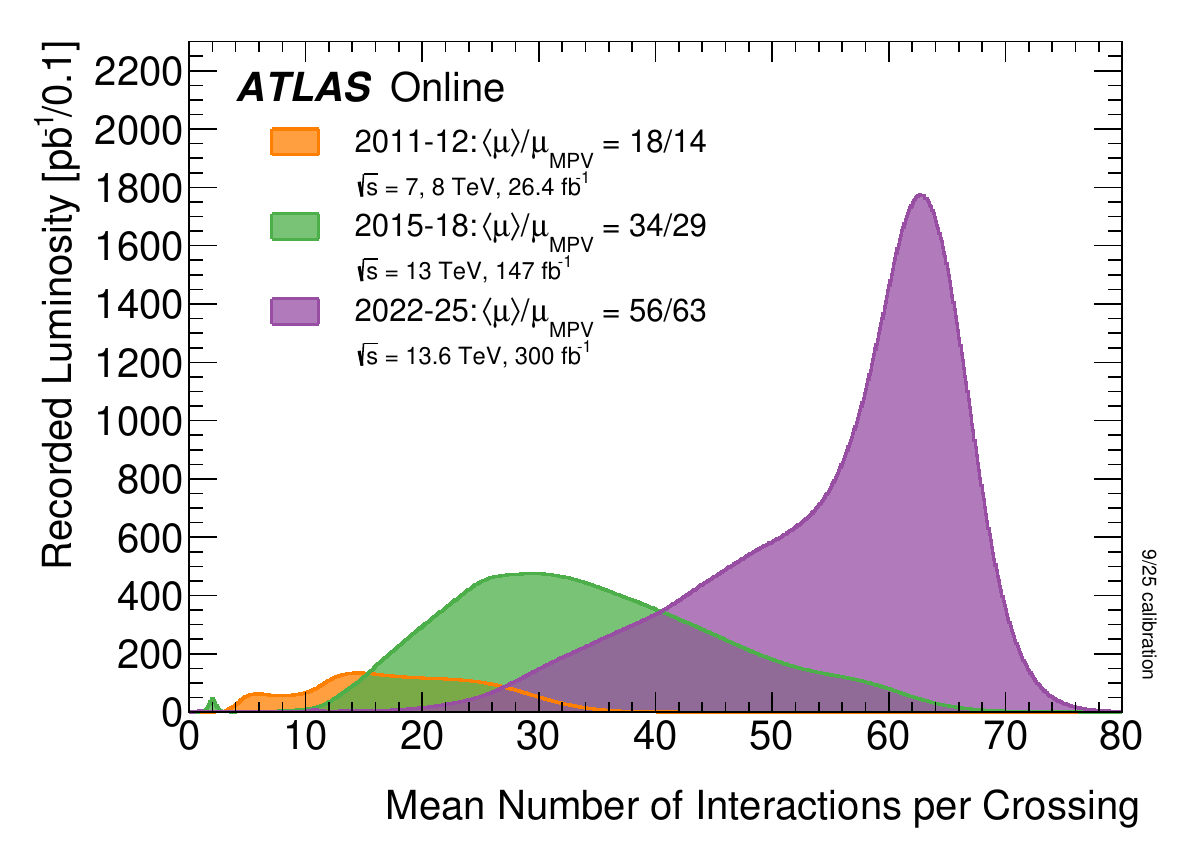}
\caption{LHC luminosity (Run-3, 2022--2025), and mean number of interactions per bunch crossing for Run-1 (7 and 8\,TeV), Run-2 (13\,TeV), and Run-3 (13.6\,TeV). From Ref.~\cite{lumi}.}
\label{fig:lumi}
\vspace{-3mm}
\end{figure}

{The LHC proton-proton collision program} took place at Run-1 (2010--2012) at center-of-mass energies of 7\,TeV and 8\,TeV, and at Run-2 (2015--2018) at 13\,TeV. Run-3 (2022--{2026}) {is taking proton-proton collision data} at 13.6\,TeV{, and is expected to complete data-taking by the end of June 2026. In addition to proton-proton collisions, the LHC also operates with heavy-ion beams (e.g.\ PbPb, XeXe, OO, NeNe, pPb), though those programs are not the subject of this review.}
Most results presented in this review are from Run-2 data-taking at 13\,TeV, and a highlight is the first result incorporating Run-3 (2022 and 2023) data at 13.6\,TeV. ATLAS and CMS have recorded similar numbers of proton-proton collisions. The recorded luminosity of Run-{3} until 2025 and the mean number of interactions per bunch crossing for Run-1, Run-2 and Run-3 are shown in Fig.~\ref{fig:lumi}.

\subsection{Analysis methodology}

The searches reviewed here share a common experimental strategy, briefly outlined below for the benefit of readers less familiar with LHC data analysis.

Events are classified into signal regions (SRs) and control regions (CRs). SRs are kinematic selections designed to be enriched in potential signal events; CRs are signal-depleted phase-space regions used to validate and constrain the background model. The background normalization and, in some analyses, its shape are determined by a simultaneous maximum-likelihood fit to the data in the CRs and SRs. Distributions shown \emph{before} this fit (``pre-fit'') reflect the prediction based purely on simulation and theoretical calculations. Distributions shown \emph{after} the fit (``post-fit'') have been adjusted so that the background model agrees with the data within the fitted nuisance parameters (systematic uncertainties).

In the kinematic distributions, shaded bands represent the total uncertainty in the background prediction after the fit, combining statistical and systematic components in quadrature. In upper-limit plots, the inner (green) band and the outer (yellow) band indicate the regions containing 68\% and 95\%, respectively, of the distribution of limits expected under the background-only hypothesis. These correspond to the $\pm 1\sigma$ and $\pm 2\sigma$ intervals around the median expected limit. All exclusion limits in this review are quoted at 95\% confidence level (CL) using the CL$_s$ method.

Machine-learning classifiers---boosted decision trees (BDTs) and parametric neural networks (PNNs)---are used in several analyses to improve the separation of signal from background. Their output score serves as the final discriminant in the statistical analysis.

\section{Resonant Production of a Heavy Scalar $X$ Decaying into a SM Higgs Boson and a Lighter Scalar $S$
}

Models with extended scalar sectors, such as the Next-to-Minimal Supersymmetric Standard Model (NMSSM)~\cite{Hayrapetyan:2960489}, predict the existence of additional neutral scalar bosons beyond the single Higgs boson of the SM. In the NMSSM, a gauge-singlet superfield is added to the MSSM superpotential, yielding a richer Higgs sector with three CP-even, two CP-odd, and two charged physical states. A characteristic signature of such extended sectors is a \emph{cascade decay}: a heavy scalar $X$ is produced (for instance via gluon-gluon fusion) and subsequently decays into two lighter scalars, one of which may be identified with the SM-like Higgs boson $H$ at 125\,GeV and the other is a new BSM scalar $S$ (or $Y$). Discovering such a two-step decay chain would provide strong evidence for an extended scalar sector. In this section we review searches for the process $X\rightarrow S H$ in four distinct final states, using Run-2 and early Run-3 data at 13 and 13.6\,TeV, respectively.

\subsection{$X\to S(\to b\bar{b})H(\to \gamma\gamma)$, SM H}

\begin{figure}[H]
\vspace{-5mm}
\centering
\includegraphics[width=0.6\linewidth]{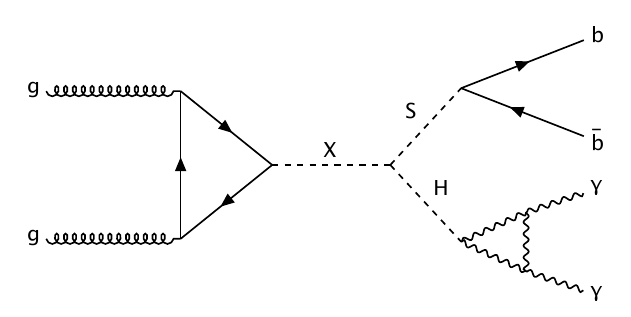}
\vspace{-3mm}
\caption{Feynman diagram for $X\rightarrow S(\rightarrow b\bar{b})H(\rightarrow \gamma\gamma)$ production via gluon-gluon fusion.  From Ref.~\cite{2026140425}.}
\label{fig:bbgg}
\vspace{-5mm}
\end{figure}

A search for a new heavy scalar $X$ in $X\rightarrow S(\rightarrow b\bar{b})H(\rightarrow \gamma\gamma)$ production via gluon-gluon fusion was conducted in the final state with two $b$-quarks and two photons, based on the full ATLAS Run-2 dataset (140\,fb$^{-1}$ at 13\,TeV) plus the 2022--2023 Run-3 data (58.6\,fb$^{-1}$ at 13.6\,TeV)~\cite{2026140425}. The corresponding Feynman diagram is shown in Fig.~\ref{fig:bbgg}. In this topology, the SM Higgs boson $H$ decays into the experimentally clean $\gamma\gamma$ final state (exploiting the excellent ATLAS electromagnetic calorimeter resolution), while the BSM scalar $S$ decays into a $b\bar{b}$ pair. The di-photon invariant mass provides a narrow, well-reconstructed peak at 125\,GeV, making this channel particularly robust against continuum backgrounds.

\begin{figure}[H]
\vspace{-6mm}
\centering
\includegraphics[width=0.46\linewidth]{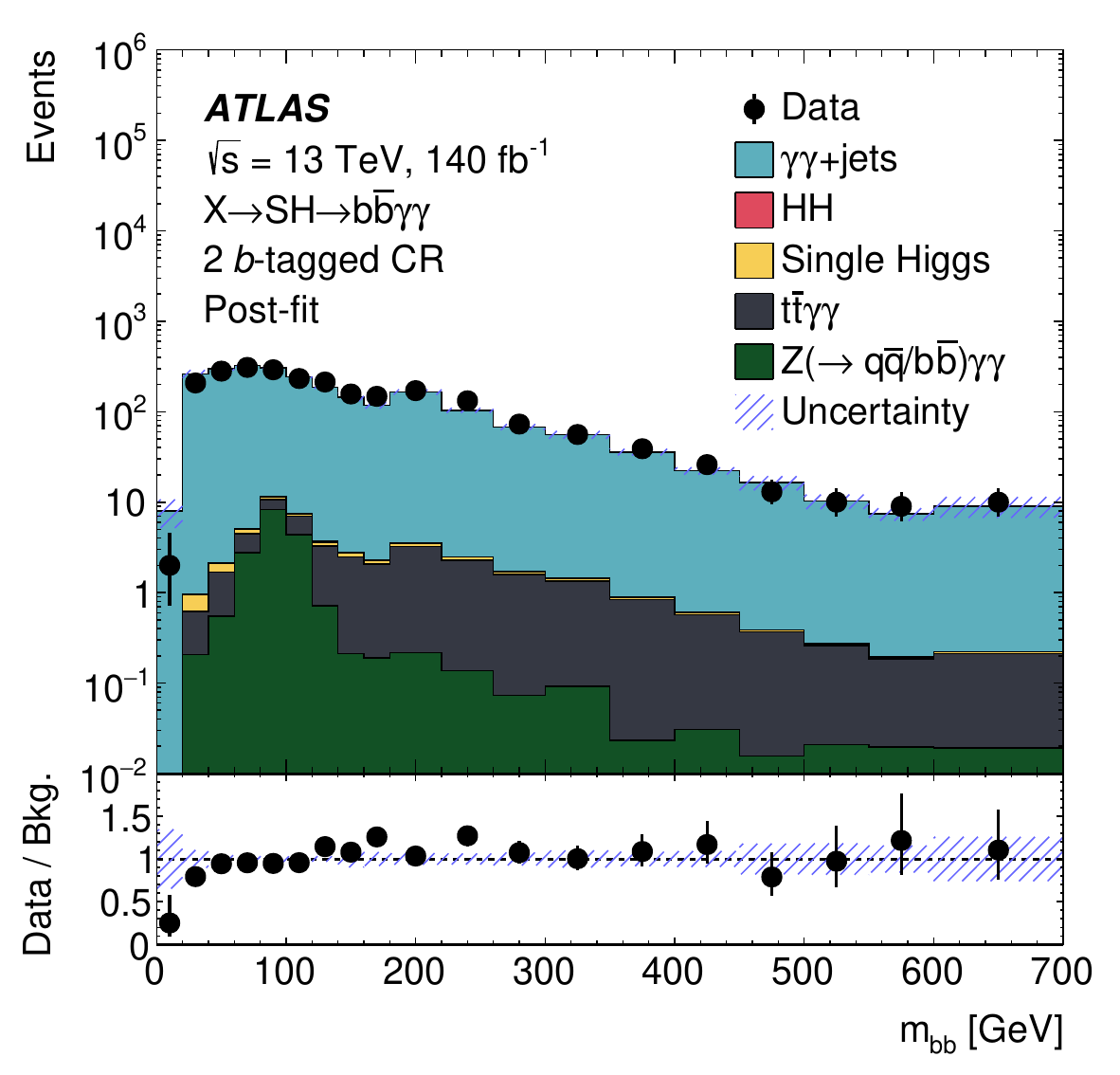}
\includegraphics[width=0.53\linewidth]{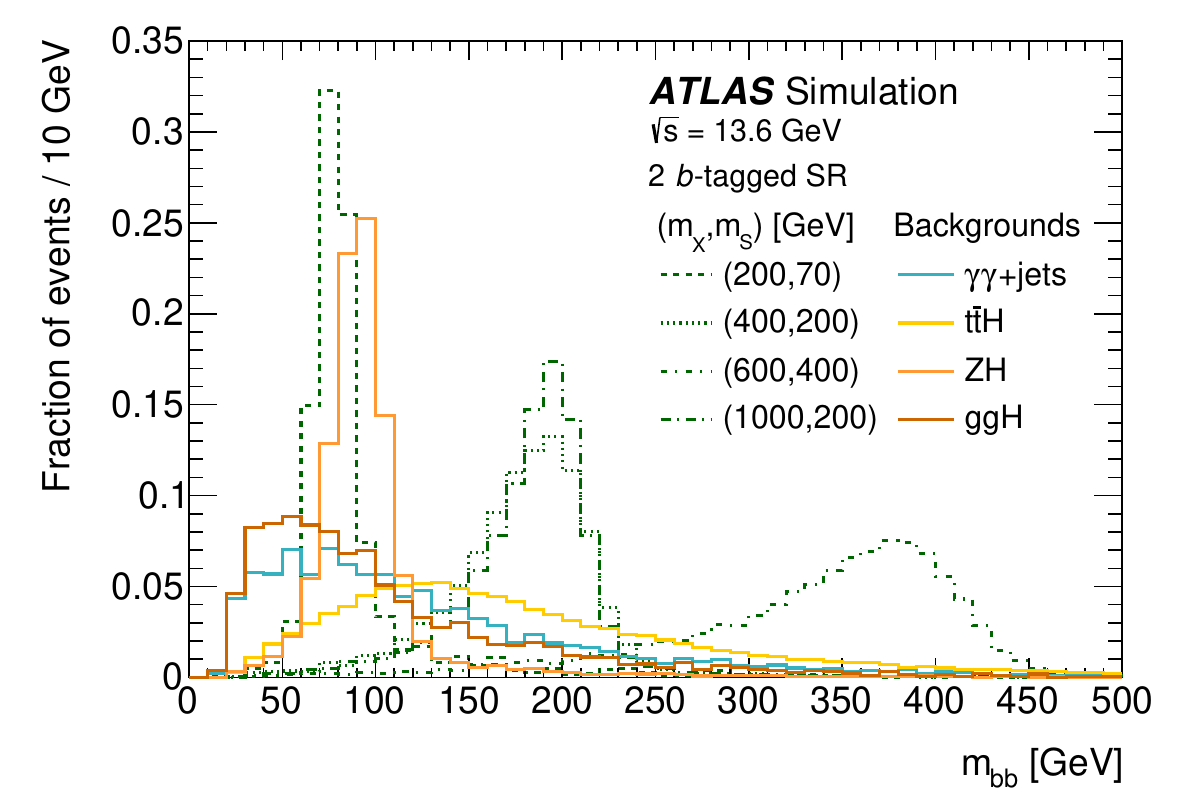}
\caption{Left: Distribution of $m_{bb}$ in data and in the post-fit background-only prediction in the control region (CR) of the 2 $b$-tagged category (Run-2, 13\,TeV). The $\gamma\gamma$ + jets category represents the sum of $\gamma\gamma$ + jets, $\gamma$ + jets and dijet backgrounds. The error band corresponds to the dominant uncertainty from the non-resonant diphoton background.
Right: Distribution of $m_{bb}$ for the 2 $b$-tagged category for a subset of signal mass points and main background processes, normalized to unity.
From Ref.~\cite{2026140425}.}
\label{fig:ATLAS_bb_CR_run2}
\vspace{-6mm}
\end{figure}

\begin{figure}[H]
\vspace{-1mm}
\centering
\includegraphics[width=0.46\linewidth]{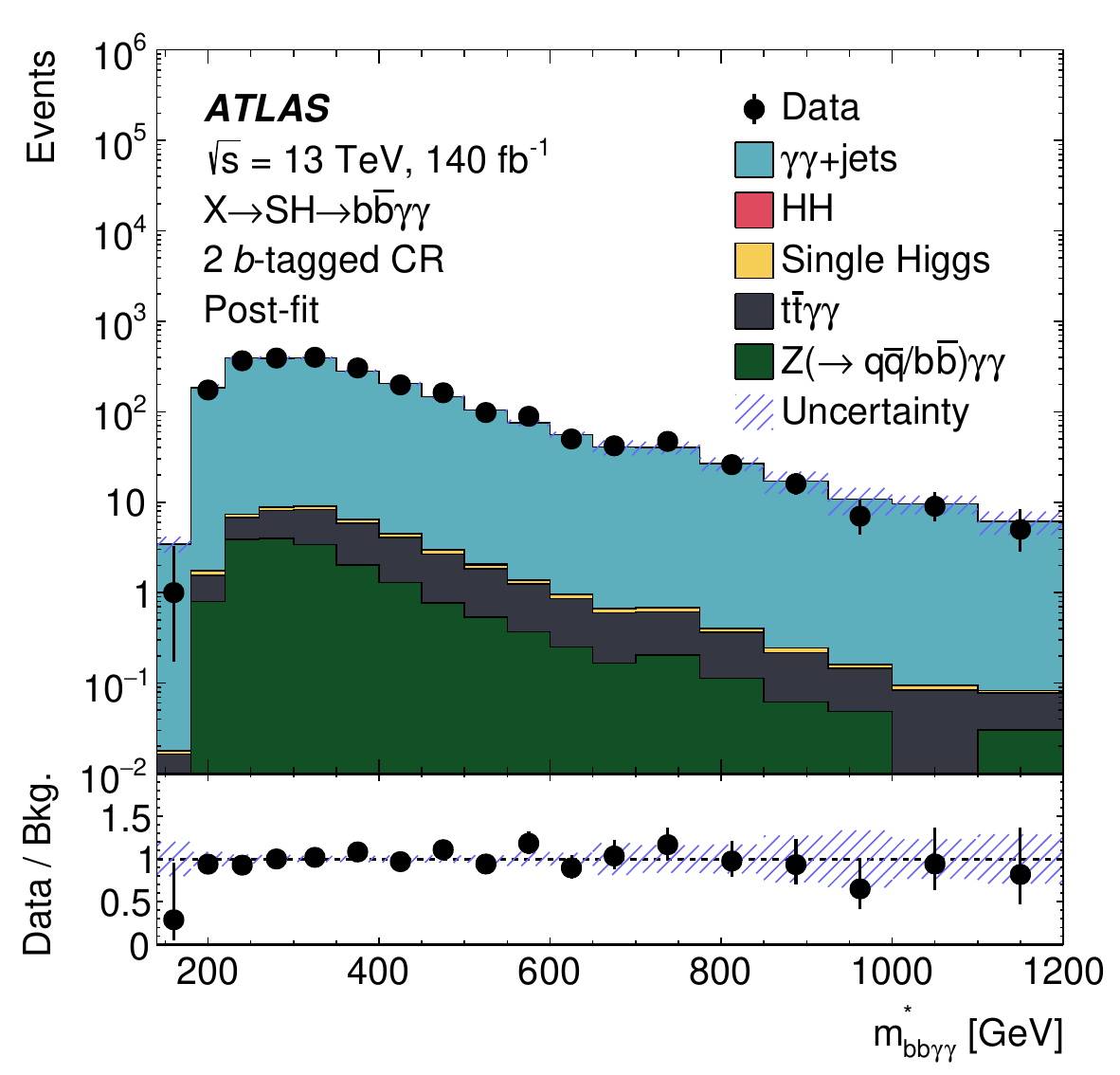}
\includegraphics[width=0.53\linewidth]{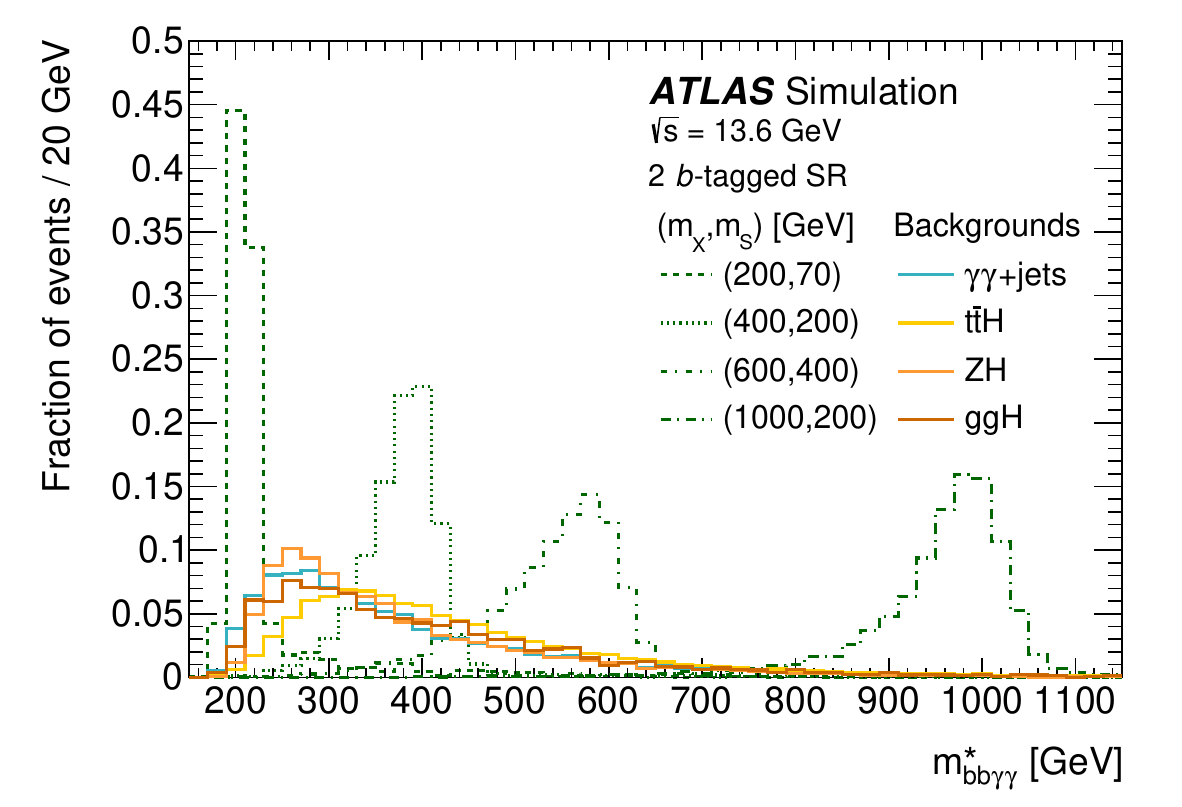}
\caption{Left: Distribution of $m_{bb\gamma\gamma}^*(= m_{bb\gamma\gamma} - (m_{bb} - 125~{\text{GeV}}))$ in data and in the post-fit background-only prediction in the signal region (SR) of the 2 $b$-tagged category (Run-2, 13\,TeV). The error band corresponds to the dominant uncertainty from the non-resonant diphoton background.
Right: Distributions of $m_{bb}$ for the 2 $b$-tagged category for selected signal mass points and main background processes, normalized to unity.
From Ref.~\cite{2026140425}.}
\label{fig:ATLAS_bb_SR_run2}
\vspace{-2mm}
\end{figure}

The mass of the scalar $S$ and the mass of the scalar $X$ correspond to the invariant mass of the $b$-quark pair ($b$-tagged jets), and the invariant mass of the $b\bar{b}\gamma\gamma$ final state, respectively. Events are classified according to the number of $b$-tagged jets (1 or 2). A control region (CR) is used to validate the non-resonant diphoton background model, which dominates the systematic uncertainty. The measured and simulated $m_{bb}$ mass distributions for the CR and SR of the 2 $b$-tagged category are shown in Figs.~\ref{fig:ATLAS_bb_CR_run2} and~\ref{fig:ATLAS_bb_SR_run2}, respectively, for Run-2 data.

\begin{figure}[H]
\centering
\includegraphics[width=0.49\linewidth]{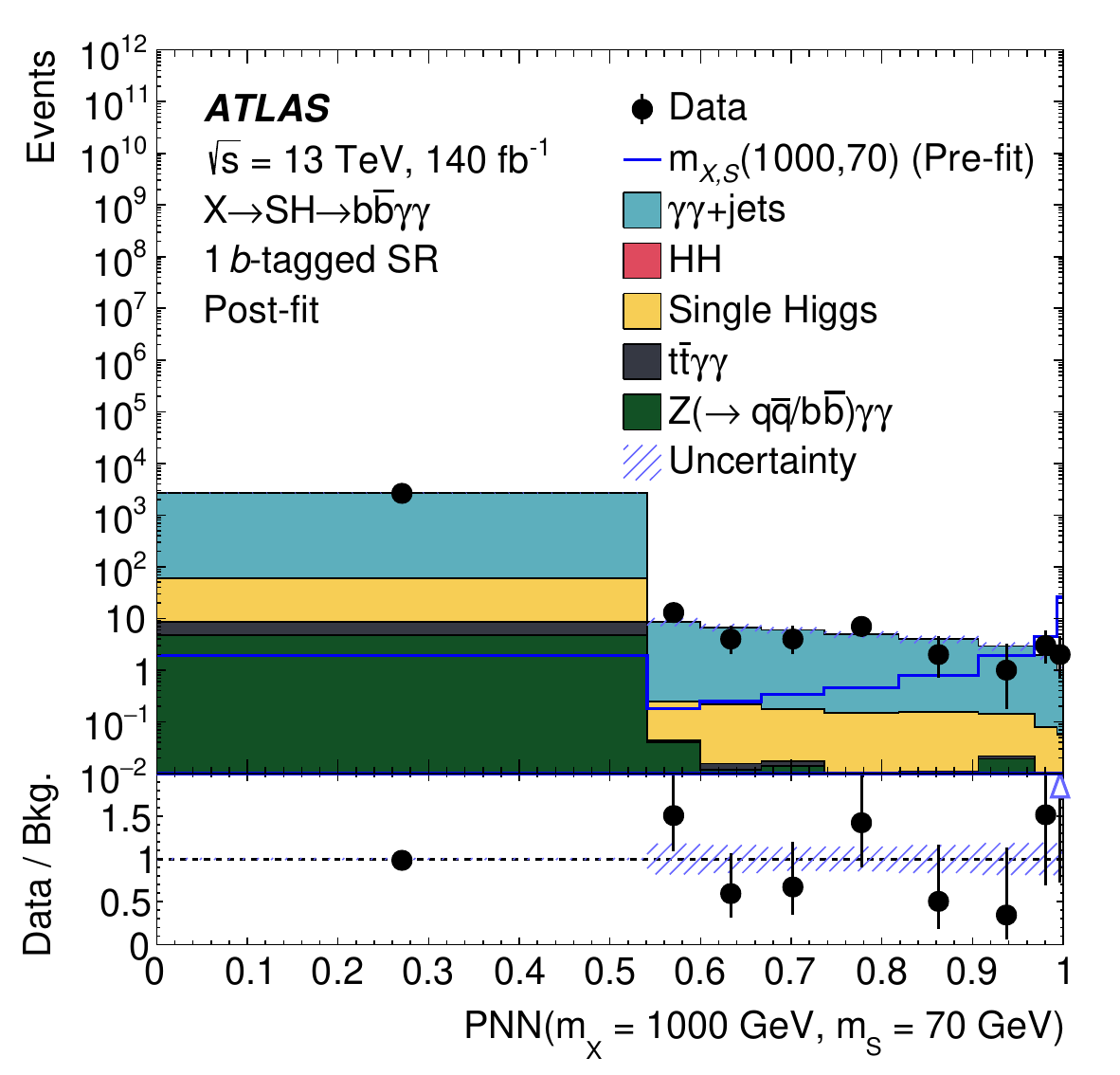}
\includegraphics[width=0.49\linewidth]{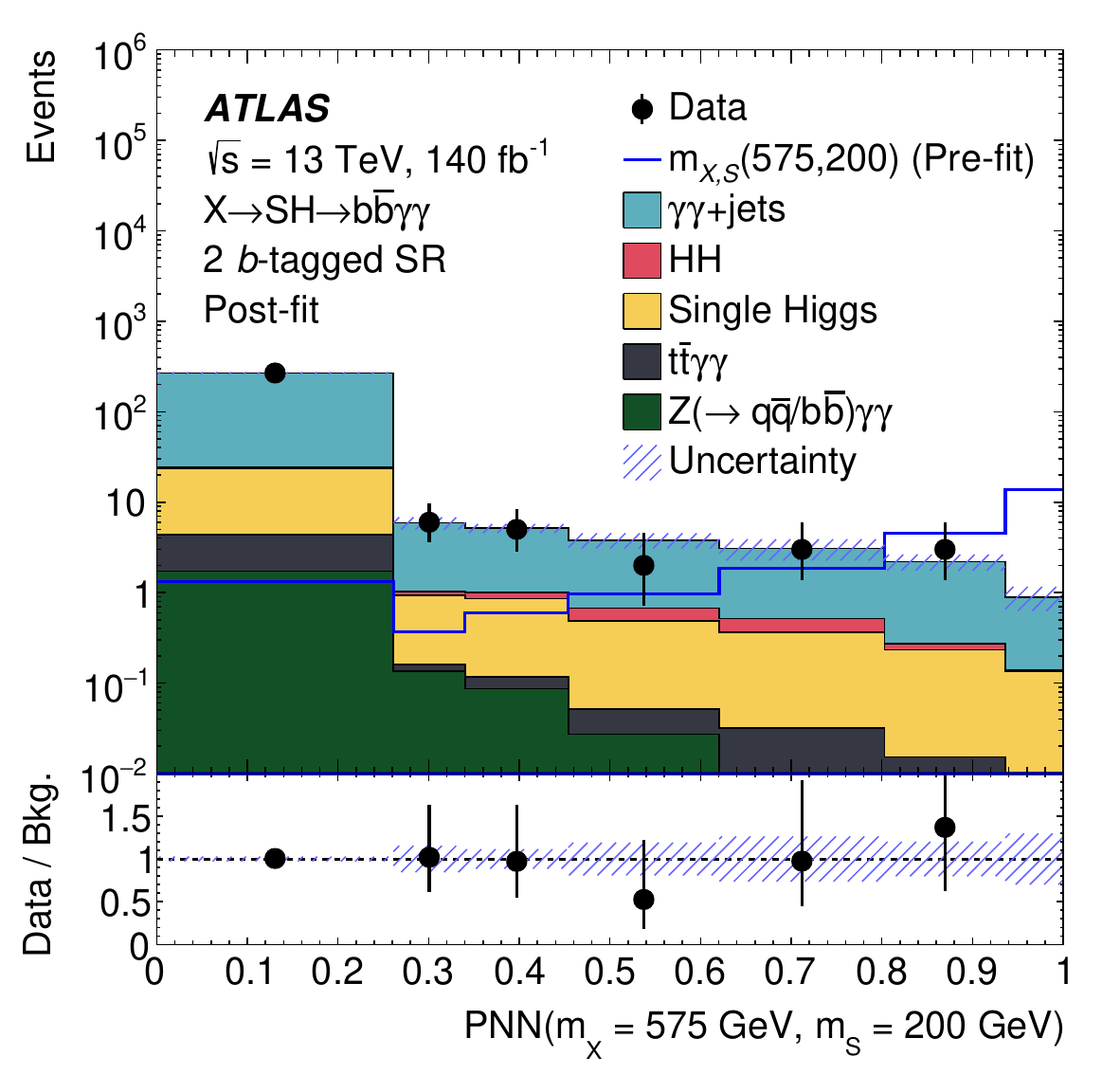}
\caption{Distribution of the PNN discriminant output after the profile-likelihood fit in the signal region (SR) for Run-2 (13\,TeV). Left: 1 $b$-tagged SR for $m_X = 1000$\,GeV, $m_S = 70$\,GeV. Right: 2 $b$-tagged SR for $m_X = 575$\,GeV, $m_S = 200$\,GeV. The $\gamma\gamma$+jets category represents the sum of $\gamma\gamma$+jets, $\gamma$+jets and dijet processes. The shaded band represents the total post-fit systematic uncertainty. The blue line represents the signal normalized to an arbitrary cross-section of 1\,fb.
From Ref.~\cite{2026140425}.}
\label{fig:ATLAS_PNN_run2}
\end{figure}

A PNN is trained to simultaneously separate signal and background events across the two-dimensional signal hypothesis $(m_X, m_S)$. The network takes as input kinematic variables together with the signal mass parameters, providing a single discriminant that is sensitive to signals over the full mass plane. Figure~\ref{fig:ATLAS_PNN_run2} shows the PNN output distributions in two representative SRs for Run-2 data, corresponding to two different signal hypotheses. The background prediction is shown post-fit after a profile-likelihood fit to the data. Good agreement between data and the background model is observed.

\begin{figure}[H]
\centering
\includegraphics[width=0.49\linewidth]{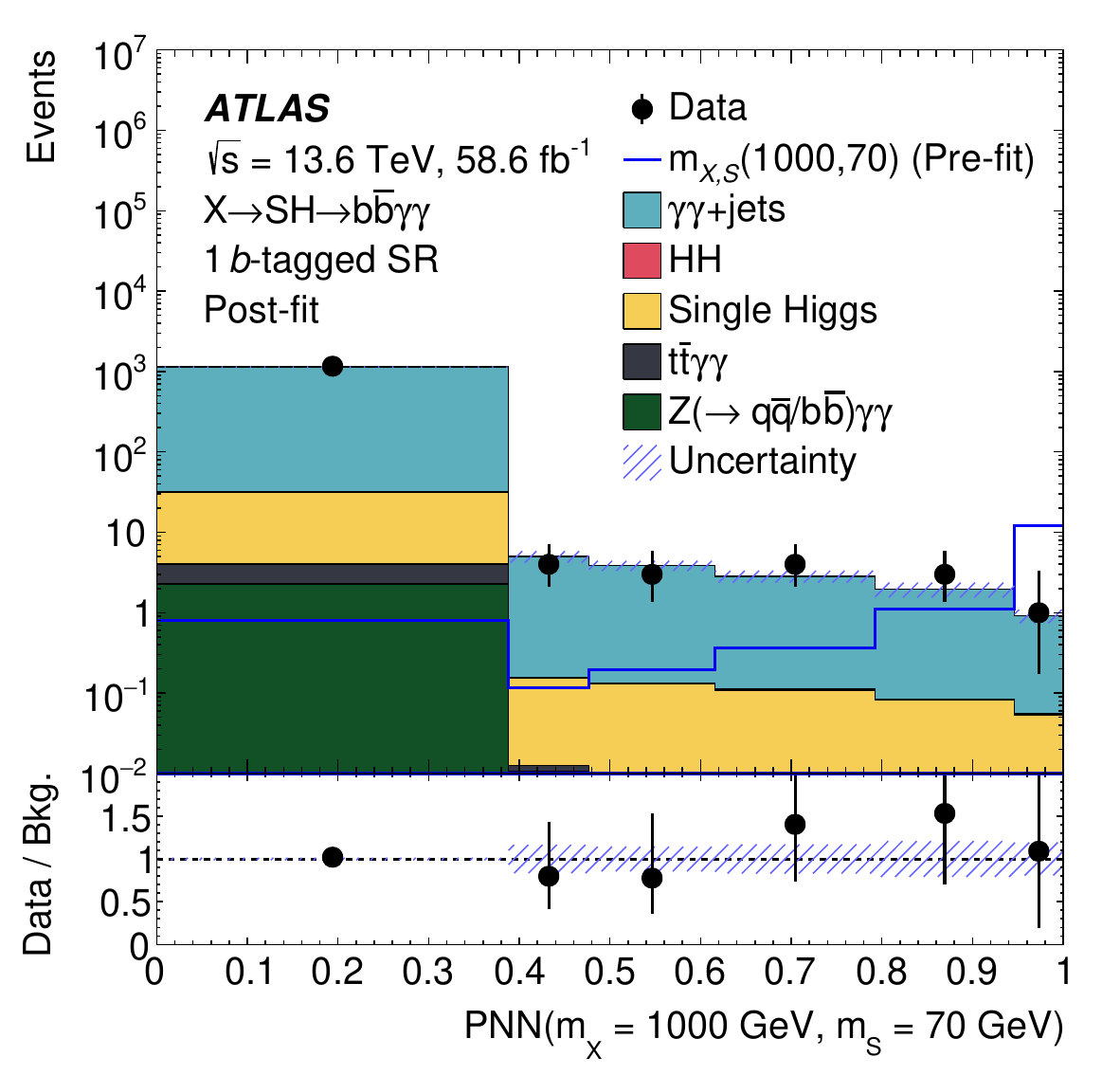}
\includegraphics[width=0.49\linewidth]{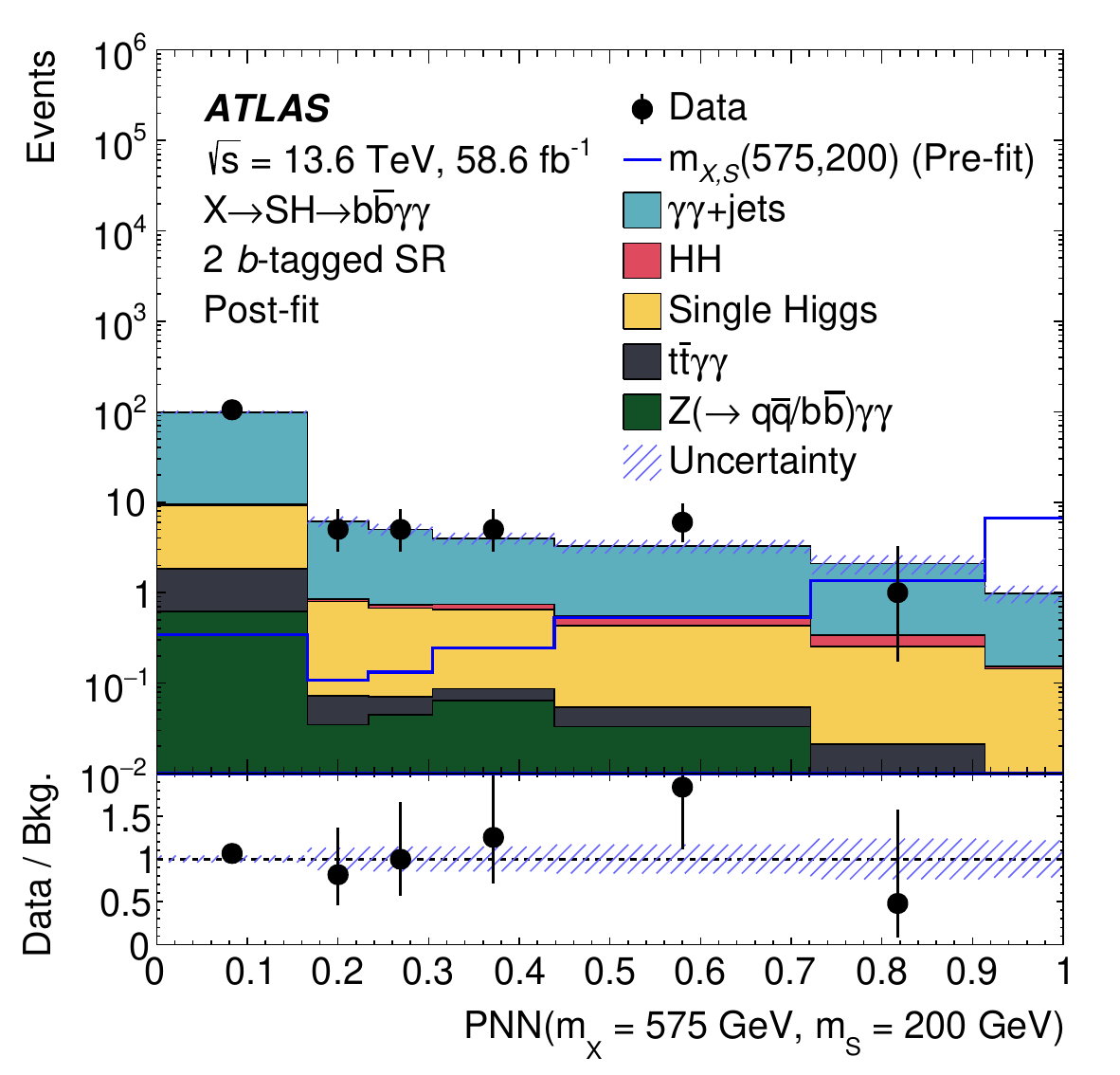}
\caption{Distributions of the PNN discriminant output after the profile-likelihood fit in the SRs for Run-3 (2022--2023, 13.6\,TeV), shown for the same signal hypotheses as Fig.~\ref{fig:ATLAS_PNN_run2}: 1 $b$-tagged SR for $m_X = 1000$\,GeV, $m_S = 70$\,GeV (left), and 2 $b$-tagged SR for $m_X = 575$\,GeV, $m_S = 200$\,GeV (right). Conventions are as in Fig.~\ref{fig:ATLAS_PNN_run2}.
From Ref.~\cite{2026140425}.}
\label{fig:ATLAS_PNN_run3}
\end{figure}

The analysis was also applied to Run-3 data (2022--2023, 13.6\,TeV). The PNN output distributions for the same representative signal hypotheses in Run-3 data are shown in Fig.~\ref{fig:ATLAS_PNN_run3}. Comparing Figs.~\ref{fig:ATLAS_PNN_run2} and~\ref{fig:ATLAS_PNN_run3} illustrates the consistent performance of the analysis across the two data-taking periods and different center-of-mass energies, and motivates the combination of both datasets for the final result.

\begin{figure}[H]
\centering
\includegraphics[width=0.49\linewidth]{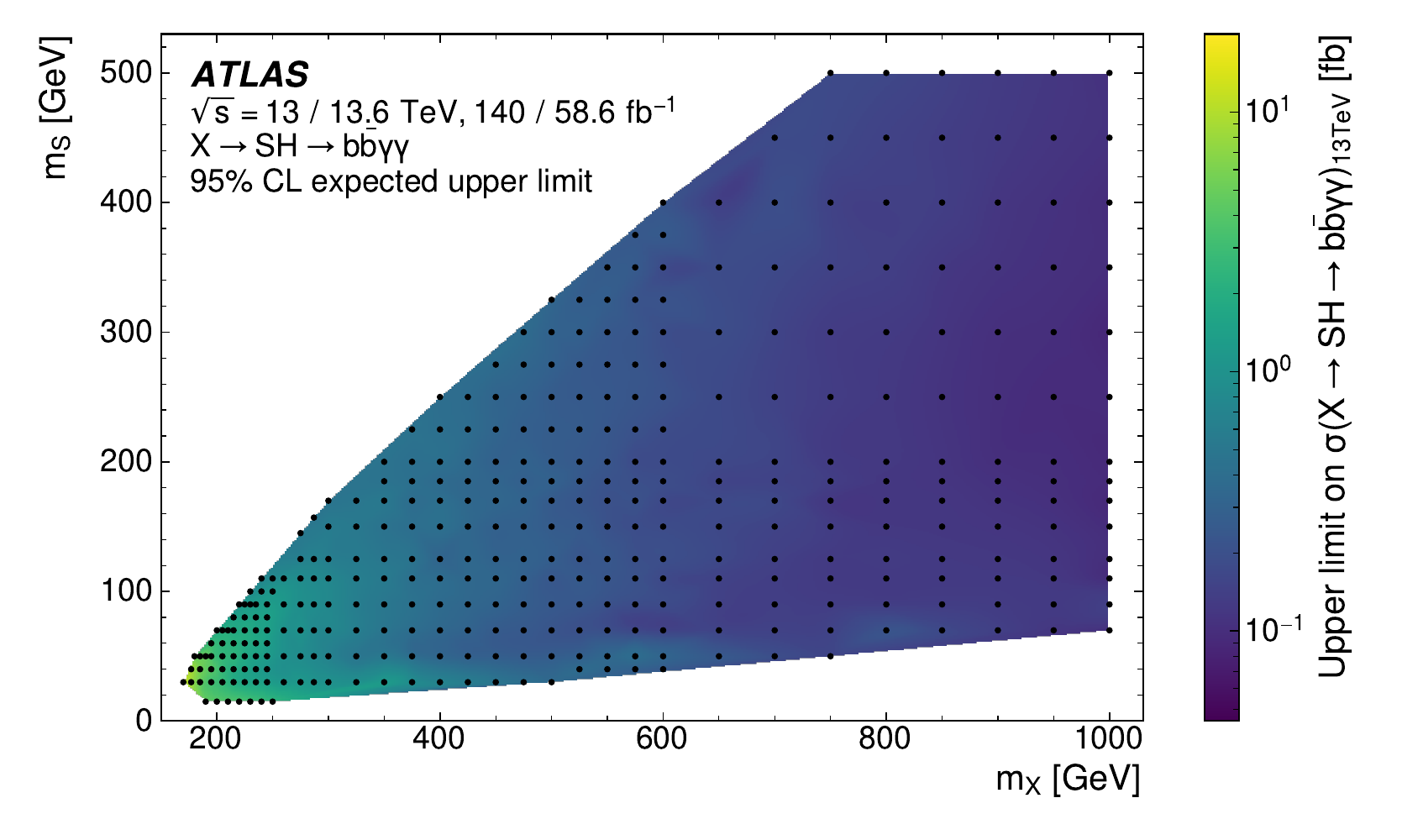}
\includegraphics[width=0.49\linewidth]{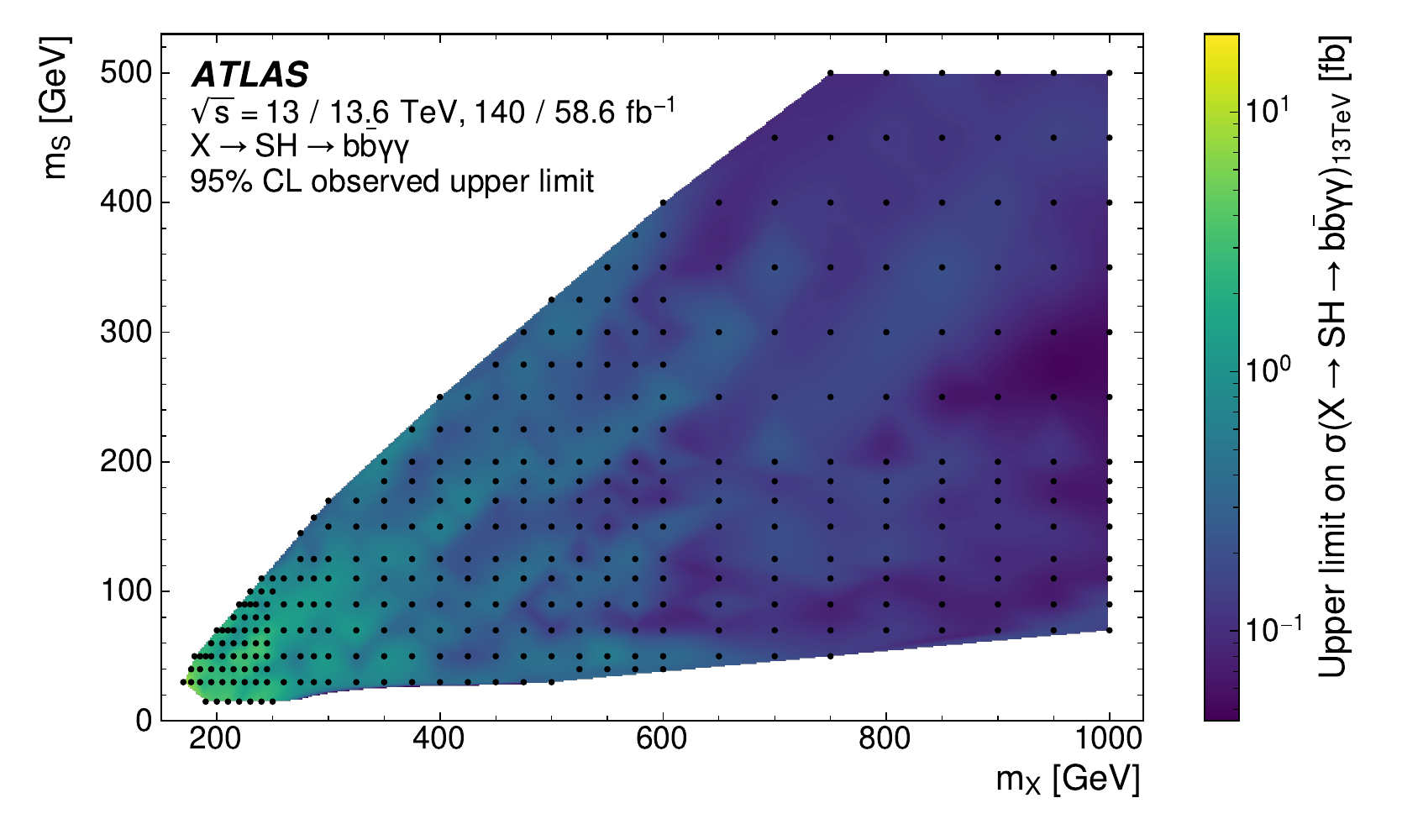}
\caption{Expected (left) and observed (right) 95\% CL upper limits on the $X\rightarrow S(\rightarrow b\bar{b})H(\rightarrow \gamma\gamma)$ cross-section, combining both the 1 $b$-tagged and 2 $b$-tagged categories and both Run-2 (13\,TeV) and Run-3 (13.6\,TeV) datasets, in the full $(m_X, m_S)$ plane.
From Ref.~\cite{2026140425}.}
\label{fig:ATLAS_bbgg_limits}
\end{figure}

Expected and observed limits using combined data from Run-2 and Run-3 are shown in Fig.~\ref{fig:ATLAS_bbgg_limits}. The limits are presented in the full two-dimensional $(m_X, m_S)$ plane, combining the 1 $b$-tagged and 2 $b$-tagged categories. No significant excess above the SM background is observed.

\subsection{$X\rightarrow Y(\rightarrow \gamma\gamma )H(\rightarrow b\bar{b})$, SM H}

\begin{figure}[H]
\vspace{-5mm}
\centering
\includegraphics[width=0.49\linewidth]{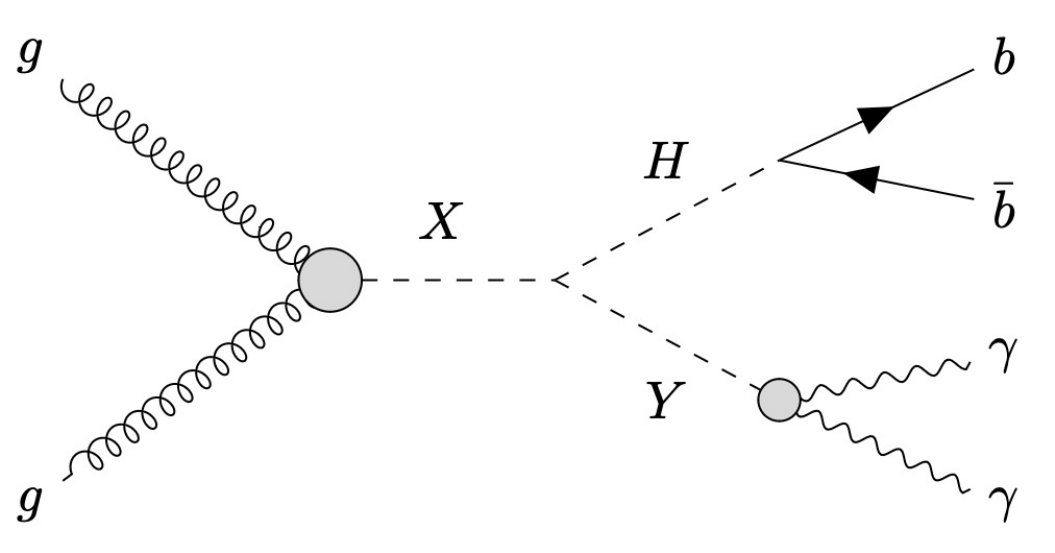}
\caption{Feynman diagram for the production of the BSM resonance $X$ decaying to two scalars: the SM Higgs boson $H$ (with $H\rightarrow b\bar{b}$) and the BSM scalar $Y$ (with $Y\rightarrow \gamma\gamma$). From Ref.~\cite{CMS:2025qit}.}
\label{fig:CMS_bbgg}
\end{figure}

The process
$X\rightarrow Y(\rightarrow \gamma\gamma )H(\rightarrow b\bar{b})$ has been searched for in the final state with two $b$-quarks and two photons, based on Run-2 data at 13\,TeV~\cite{CMS:2025qit}. This search is complementary to the one described in Sec.~2.1: here the SM Higgs boson $H$ decays into a pair of $b$-quarks, while the BSM scalar $Y$ decays into a di-photon pair. The di-photon invariant mass provides a narrow peak at the unknown mass $m_Y$, which is used as the primary discriminating variable.
The corresponding Feynman diagram is shown in Fig.~\ref{fig:CMS_bbgg}.

\begin{figure}[H]
\vspace{-5mm}
\centering
\includegraphics[width=0.49\linewidth]{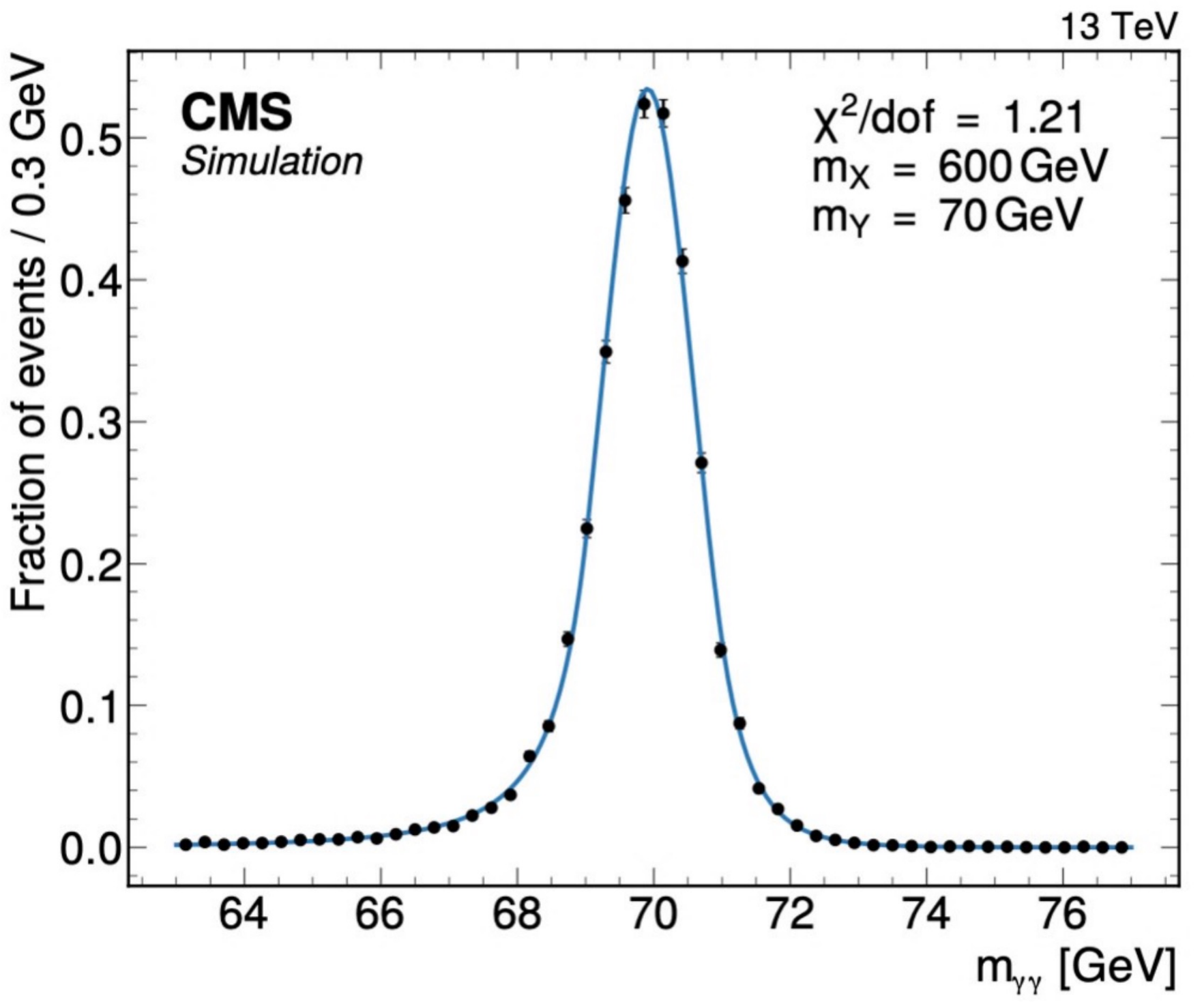}\includegraphics[width=0.49\linewidth]{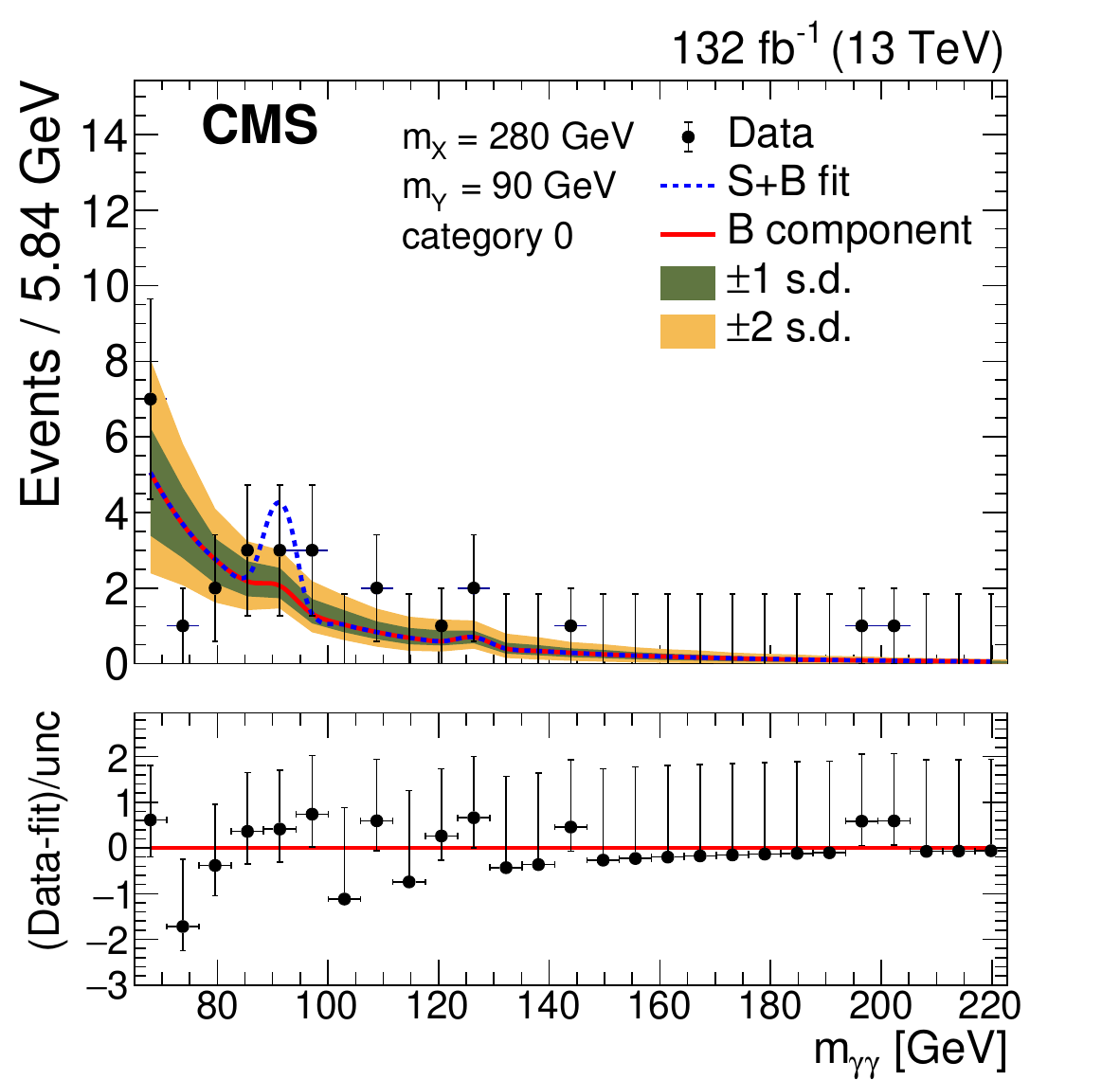}
\caption{Left:
Simulated di-photon invariant mass distribution for the signal process at $m_X=600$\,GeV and $m_Y=70$\,GeV in the most sensitive SR (Run-2, 13\,TeV), normalized to unity. The acronym ``dof'' denotes the number of degrees of freedom of the parametric model.
Right: Background-only fit (solid red) and signal+background fit (dashed blue) for the mass hypothesis $m_X=280$\,GeV, $m_Y=90$\,GeV in the second most sensitive SR. Points in the lower panel show the difference between data and the background-only fit, divided by the average data uncertainty. The background functional form is determined by the maximum-likelihood fit.
From Ref.~\cite{CMS:2025qit}.}
\label{fig:CMS_gg_mass}
\vspace{-5mm}
\end{figure}

The mass of the scalar $Y$ corresponds to the invariant mass of the photon pair. The signal model for the di-photon invariant mass is parameterized analytically, enabling sensitivity across a continuous range of $m_Y$ hypotheses. The parameterized signal model and a representative background-plus-signal fit to data are shown in Fig.~\ref{fig:CMS_gg_mass}. The signal model is shown for the hypothesis $m_X=600$\,GeV, $m_Y=70$\,GeV (left), and the background-plus-signal fit for $m_X=280$\,GeV, $m_Y=90$\,GeV (right). The background functional form is determined by the maximum-likelihood fit to data.

\begin{figure}[H]
\centering
\includegraphics[width=0.49\linewidth]{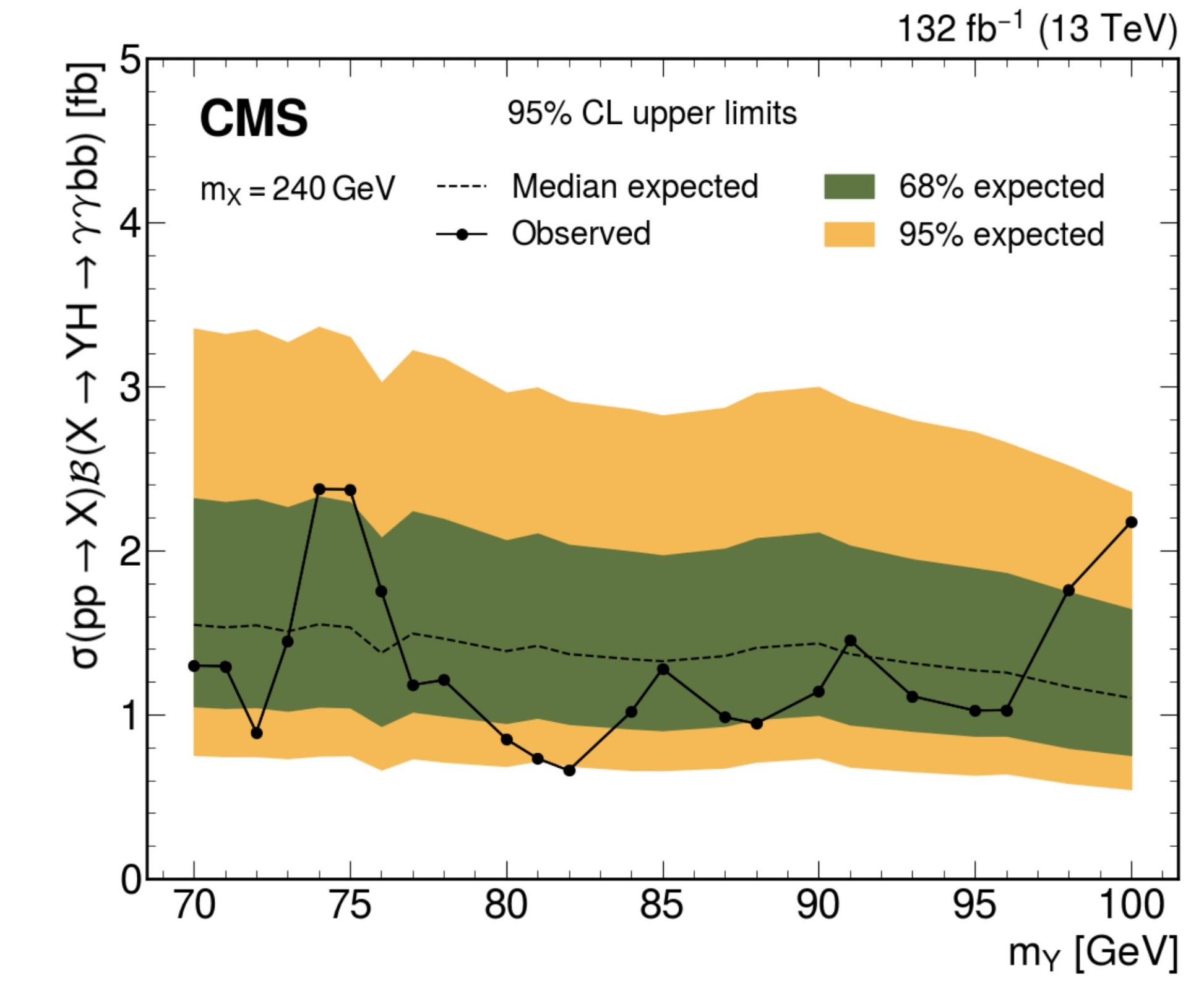}
\includegraphics[width=0.49\linewidth]{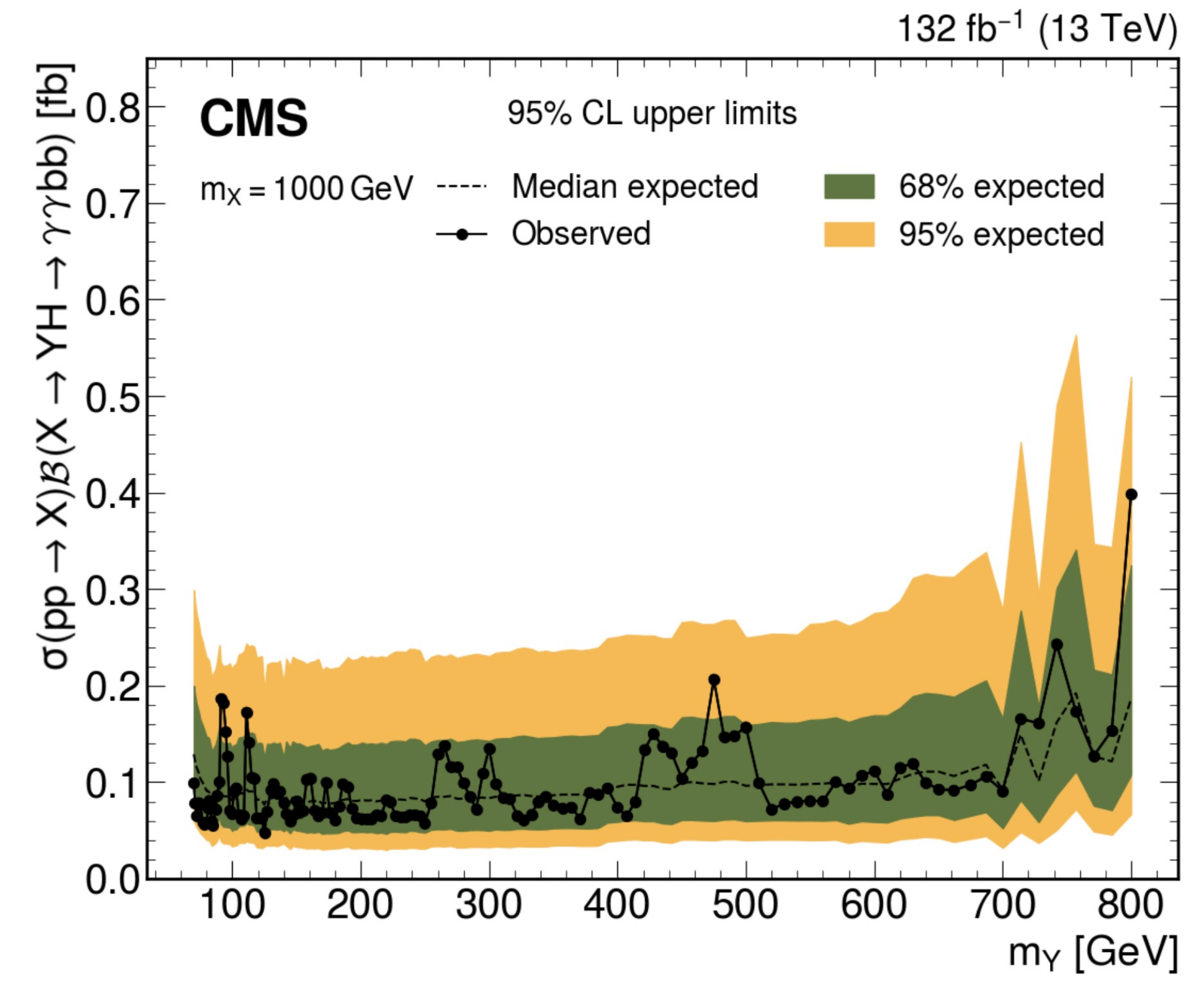}
\caption{Observed (solid lines) and expected (dashed lines) 95\% CL upper limits on the product $\sigma\mathcal{B}$
for the $X\rightarrow Y(\rightarrow \gamma\gamma )H(\rightarrow b\bar{b})$ signal, shown for different $m_Y$ hypotheses at the lowest mass $m_X = 240$\,GeV (left) and highest mass $m_X = 1000$\,GeV (right) (Run-2, 13\,TeV). The inner (green) band and the outer (yellow) band indicate the 68\% and 95\% intervals of the expected limit distribution under the background-only hypothesis.
{Each curve corresponds to a fixed $m_Y$ hypothesis. The full set of  $m_Y$ values is scanned continuously using the parameterized signal model. No significant excess above the SM background expectation is observed across the explored mass plane.}
From Ref.~\cite{CMS:2025qit}.}
\label{fig:CMS_bbgg_limits}
\end{figure}

%\clearpage
Expected and observed limits are shown in Fig.~\ref{fig:CMS_bbgg_limits} as a function of $m_X$ for two $m_Y$ mass ranges: 70--100\,GeV and 100--800\,GeV{, evaluated at the two extreme values $m_X = 240$\,GeV (left) and $m_X = 1000$\,GeV (right)}. No significant excess is observed.

\subsection{$X\rightarrow S(\rightarrow VV)H(\rightarrow \tau\tau)$, SM H}

A search for the process
$X\rightarrow S(\rightarrow VV)H(\rightarrow \tau\tau)$ was performed in the final state with two vector bosons ($W$ or $Z$) and two tau leptons~\cite{ATLAS:2023tkl}, using the full ATLAS Run-2 dataset at 13\,TeV. In this search the SM Higgs boson decays into a pair of tau leptons and the BSM scalar $S$ decays into two vector bosons. Six sub-channels are investigated, targeting different leptonic and hadronic decays of the $W$ and $Z$ bosons and the tau pairs: $V=W$ with 1-lepton ($e,\mu$), $V=W$ with 2-leptons, and $V=Z$ with 2-leptons, each combined with di-hadronic tau decays ($\tau_{\text{had}}\tau_{\text{had}}$). A BDT is trained in each sub-channel to exploit the correlations among the input kinematic variables.

\begin{figure}[H]
\vspace{-5mm}
\centering
\includegraphics[width=0.49\linewidth]{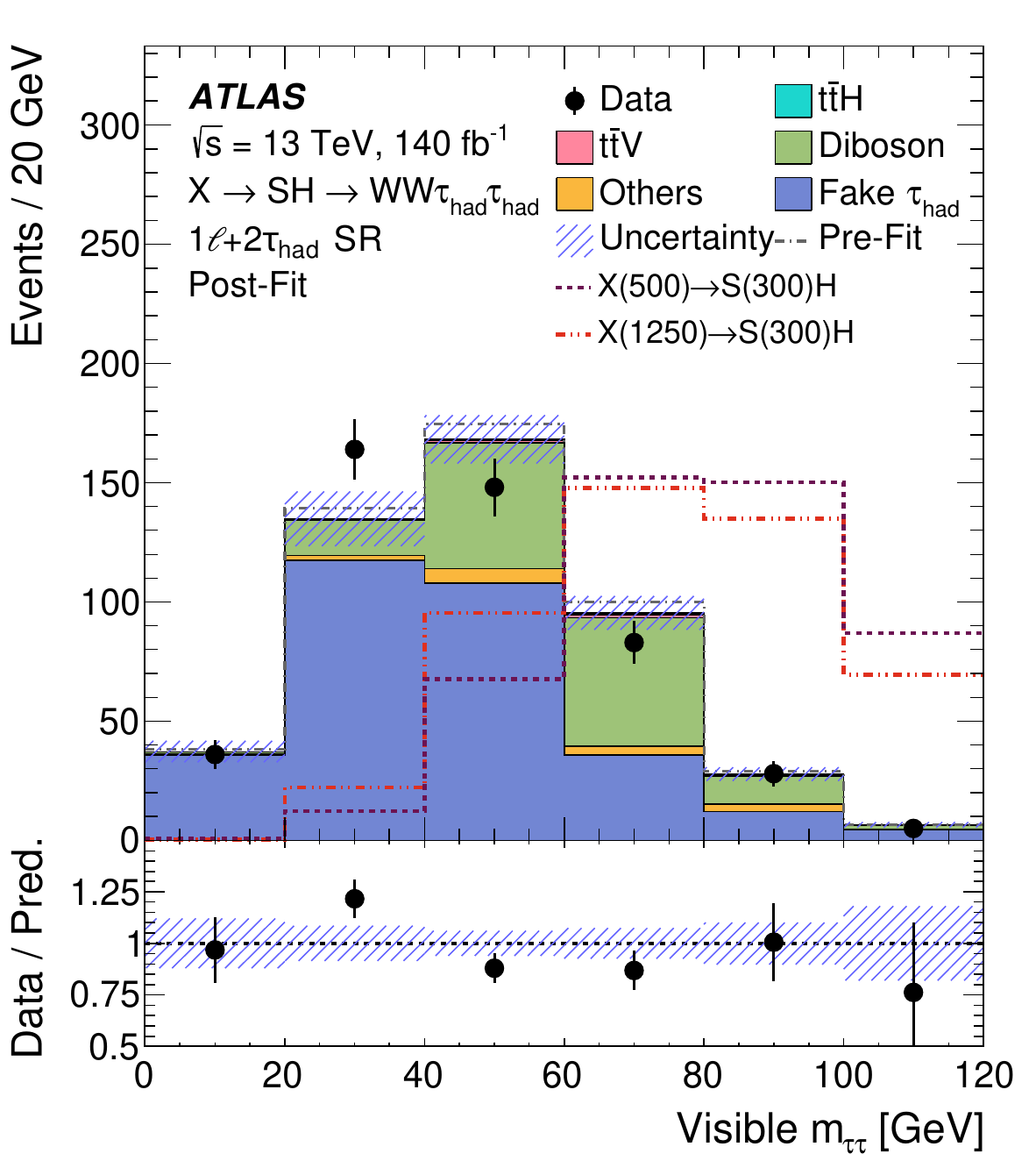}
\includegraphics[width=0.49\linewidth]{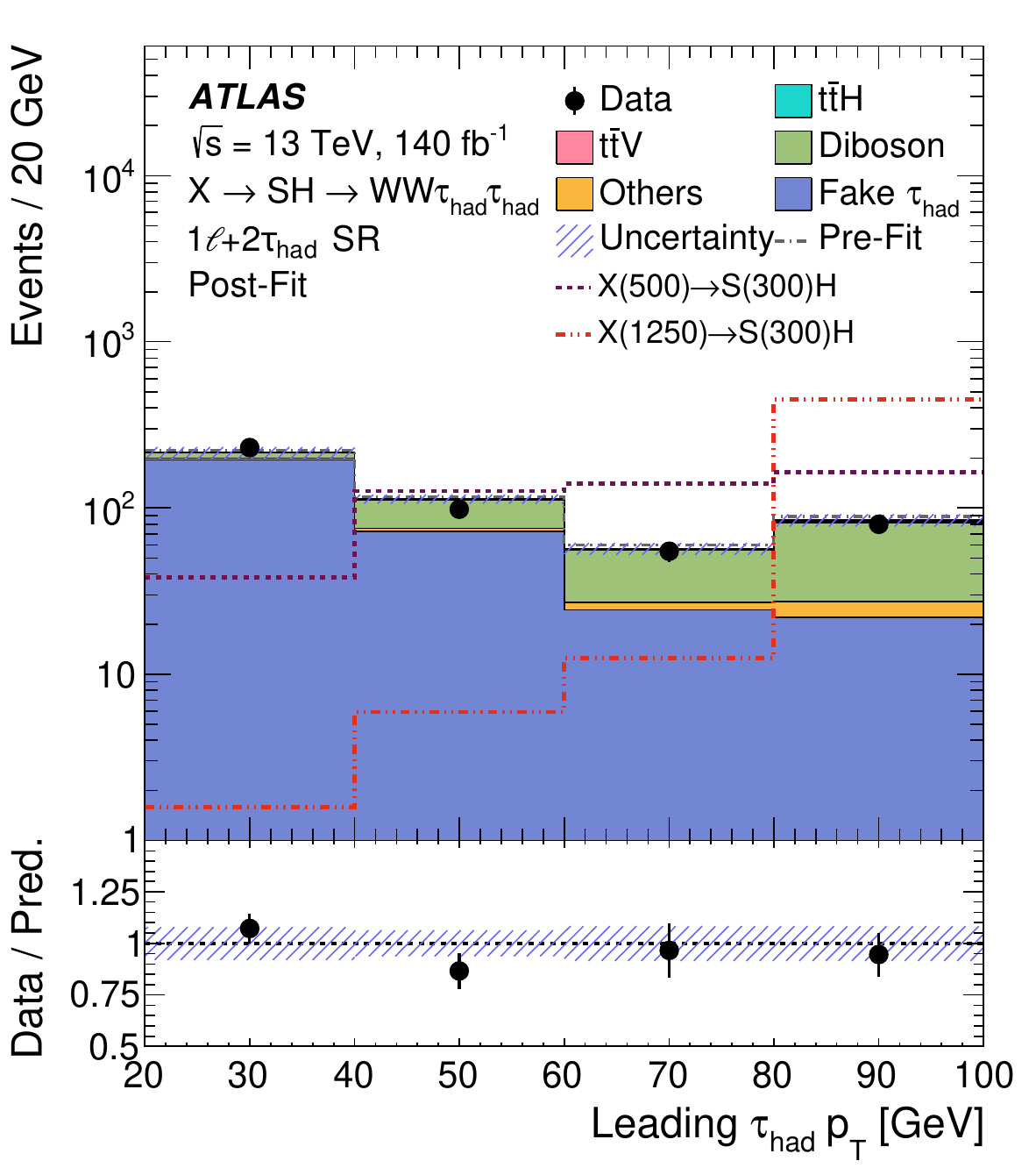}
\vspace{-2mm}
\caption{Post-fit kinematic distributions in the
$\text{WW}\,1\ell\,2\tau_{\text{had}}$ signal region (Run-2, 13\,TeV): $m_{\tau\tau}$ (left), leading
$\tau_{\text{had}}$ $p_{\text{T}}$ (right).
The purple (red) dashed line represents the signal for $m_X = 500\,\text{GeV}$, $m_S = 300\,\text{GeV}$ ($m_X = 1250\,\text{GeV}$, $m_S = 300\,\text{GeV}$),
normalized to the total background.
The uncertainty band includes both systematic and statistical uncertainties.
The lower panel shows the ratio of data to the post-fit background prediction.
From Ref.~\cite{ATLAS:2023tkl}}
\label{fig:ATLAS_VV_lepton}
\end{figure}

In the channel $V=W$ and 1-lepton ($e,\mu$), the reconstructed invariant mass of the tau pairs ($m_{\tau\tau}$) and the leading tau transverse momentum is shown in Fig.~\ref{fig:ATLAS_VV_lepton}.

\begin{figure}[H]
\vspace{-5mm}
\centering
\includegraphics[width=0.49\linewidth]{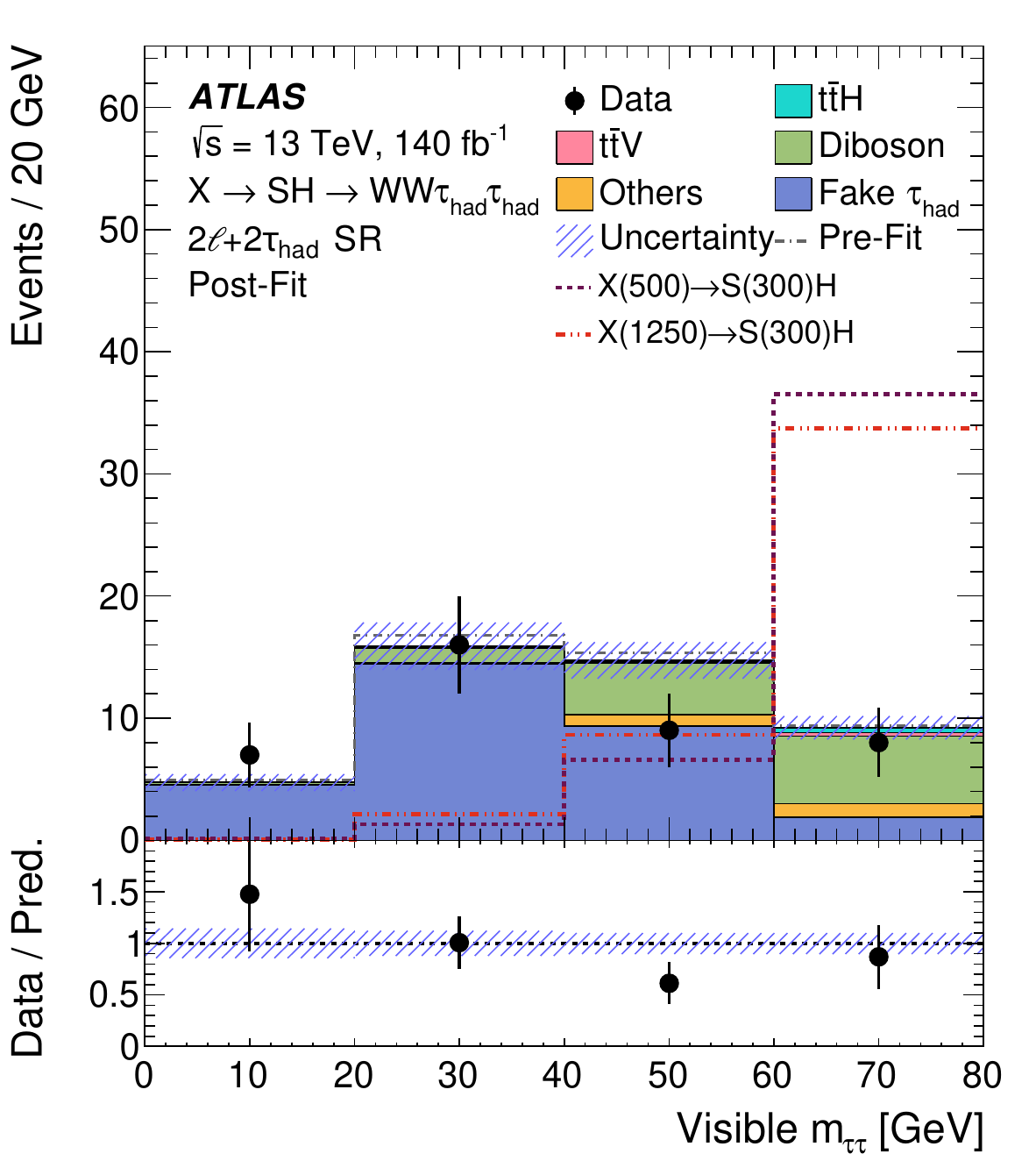}
\includegraphics[width=0.49\linewidth]{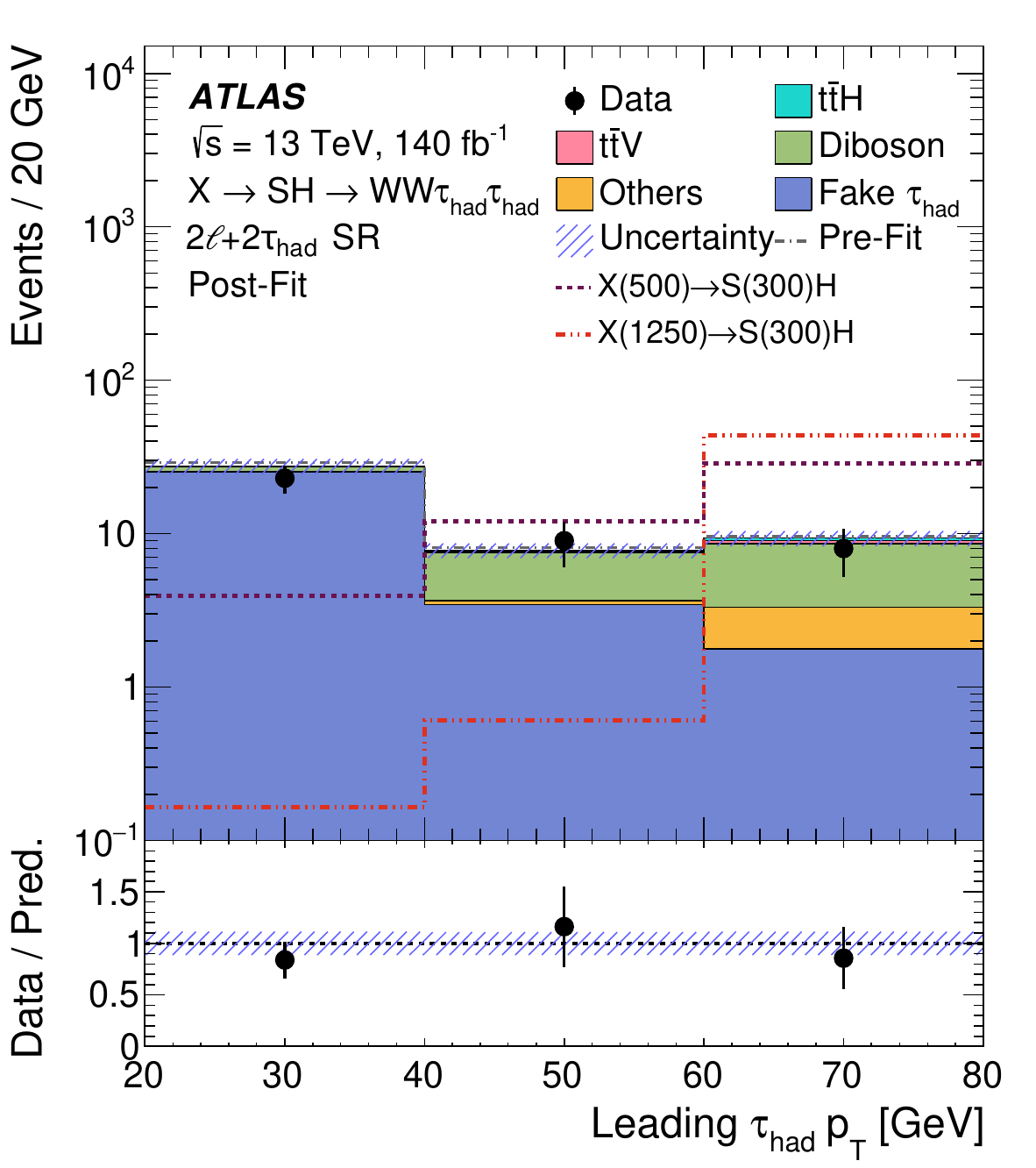}
\vspace{-2mm}
\caption{
Post-fit kinematic distributions in the
$\text{WW}\,2\ell\,2\tau_{\text{had}}$ signal region (Run-2, 13\,TeV): $m_{\tau\tau}$ (left), leading
$\tau_{\text{had}}$ $p_{\text{T}}$ (right).
Conventions are as in Fig.~\ref{fig:ATLAS_VV_lepton}.
From Ref.~\cite{ATLAS:2023tkl}}
\label{fig:ATLAS_VV_lepton2}
\end{figure}

In the channel $V=W$ and 2-lepton ($e,\mu$), the corresponding distributions are shown in Fig.~\ref{fig:ATLAS_VV_lepton2}.

\begin{figure}[H]
\vspace{-4mm}
\centering
\includegraphics[width=0.49\linewidth]{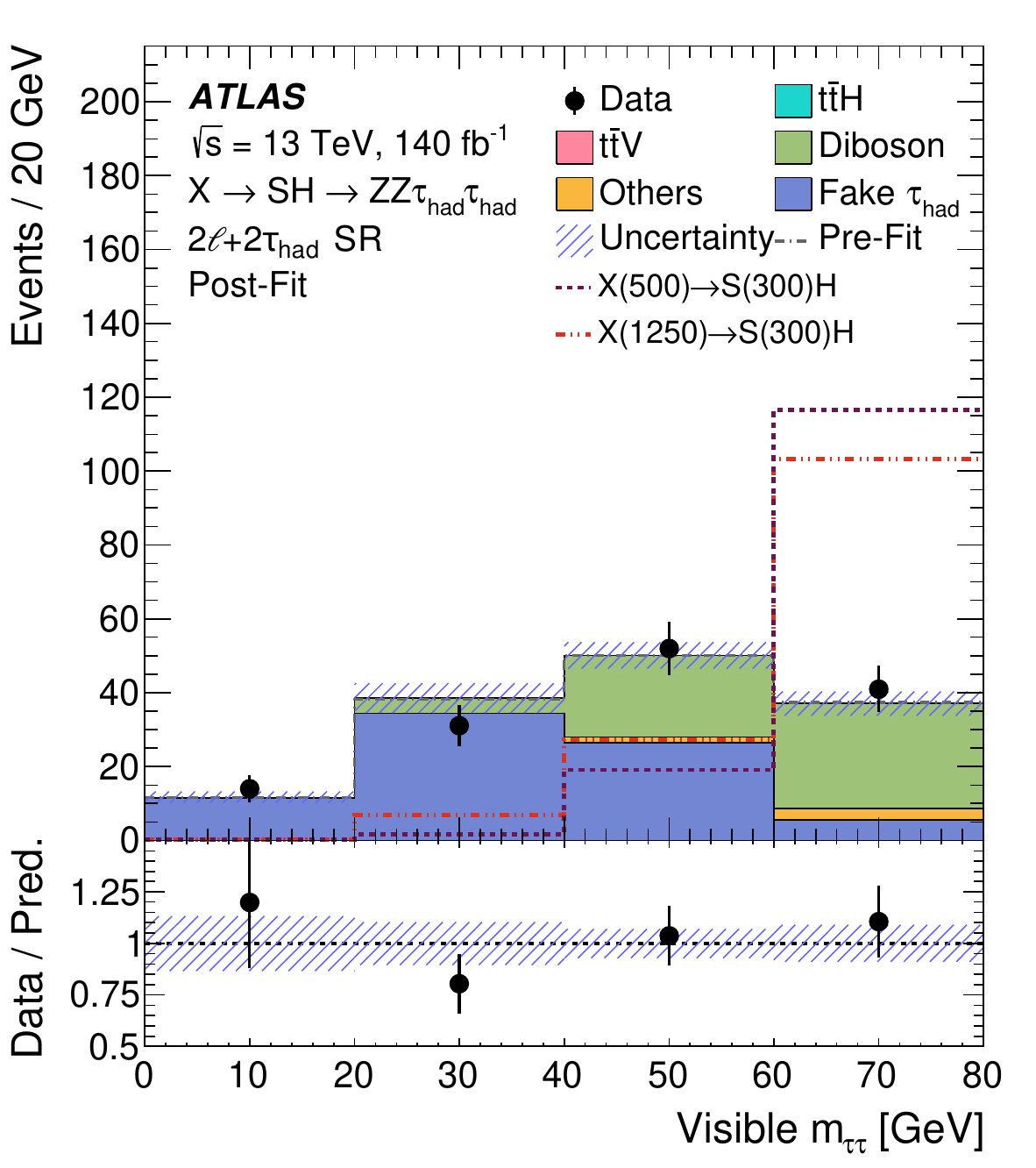}
\includegraphics[width=0.49\linewidth]{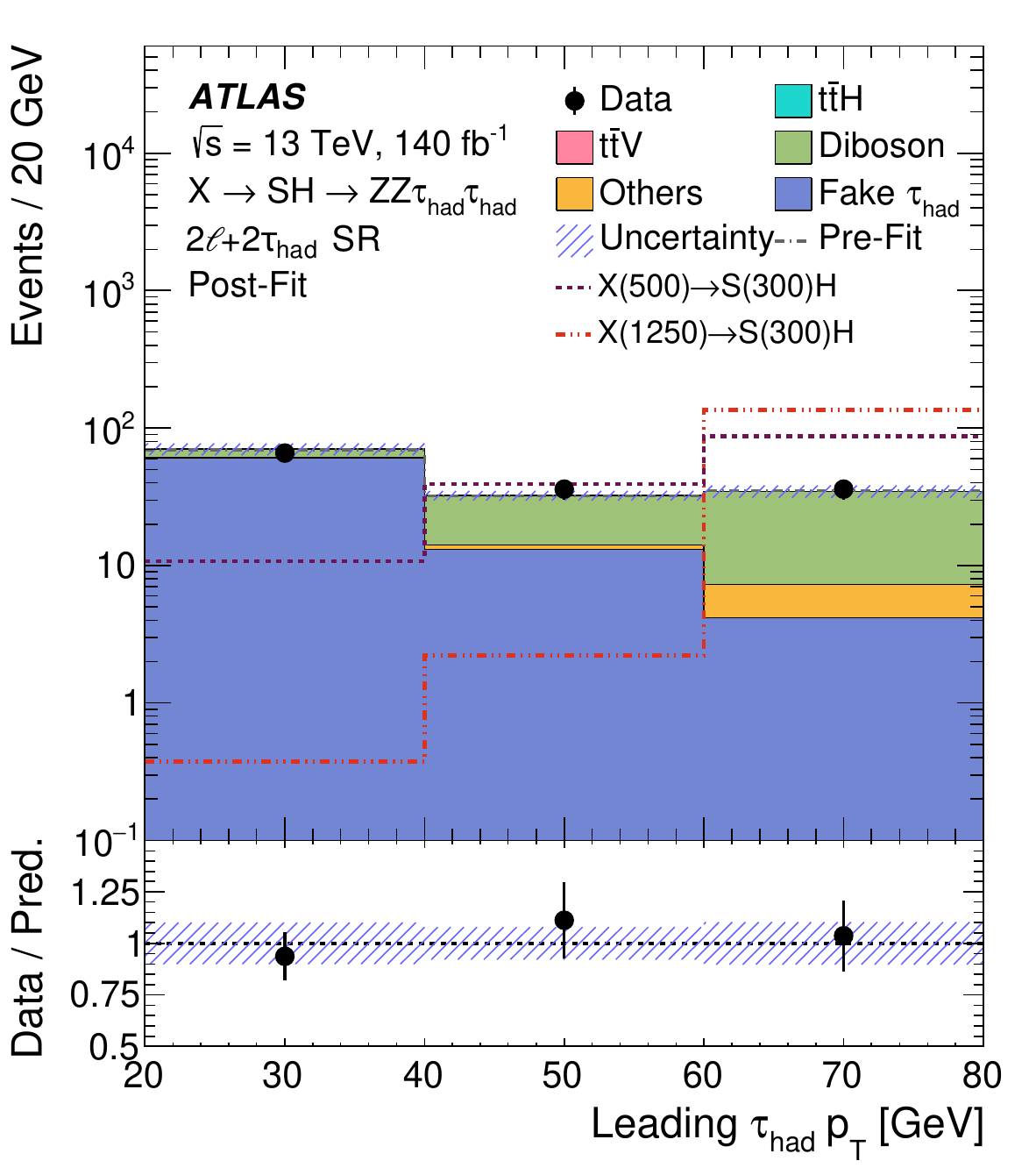}
\caption{
Post-fit kinematic distributions in the
$\text{ZZ}\,2\ell\,2\tau_{\text{had}}$ signal region (Run-2, 13\,TeV): $m_{\tau\tau}$ (left), leading
$\tau_{\text{had}}$ $p_{\text{T}}$ (right).
Conventions are as in Fig.~\ref{fig:ATLAS_VV_lepton}.
From Ref.~\cite{ATLAS:2023tkl}}
\label{fig:ATLAS_VV_lepton3}
\vspace{-5mm}
\end{figure}

In the channel $V=Z$ and 2-lepton ($e,\mu$), the corresponding distributions are shown in Fig.~\ref{fig:ATLAS_VV_lepton3}.

\begin{figure}[H]
\vspace{-5mm}
\centering
\includegraphics[width=0.49\linewidth]{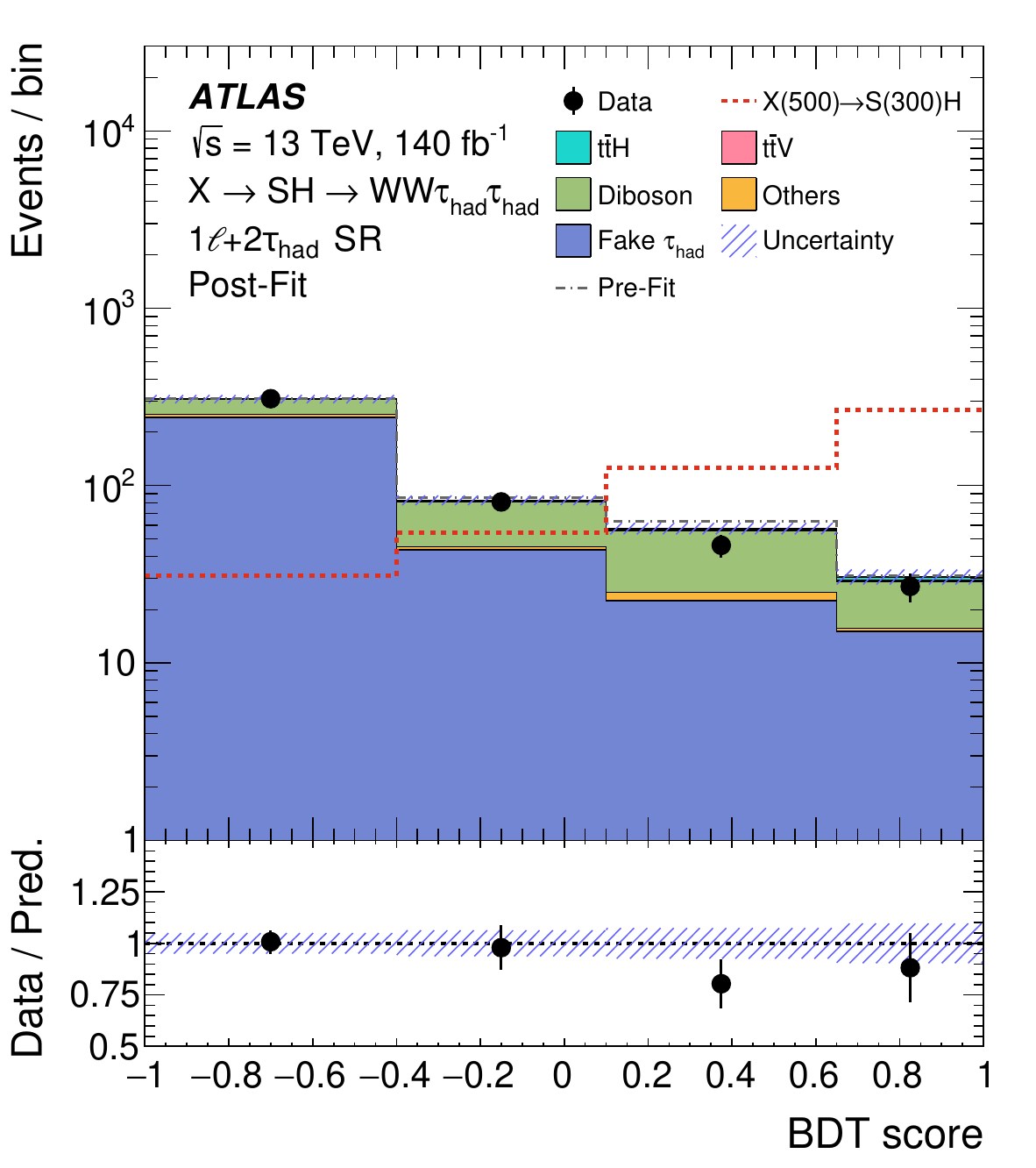}
\includegraphics[width=0.49\linewidth]{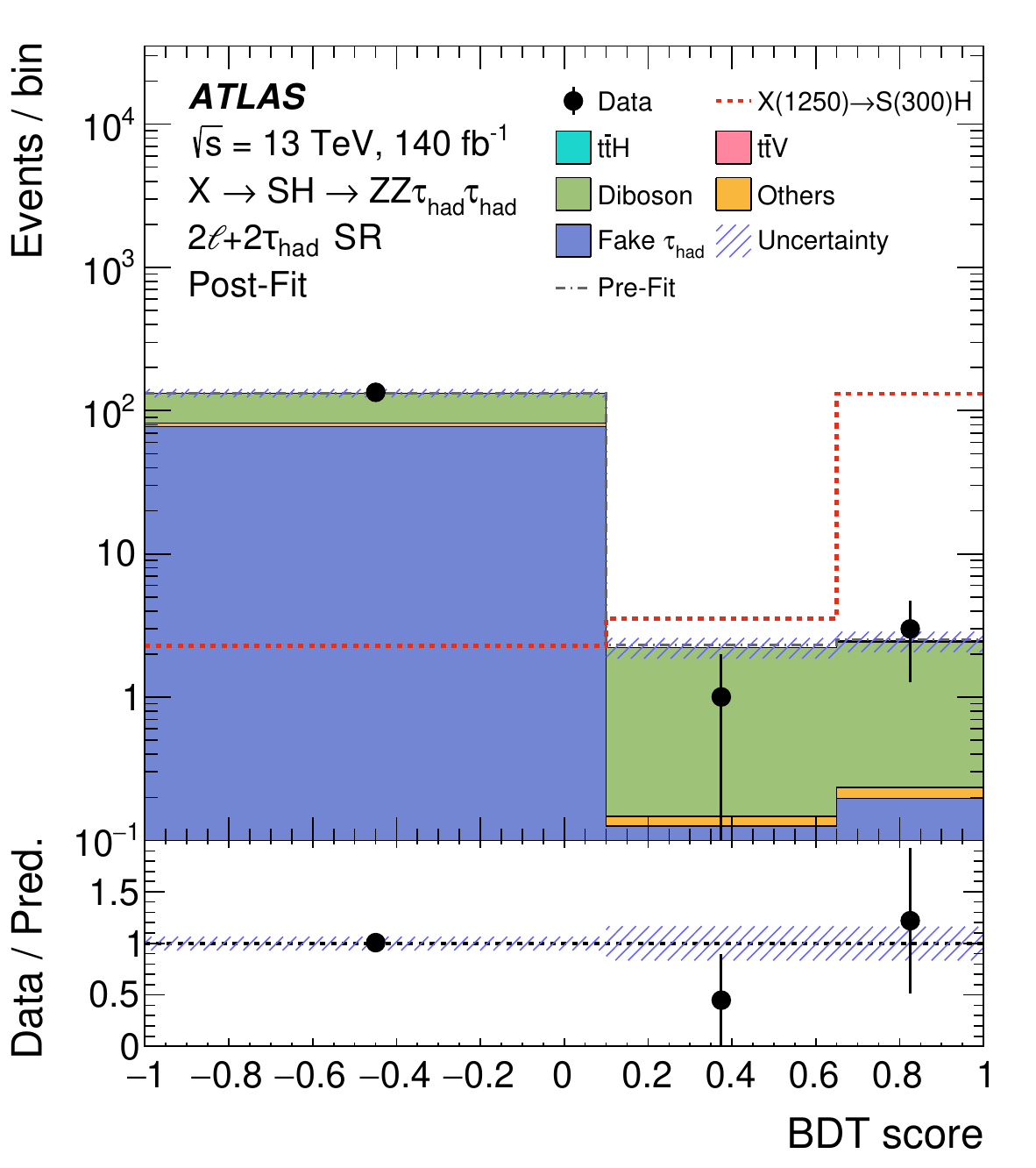}
\caption{BDT output distributions from a background-only fit
to data for the $X \to SH$ search (Run-2, 13\,TeV), shown for
$m_X = 1250\,\text{GeV}$, $m_S = 300\,\text{GeV}$ in the
(left) $\text{WW}\,1\ell\,2\tau_{\text{had}}$ and
(right) $\text{ZZ}\,2\ell\,2\tau_{\text{had}}$ channels.
The red dashed line represents the signal normalized to the total background.
The uncertainty band includes both systematic and statistical uncertainties.
From Ref.~\cite{ATLAS:2023tkl}}
\label{fig:ATLAS_VV_BDT}
\vspace{5mm}
\end{figure}

Out of the six investigated channels, the BDT output is shown in Fig.~\ref{fig:ATLAS_VV_BDT} for two representative cases: $V=W$ with 1-lepton ($e,\mu$) and $V=Z$ with 2-leptons. The BDT combines multiple input kinematic variables into a single discriminant, providing improved signal-to-background separation compared to any single variable.

\begin{figure}[H]
\centering
\includegraphics[width=0.7\linewidth]{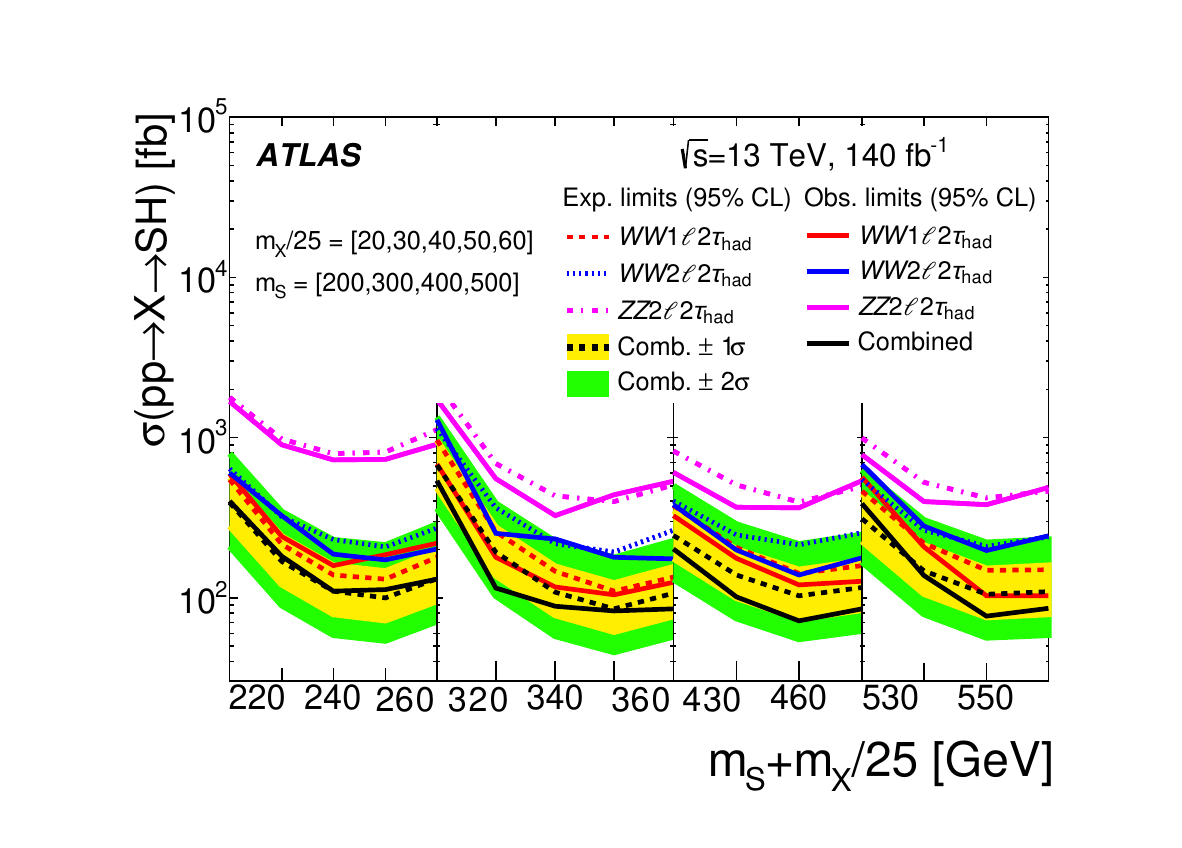}
\vspace{-6mm}
\caption{Observed and expected 95\% CL upper limits on
$\sigma(pp \to X \to SH)$ (Run-2, 13\,TeV), obtained from the three individual channels and their
combination.
From Ref.~\cite{ATLAS:2023tkl}}
\label{fig:ATLAS_VV_limits}
\end{figure}

Expected and observed limits from the combination of all six sub-channels are shown in Fig.~\ref{fig:ATLAS_VV_limits}. The results from the three individual lepton channels are also displayed separately to show their relative contributions to the combined sensitivity. { The $\text{WW}\,1\ell\,2\tau_{\text{had}}$ channel provides the most sensitive individual contribution. No significant excess above the SM background expectation is observed in any channel or in the combination.}

\subsection{$X\rightarrow Y(\rightarrow b\bar{b})H(\rightarrow b\bar{b})$, SM H}

A search for the process
$X\rightarrow Y(\rightarrow b\bar{b})H(\rightarrow b\bar{b})$ was performed in the four-$b$-quark final state, using CMS 2018 Run-2 data at 13\,TeV~\cite{Hayrapetyan:2960489}. Both the SM Higgs boson $H$ and the BSM scalar $Y$ decay into $b\bar{b}$ pairs, giving a highly boosted topology at large $m_X$. Events are classified into signal regions SR(4$b$) requiring four $b$-tagged jets, and a SR(3$b$) control sample with three $b$-tagged jets used to estimate the multijet background via a BDT-based reweighting procedure.

\begin{figure}[H]
\centering
\includegraphics[width=0.49\linewidth]{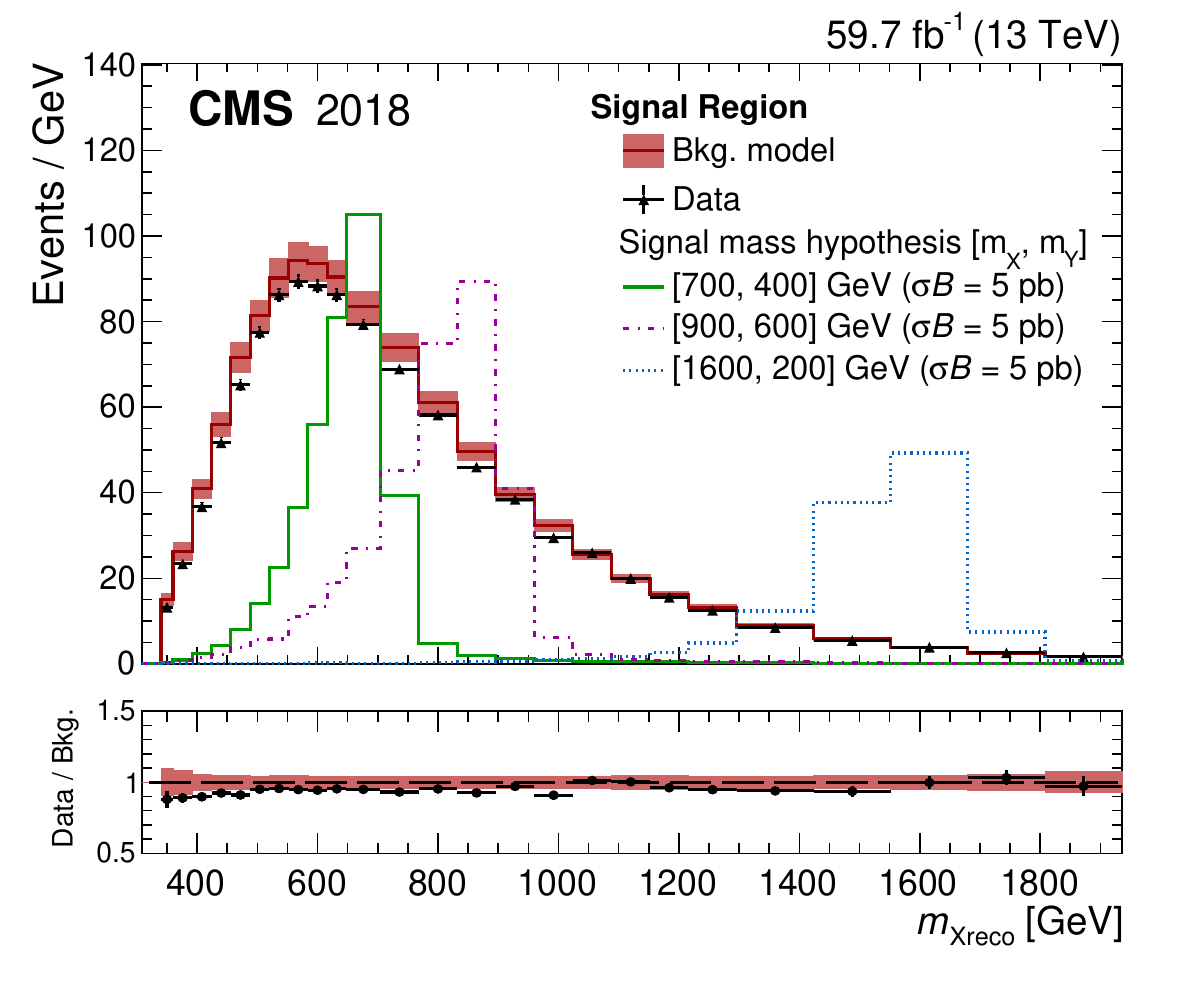}
\includegraphics[width=0.49\linewidth]{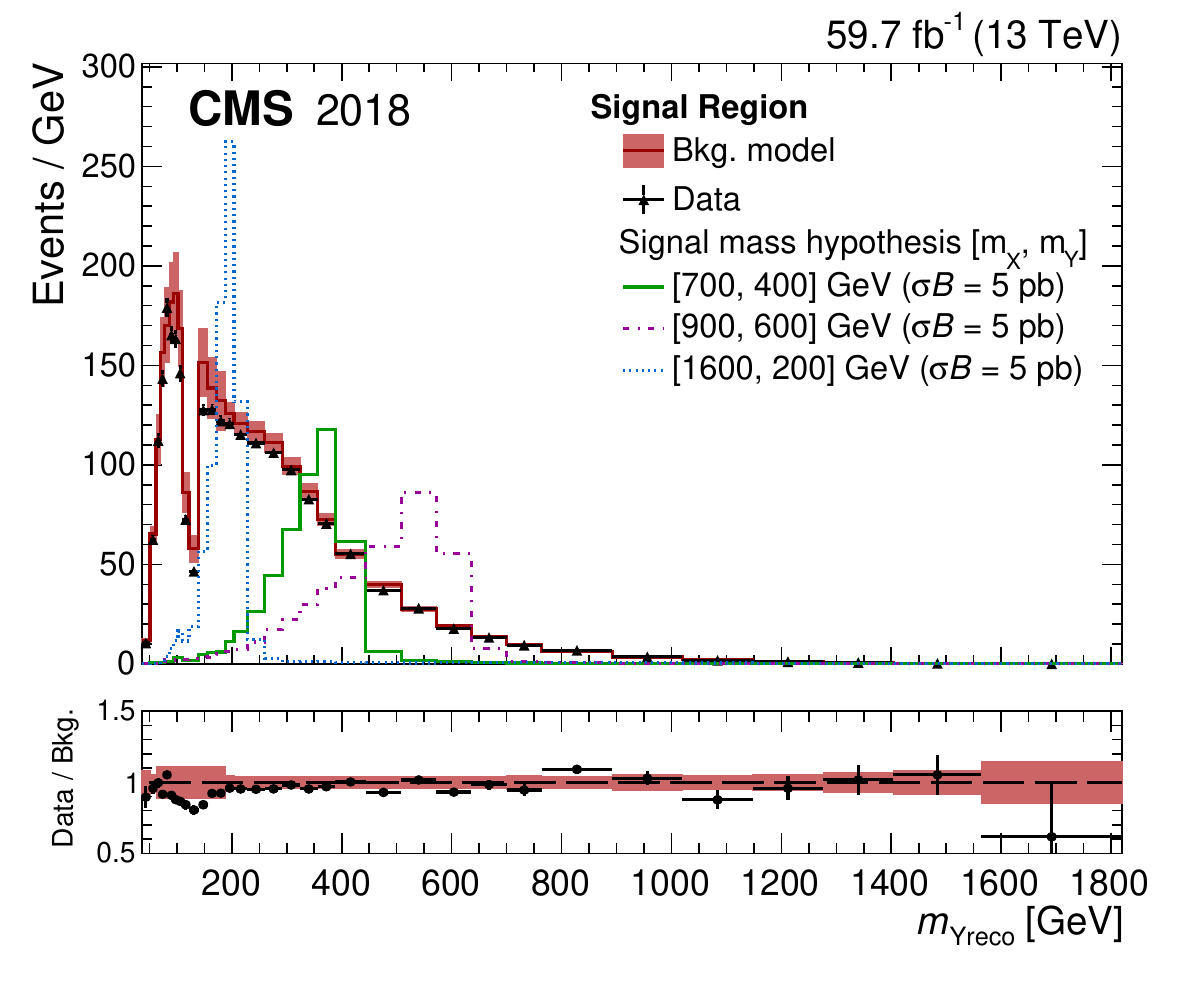}
\caption{Event distributions in the signal regions for the 2018 CMS dataset (Run-2, 13\,TeV): reconstructed $m_X^{\text{reco}}$
(left) and $m_Y^{\text{reco}}$ (right).
The SR(4$b$) data are shown in black; the BDT-reweighted
SR(3$b$) background model is shown in red.
Three signal mass hypotheses are overlaid, scaled to $\sigma = 5\,\text{pb}$.
The ratios of SR(3$b$) to SR(4$b$) are shown in the lower panels.
From Ref.~\cite{Hayrapetyan:2960489}.}
\label{fig:CMS_bbbb}
\end{figure}

\begin{figure}[H]
\centering
\includegraphics[width=0.49\linewidth]{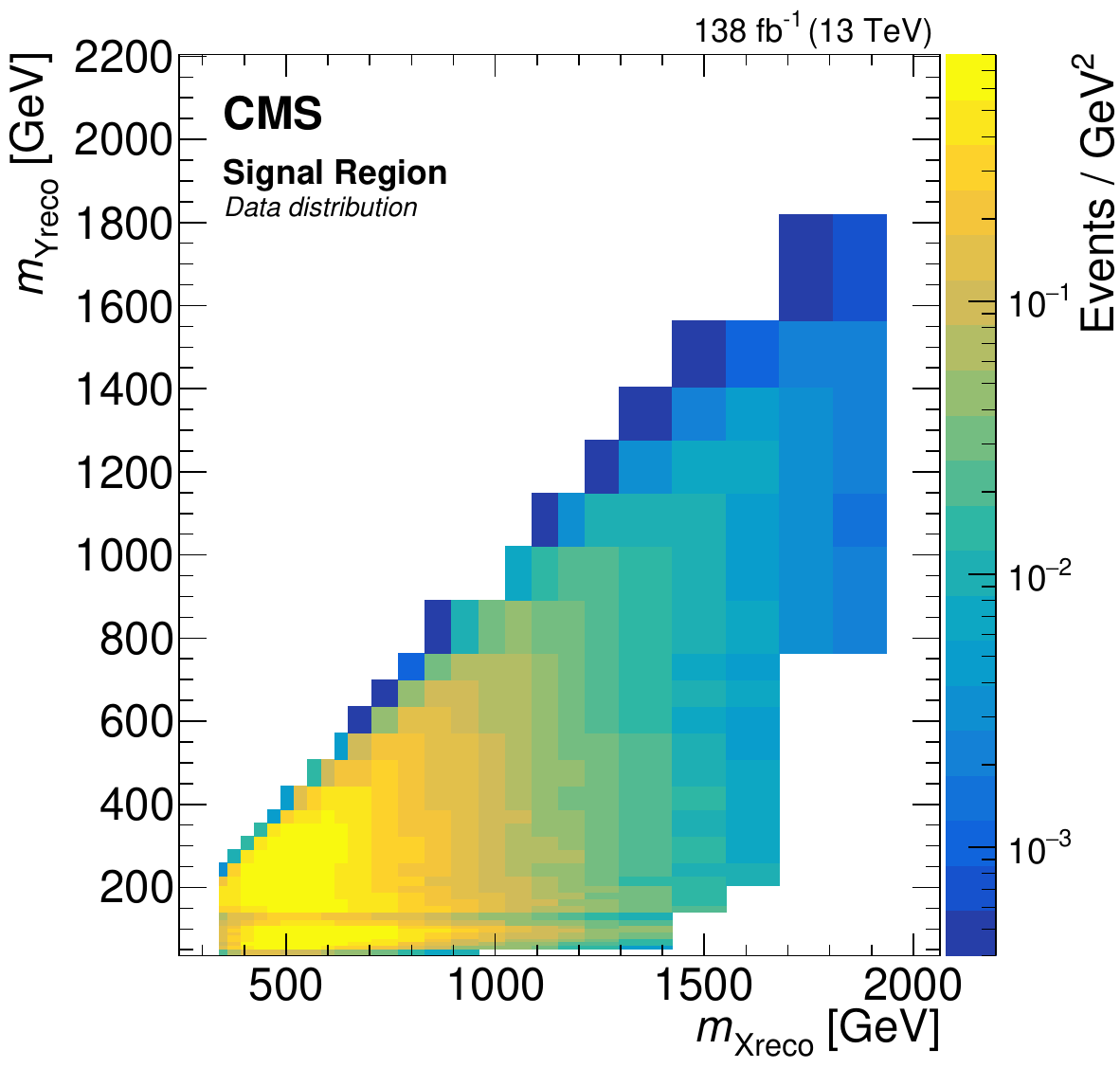}
\includegraphics[width=0.49\linewidth]{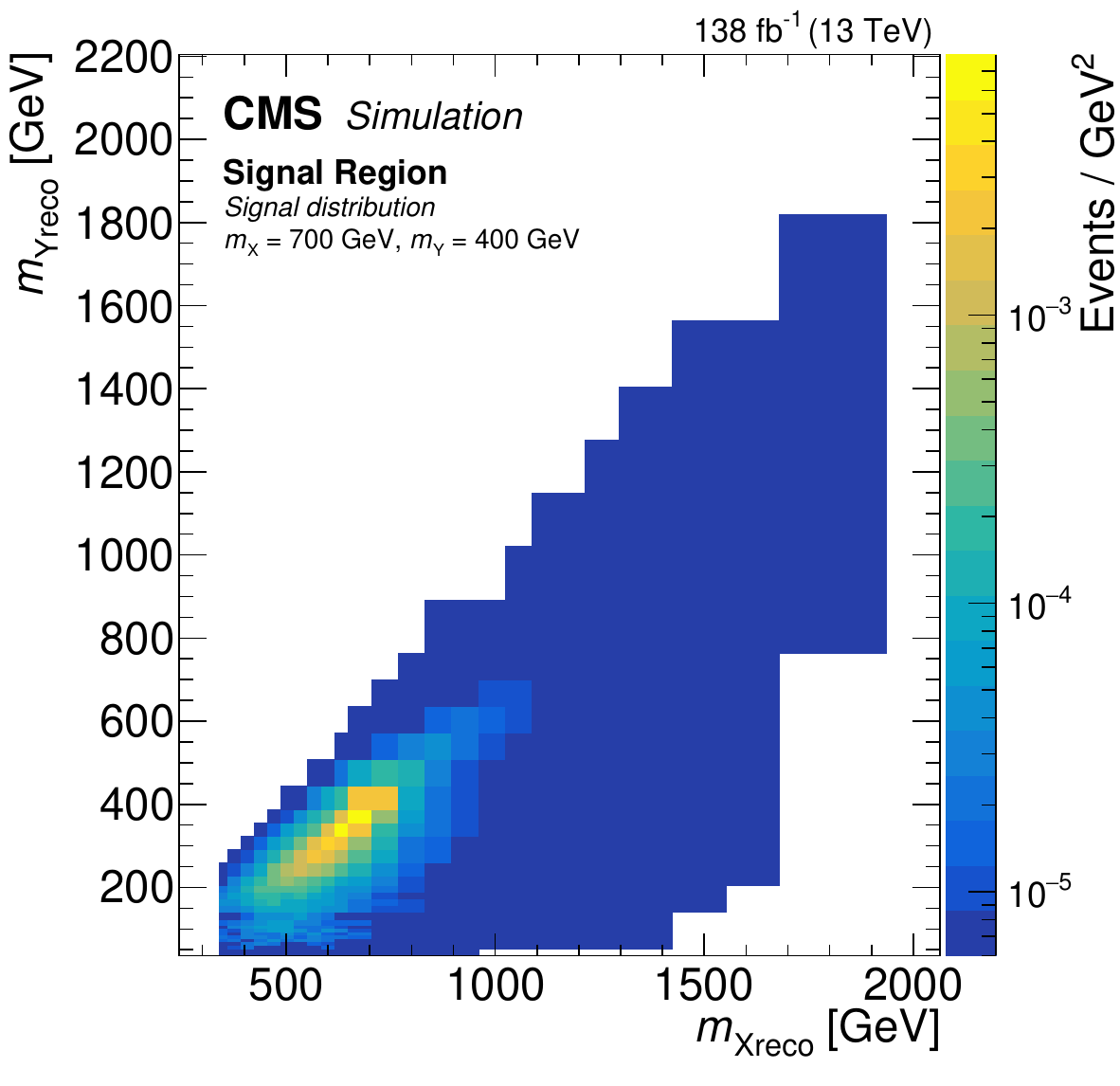}
\caption{Distributions of events in the $m_X^{\text{reco}}$--$m_Y^{\text{reco}}$ plane
observed in the SR(4$b$) for CMS data (Run-2, 13\,TeV).
Left: data. Right: expected signal for $m_X = 700\,\text{GeV}$, $m_Y = 400\,\text{GeV}$.
Empty bins at high $m_X^{\text{reco}}$ and low $m_Y^{\text{reco}}$ are excluded because events in that region are highly boosted and fall outside the resolved analysis acceptance.
From Ref.~\cite{Hayrapetyan:2960489}.}
\label{fig:CMS_bbbb2}
\end{figure}

The mass of the scalar $Y$ is reconstructed from the invariant mass of one $b$-quark pair, while the other pair reconstructs the Higgs boson mass. Figure~\ref{fig:CMS_bbbb} shows the reconstructed $m_X^{\text{reco}}$ and $m_Y^{\text{reco}}$ distributions in the SR. Figure~\ref{fig:CMS_bbbb2} shows the distribution of events in the two-dimensional $(m_X^{\text{reco}}, m_Y^{\text{reco}})$ plane for data and for a representative signal hypothesis.

\begin{figure}[H]
\centering
\includegraphics[width=0.49\linewidth]{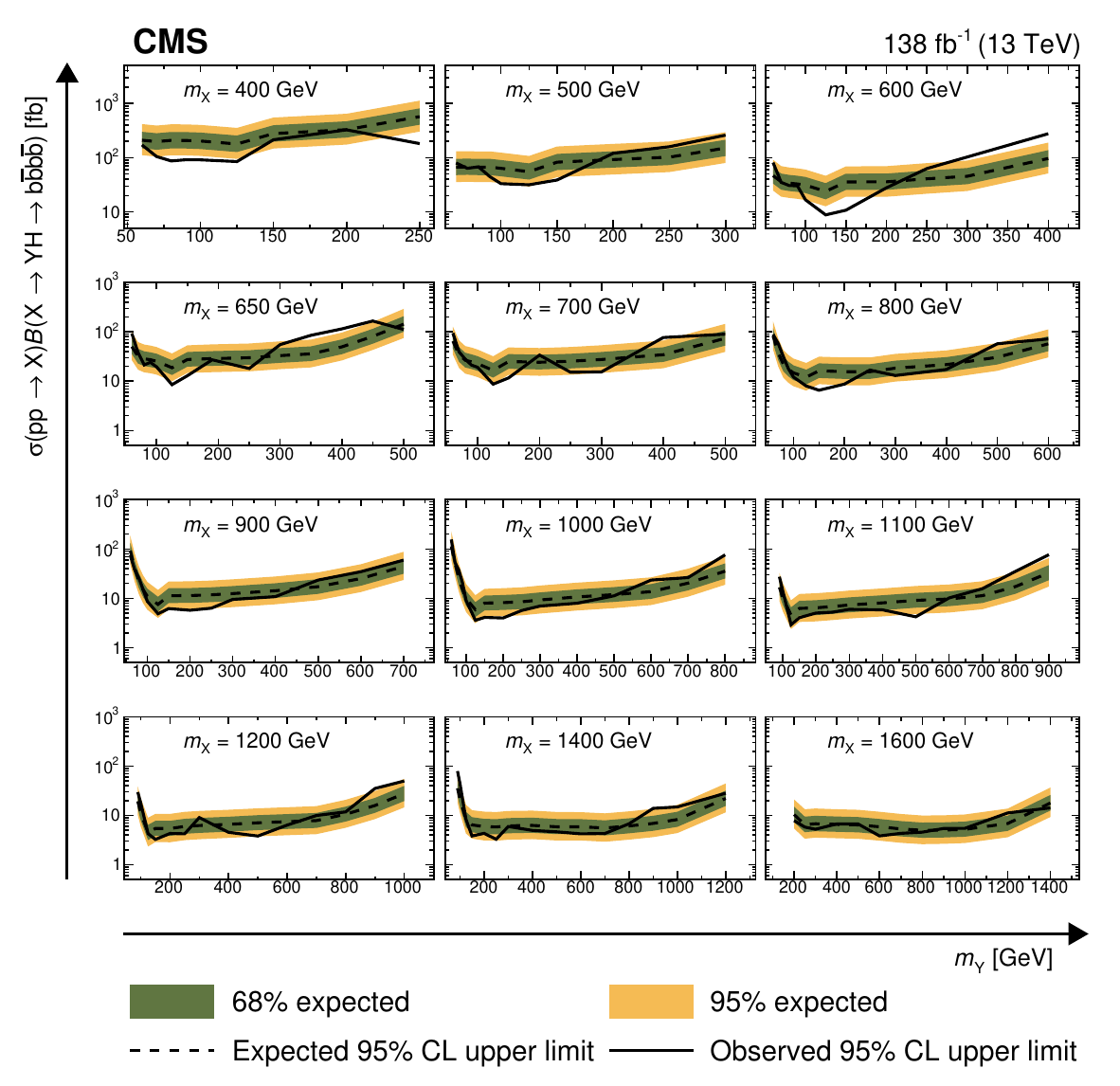}
\includegraphics[width=0.49\linewidth]{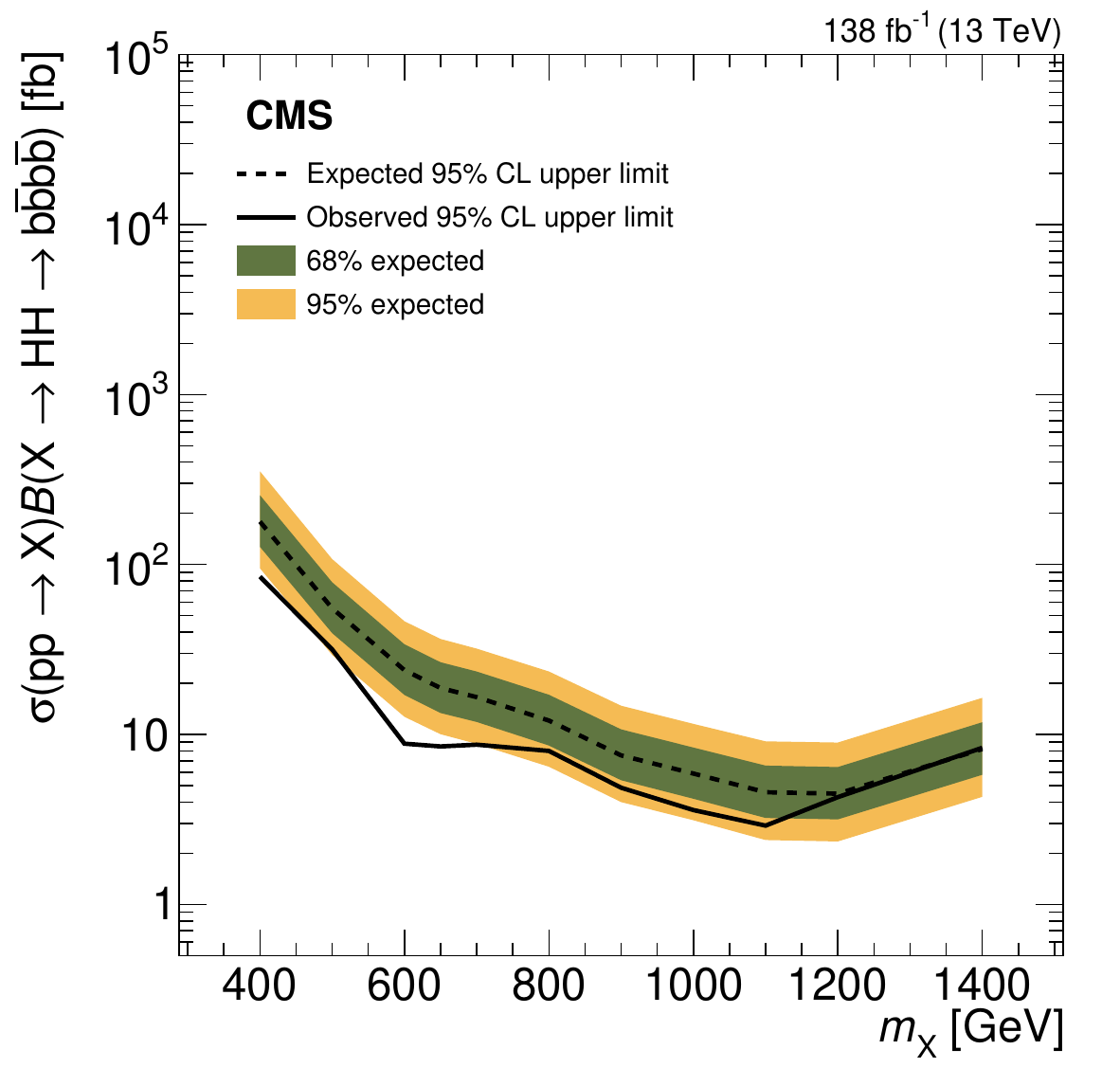}
\caption{Left: Expected (dashed) and observed (solid red) 95\% CL upper limits on $\sigma\mathcal{B}$ for the $X \to YH$ signal as a function of $m_Y$ for selected values of $m_X$ (Run-2, 13\,TeV). The inner (green) and outer (yellow) bands indicate the $\pm1\sigma$ and $\pm2\sigma$ intervals around the expected limit.
Right: Corresponding limits for the $X\to HH$ signal.
From Ref.~\cite{Hayrapetyan:2960489}.
}
\label{fig:CMS_bbbb_limits}
\end{figure}

\begin{figure}[H]
\centering
\includegraphics[width=0.49\linewidth]{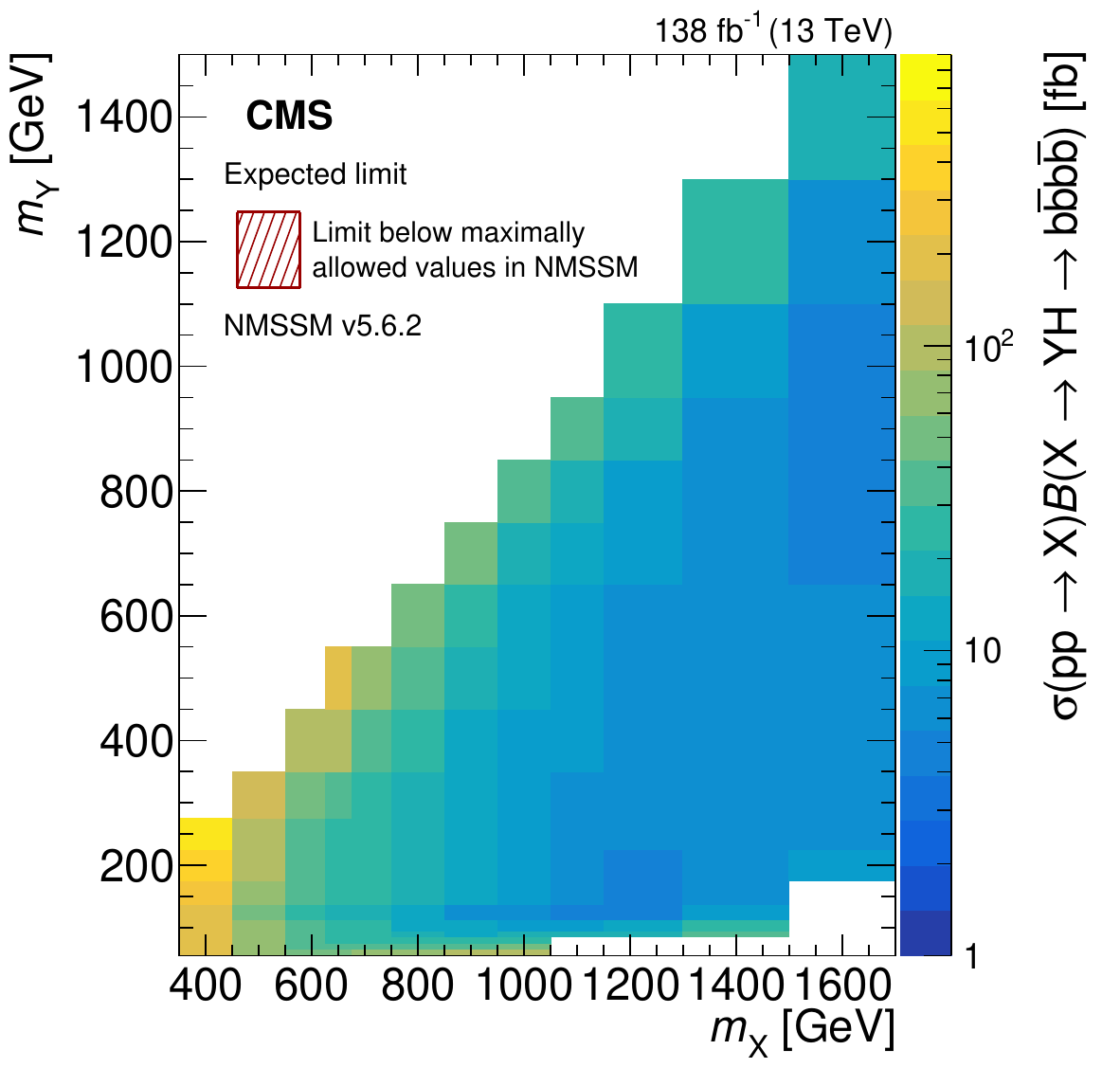}
\includegraphics[width=0.49\linewidth]{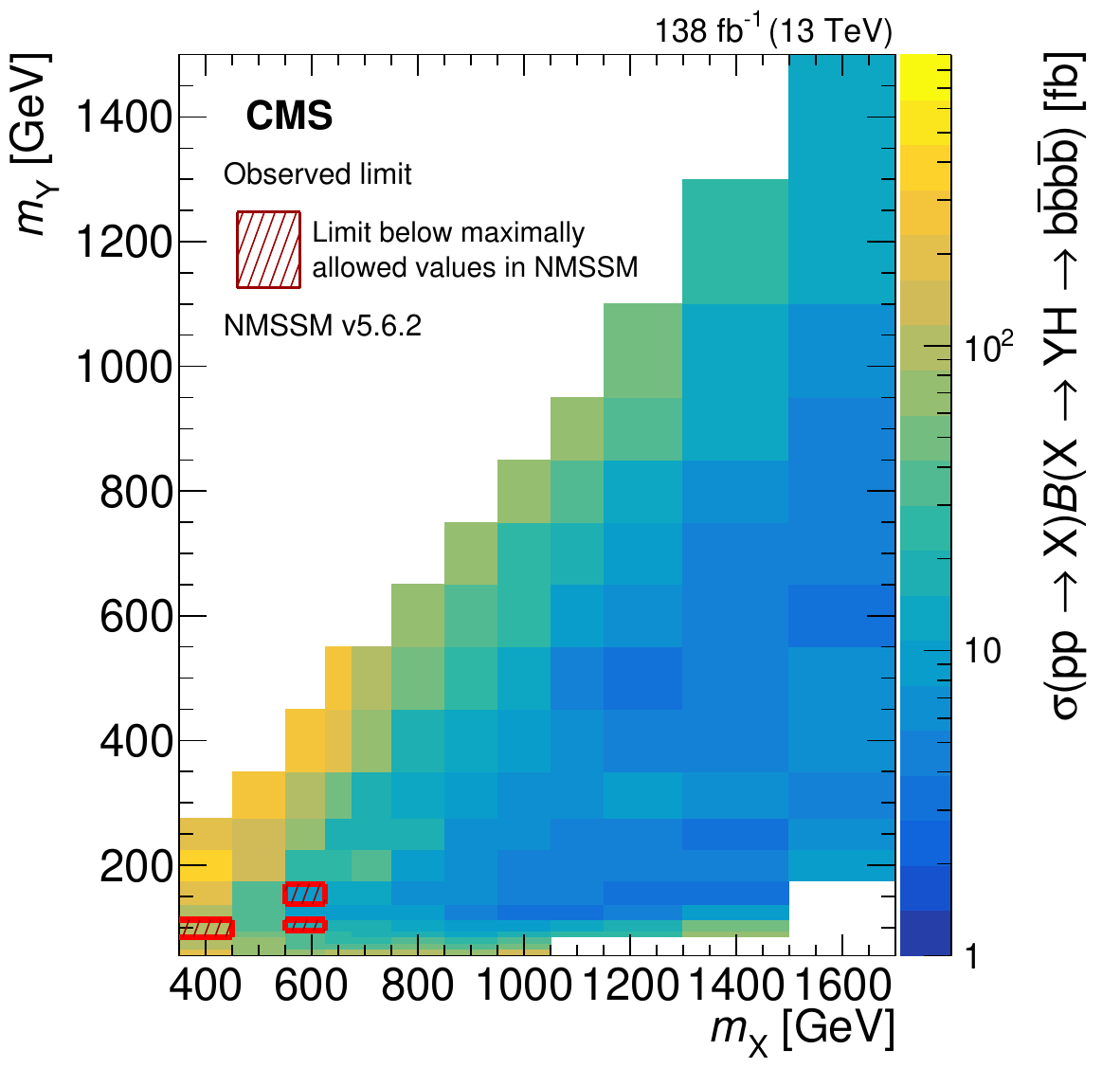}
\caption{Expected (left) and observed (right) 95\% CL upper limits on $\sigma\mathcal{B}$ for $X \to YH$ signals in the two-dimensional $m_X$--$m_Y$ plane (Run-2, 13\,TeV), compared to the maximally allowed values predicted in the NMSSM (using \texttt{NMSSMTools}~5.6.2). Red hatched areas indicate mass hypotheses where observed limits are more restrictive than the NMSSM predictions.
From Ref.~\cite{Hayrapetyan:2960489}.
}
\label{fig:CMS_bbbb_limit2}
\end{figure}

Expected and observed limits are shown in Figs.~\ref{fig:CMS_bbbb_limits} and~\ref{fig:CMS_bbbb_limit2}.
Figure~\ref{fig:CMS_bbbb_limits} shows the limits as a function of $m_Y$ for selected $m_X$ masses, and also includes limits for the special case $X\rightarrow HH \rightarrow b\bar{b}b\bar{b}$.
Limits in the two-dimensional plane $(m_X, m_Y)$ are shown in Fig.~\ref{fig:CMS_bbbb_limit2}, compared to predictions from the NMSSM. Expected and observed limits agree within their uncertainties.

\section{Heavy Neutral Higgs Boson Decaying into Another Neutral Higgs Boson and a $Z$ Boson}

A natural framework for the searches reviewed in this section is the two-Higgs-doublet model (2HDM), which is one of the simplest extensions of the SM Higgs sector. Instead of a single $\text{SU}(2)_L$ doublet, the 2HDM introduces two doublets, yielding five physical Higgs states: two neutral CP-even bosons ($h$ and $H$, with $m_h < m_H$), one neutral CP-odd (pseudoscalar) boson ($A$), and two charged bosons ($H^\pm$). The lighter CP-even state $h$ is identified with the observed SM-like Higgs boson at 125\,GeV. In the \emph{alignment limit}, characterised by $\cos(\beta-\alpha) \approx 0$, the couplings of $h$ to the SM gauge bosons and fermions approach their SM values, consistent with all current measurements. In this limit the decay $A \to ZH$ (or $A \to Zh$) is kinematically accessible for $m_A > m_H + m_Z$ and provides a distinctive BSM signature with a $Z$ boson and a heavy Higgs boson in the final state. Four independent searches for this topology are reviewed below, covering different decay modes of $H$ and the $Z$ boson, using Run-2 data at 13\,TeV.

\subsection{$A\rightarrow Z(\rightarrow \ell\ell)H(\rightarrow b\bar{b},WW)$, heavy H}

\begin{figure}[H]
\centering
\includegraphics[width=0.32\linewidth]{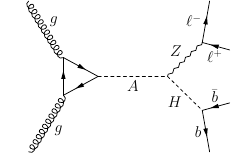}
\includegraphics[width=0.32\linewidth]{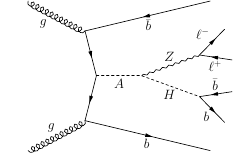}
\includegraphics[width=0.32\linewidth]{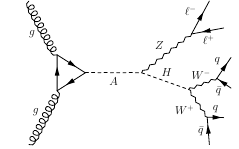}
\caption{Representative lowest-order Feynman diagrams for the three production and decay modes considered (Run-2, 13\,TeV):
(left) gluon--gluon fusion production of $A \to ZH \to \ell\ell b\bar{b}$,
(middle) $b$-associated production of $A \to ZH \to \ell\ell b\bar{b}$, and
(right) gluon--gluon fusion production of $A \to ZH \to \ell\ell WW$.
From Ref.~\cite{ATLAS:2020gxx}.
}
\label{fig:bbll}
\end{figure}

A search for $A\rightarrow Z(\rightarrow \ell\ell)H(\rightarrow b\bar{b},WW)$ production via gluon-gluon fusion was conducted in the final state with two or three $b$-quark jets, or four jets, and two leptons, using the full ATLAS Run-2 dataset at 13\,TeV~\cite{ATLAS:2020gxx}. This search targets a \emph{heavy} $H$ (i.e.\ $H$ is not the SM Higgs boson), making it sensitive to the type-I and type-II 2HDM with large mass splittings. The corresponding Feynman diagrams are shown in Fig.~\ref{fig:bbll}.

\begin{figure}[H]
\centering
\includegraphics[width=0.49\linewidth]{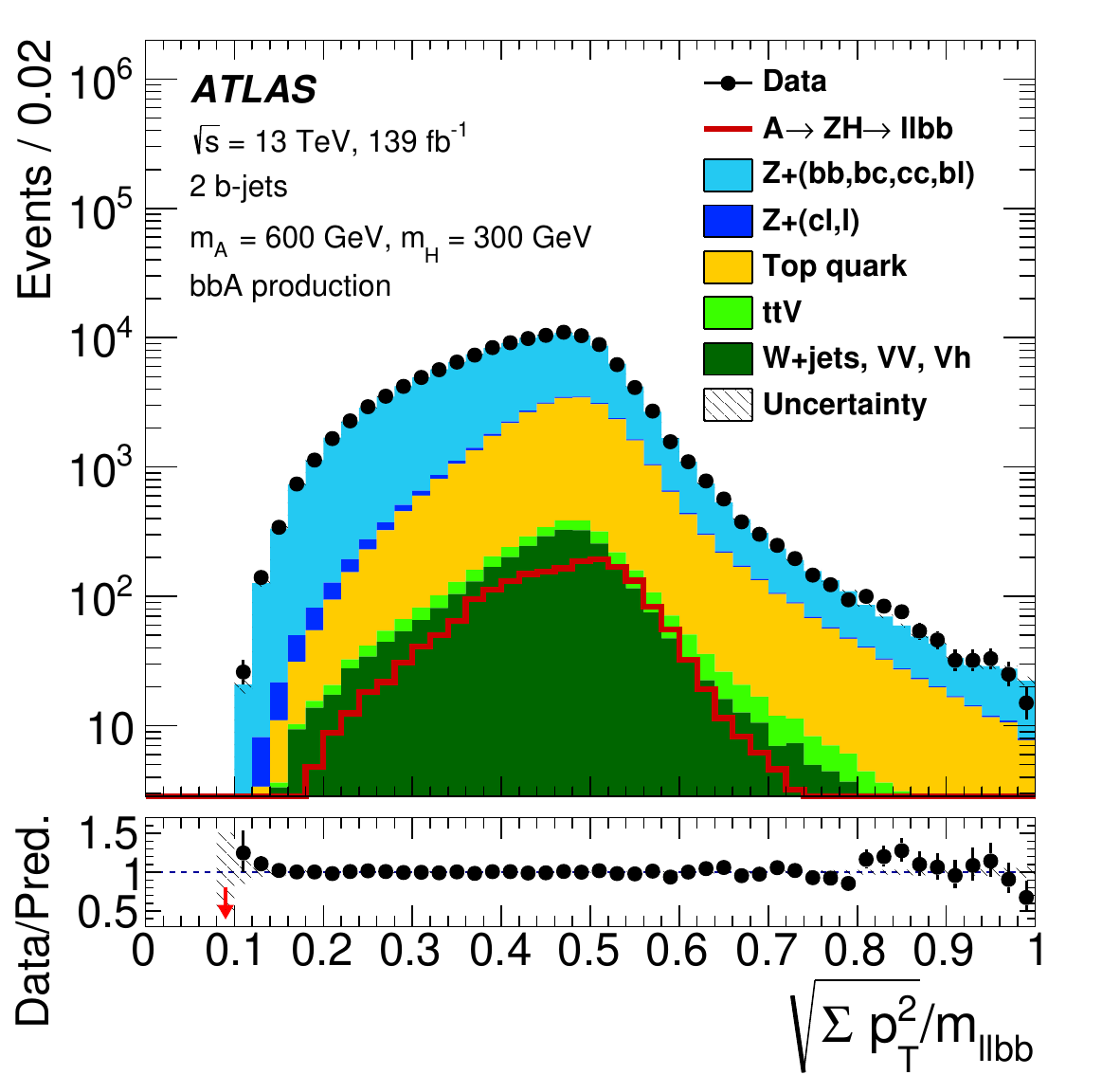}
\includegraphics[width=0.49\linewidth]{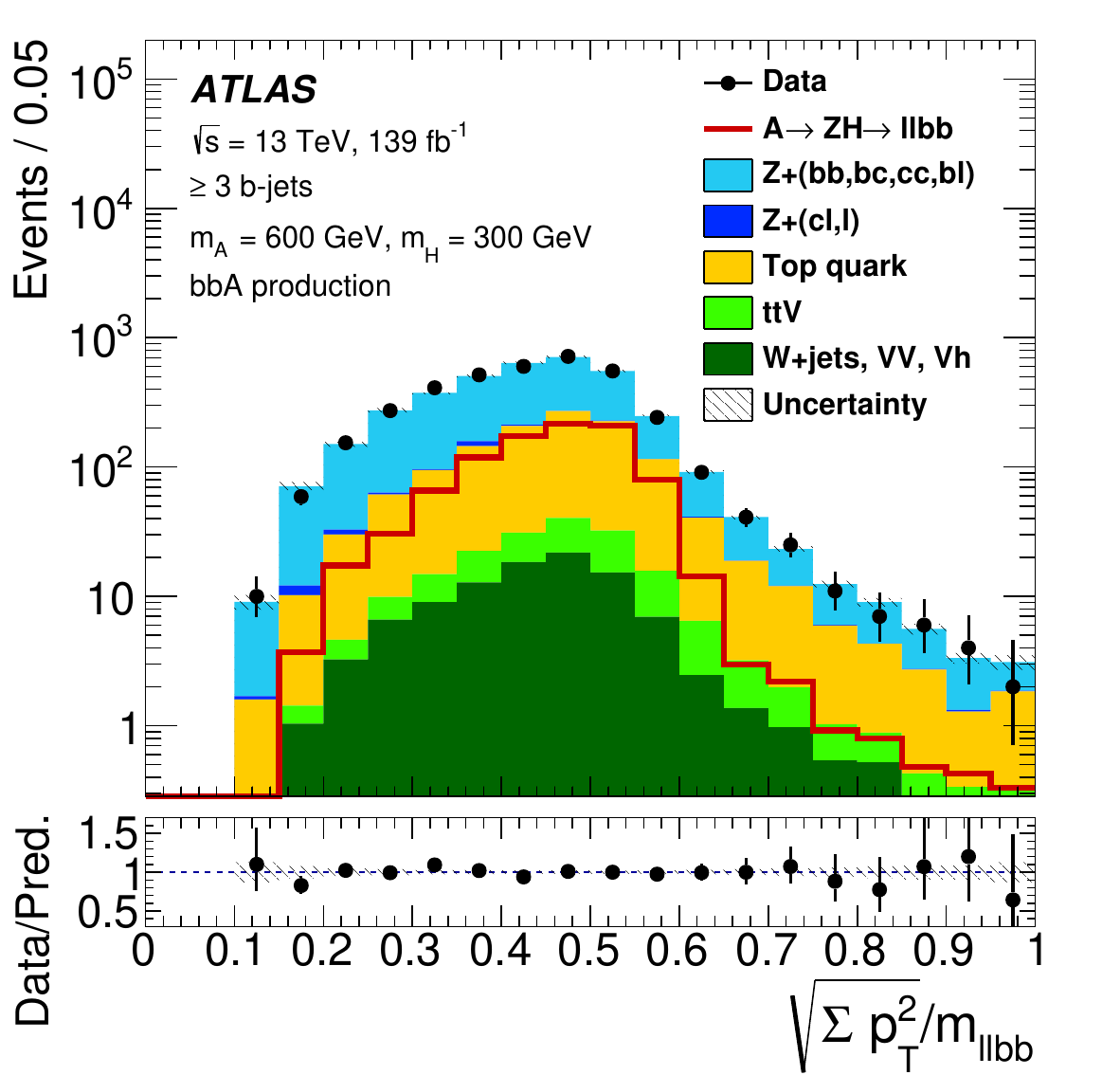}
\caption{Distributions of $\sqrt{\sum p_{\text{T}}^{2}}/m_{\ell\ell bb}$ before
the requirement on this variable is applied, for events with
(left) exactly two $b$-jets and
(right) three or more $b$-jets (Run-2, 13\,TeV).
The signal distribution for $(m_A, m_H) = (600, 300)\,\text{GeV}$ is shown normalized to $\sigma\mathcal{B}(A \to ZH)\mathcal{B}(H \to b\bar{b}) = 1\,\text{pb}$ ($b$-associated production only).
The lower panel shows the ratio of data to the background prediction (filled circles) and the relative background uncertainty (hatched area).
The notation $t\bar{t}V$, $VV$, and $Vh$ refer to $t\bar{t}$+boson, diboson, and SM $Vh$ production.
From Ref.~\cite{ATLAS:2020gxx}.
}
\label{fig:ATLAS_pt}
\end{figure}

\begin{figure}[H]
\vspace{-4mm}
\centering
\includegraphics[width=0.49\linewidth]{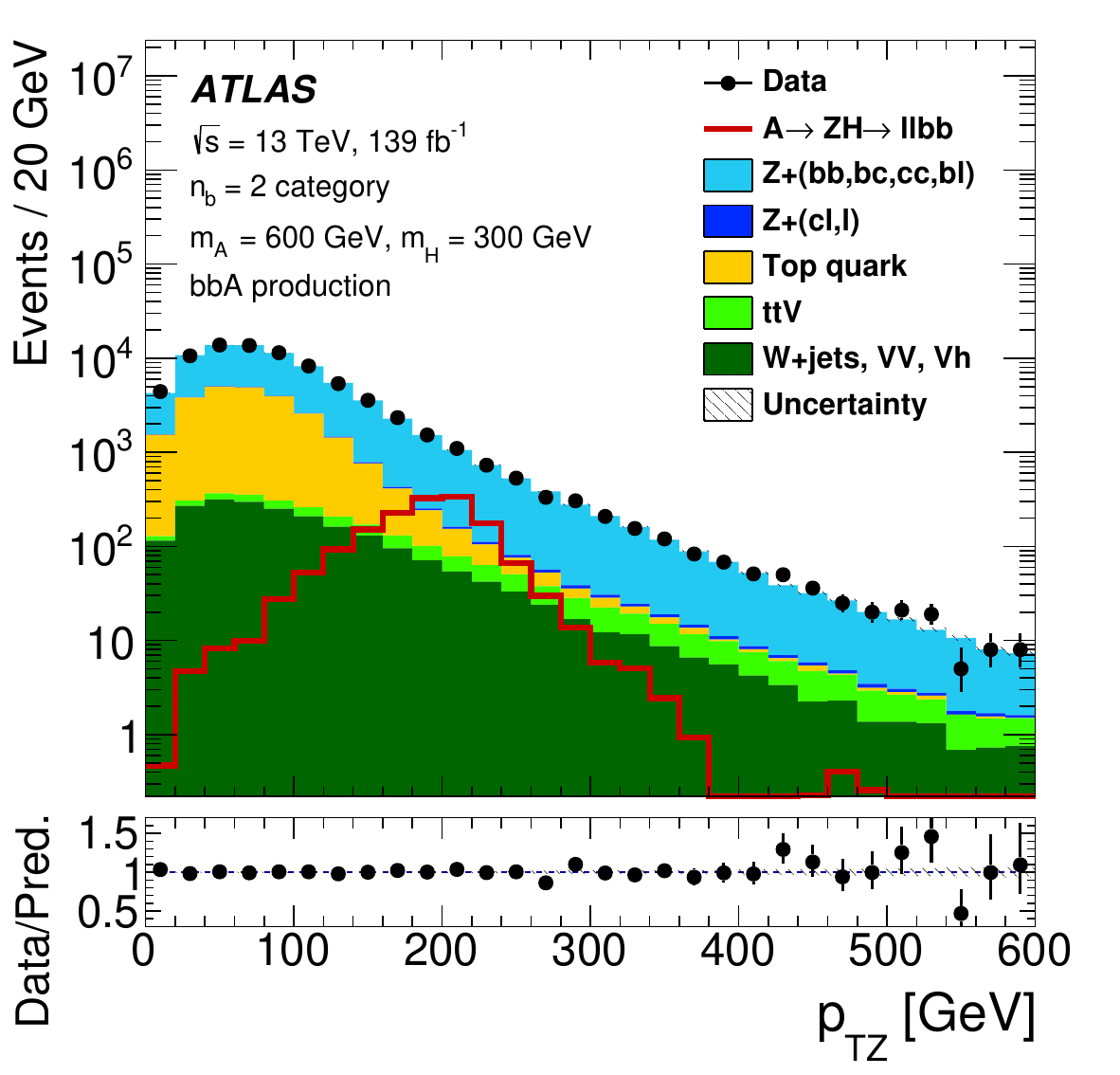}
\includegraphics[width=0.49\linewidth]
{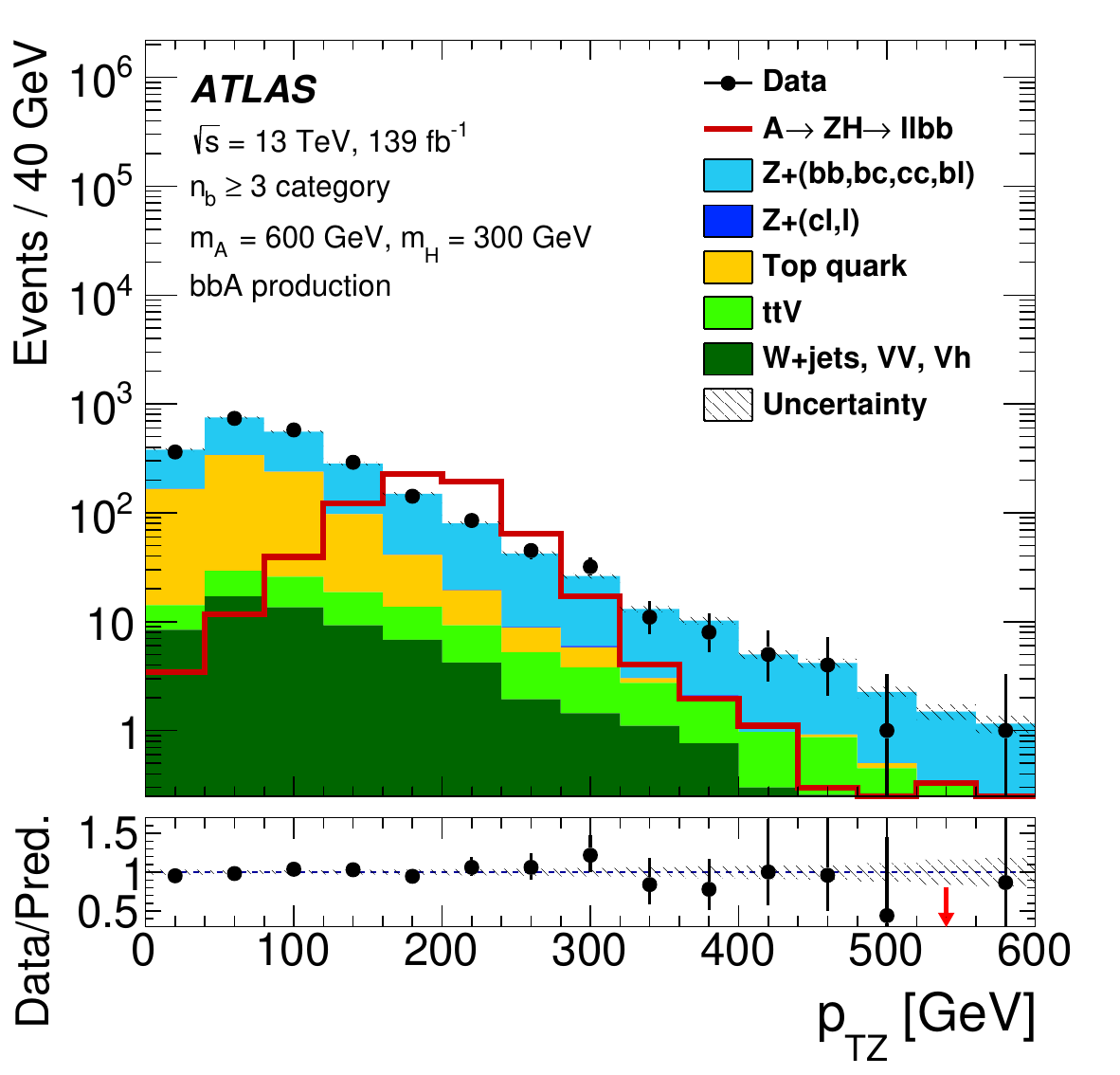}
\caption{
Distributions of
$p_{\text{T}}^{Z}$ for
(left) the $n_b = 2$ category and
(right) the $n_b \geq 3$ category (Run-2, 13\,TeV).
The same conventions as in
{\protect Fig.~\ref{fig:ATLAS_pt}}
are used.
From Ref.~\cite{ATLAS:2020gxx}.
}
\label{fig:ATLAS_pt2}
\end{figure}

Two kinematic variables are used to separate signal from background: the normalized transverse momentum sum $\sqrt{\sum p_{\text{T}}^{2}}/m_{\ell\ell bb}$ and the transverse momentum of the reconstructed $Z$ boson, $p_{\text{T}}^Z$. These exploit the fact that signal events tend to be more isotropic and have a harder $p_T^Z$ spectrum than the dominant $Z$+jets background. Figs.~\ref{fig:ATLAS_pt} and~\ref{fig:ATLAS_pt2} show examples of these distributions for the $t\bar{t}A$ production mode in the two and three $b$-quark channels.

\begin{figure}[H]
\centering
\includegraphics[width=0.49\linewidth]{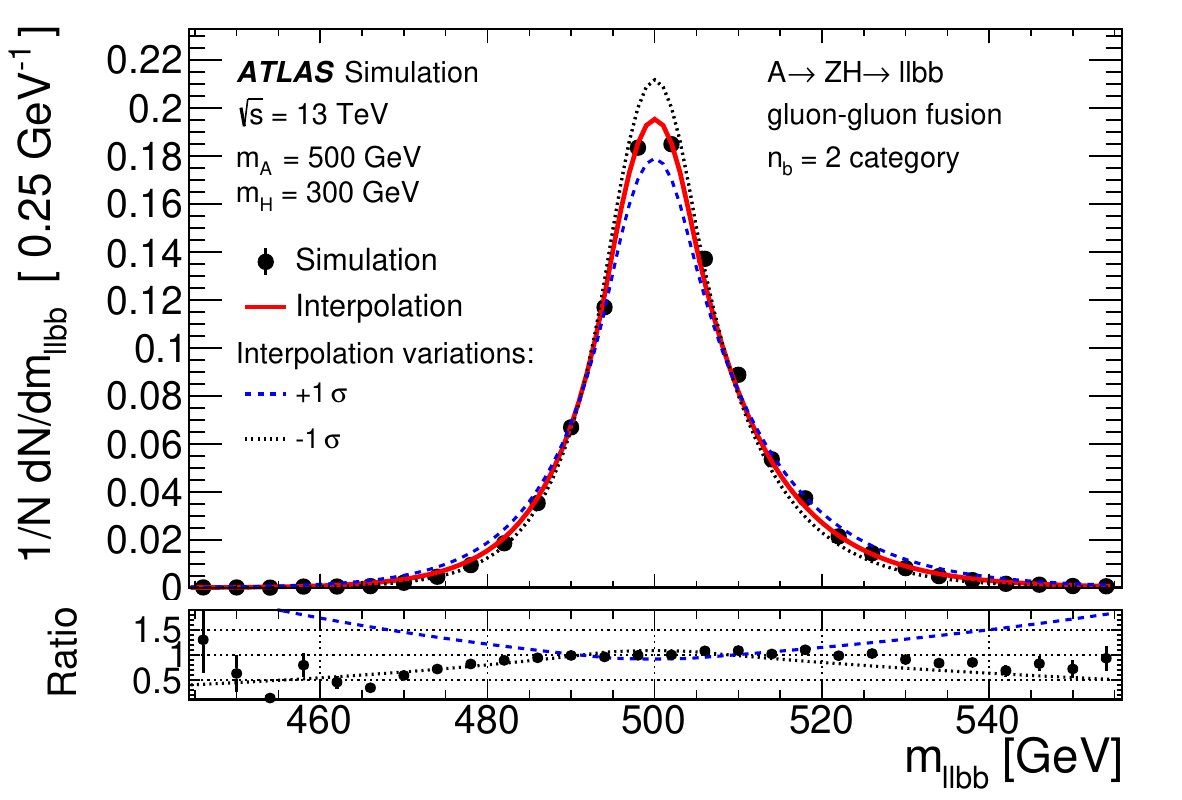}
\includegraphics[width=0.49\linewidth]{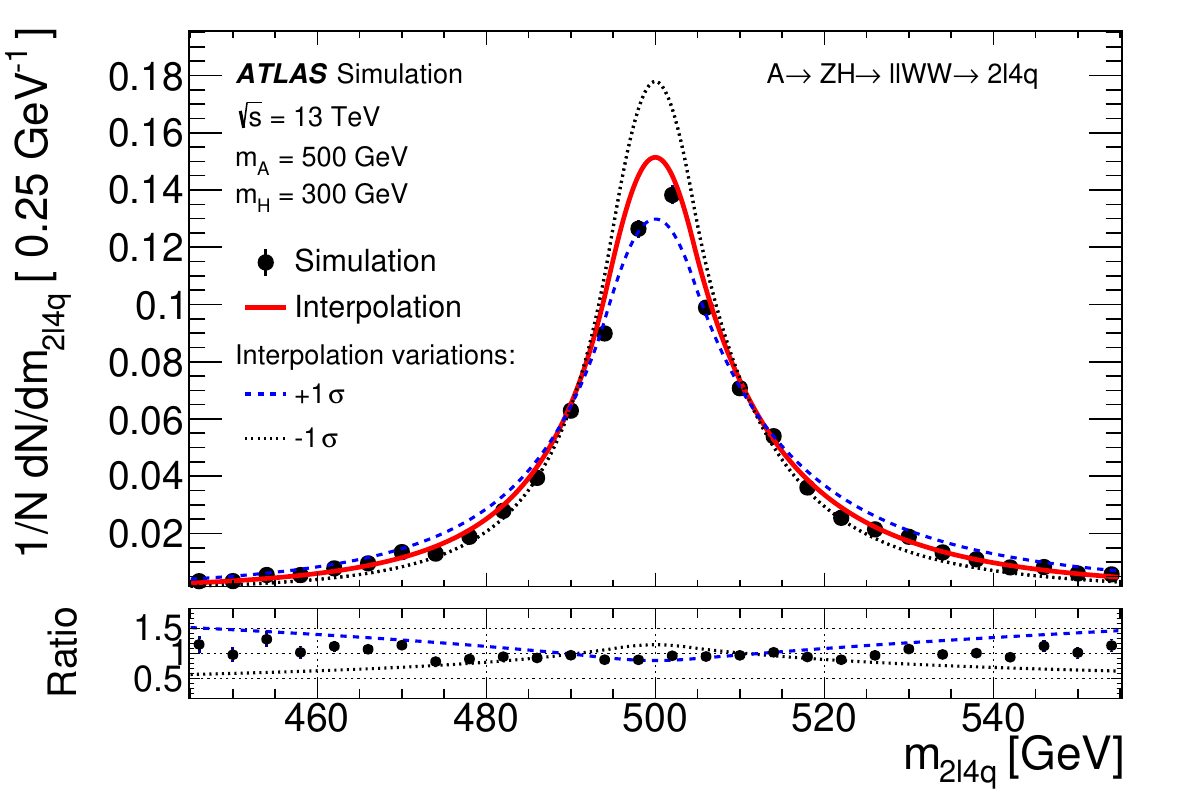}
\caption{
Signal mass distributions assuming $(m_A, m_H) = (500, 300)\,\text{GeV}$ (Run-2, 13\,TeV): (left) $m_{\ell\ell bb}$ in the $\ell\ell b\bar{b}$ channel via gluon--gluon fusion in the $n_b = 2$ category; (right) $m_{2\ell4q}$ in the $\ell\ell WW$ channel.
Black filled circles are simulated events; solid red curves show the interpolated parameterized signal. Dotted blue ($+1\sigma$) and black ($-1\sigma$) lines show shape variations. The lower panels show the ratio of simulation to the parameterized curve.
From Ref.~\cite{ATLAS:2020gxx}.
}
\label{fig:ATLAS_llbb_mass}
\end{figure}

\begin{figure}[H]
\centering
\includegraphics[width=0.49\linewidth]{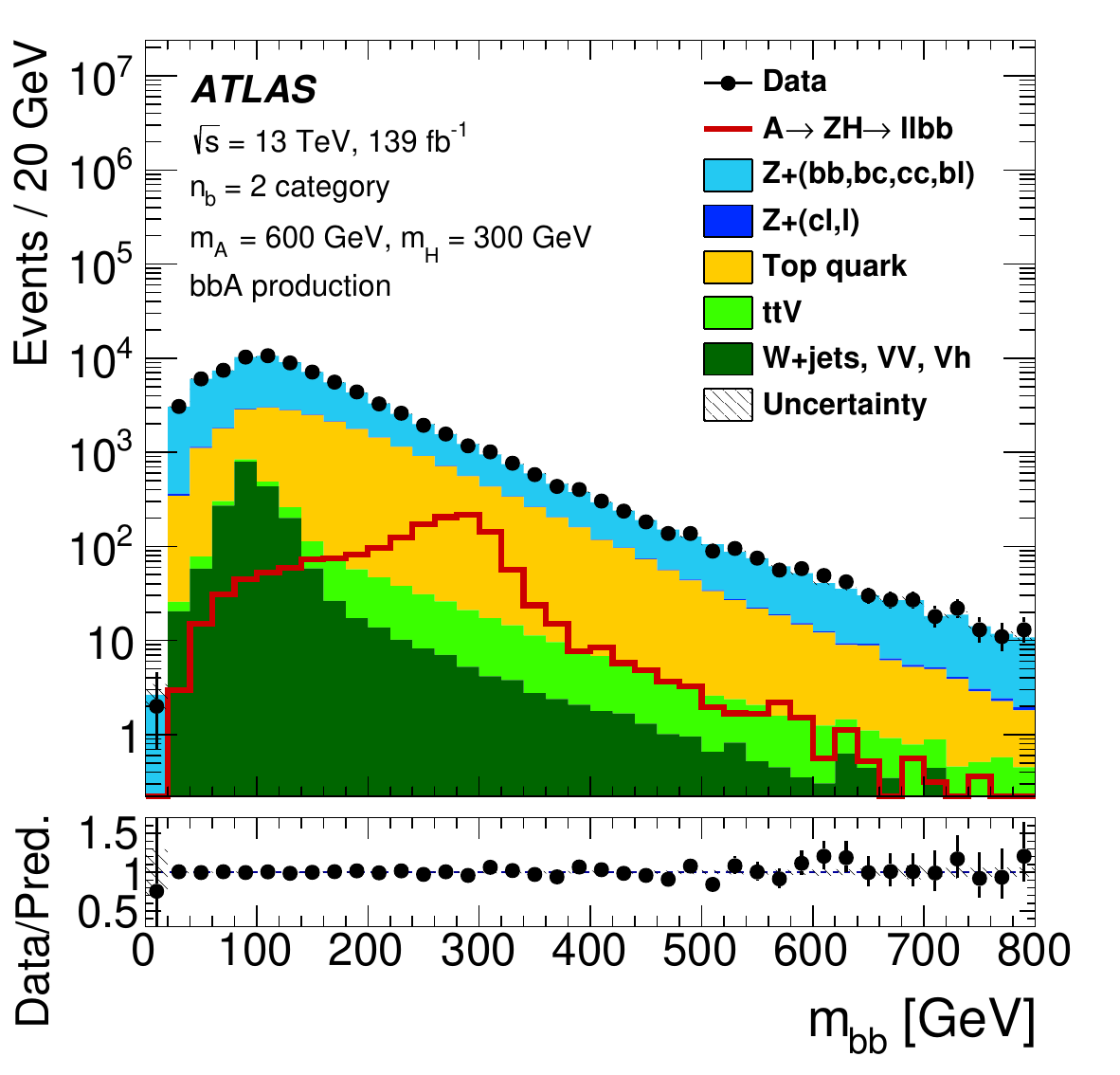}
\includegraphics[width=0.49\linewidth]{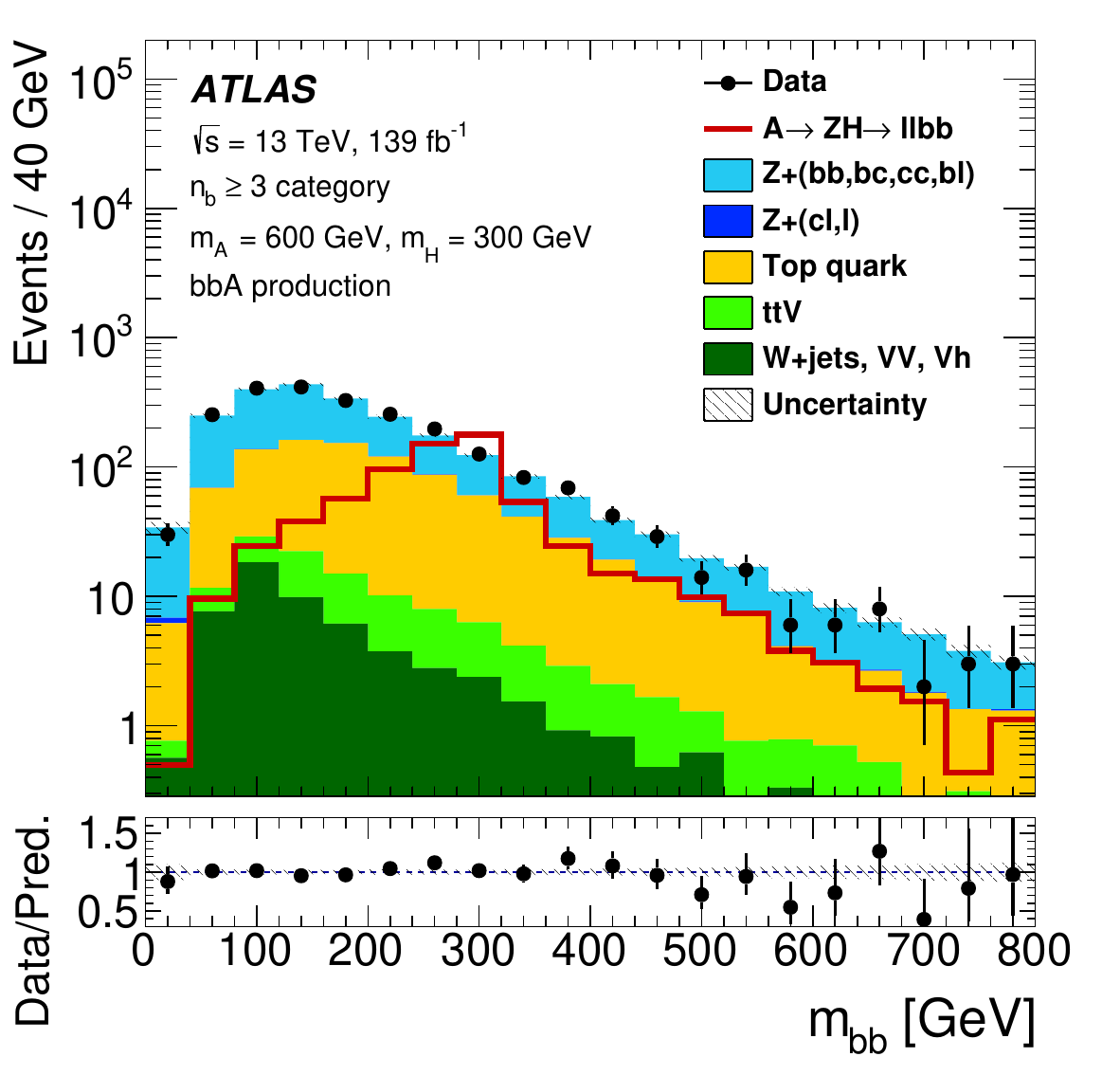}
\caption{
The $m_{bb}$ distribution before the $m_{bb}$ window requirement is applied, for
(left) the $n_b = 2$ category and
(right) the $n_b \geq 3$ category (Run-2, 13\,TeV).
The signal distribution for $(m_A, m_H) = (600, 300)\,\text{GeV}$ is shown normalized to $\sigma\mathcal{B}(A \to ZH)\mathcal{B}(H \to b\bar{b}) = 1\,\text{pb}$.
The same conventions as in {\protect Fig.~\ref{fig:ATLAS_pt}} are used.
From Ref.~\cite{ATLAS:2020gxx}.
}
\label{fig:ATLAS_llbb_mass2}
\end{figure}

The mass of the pseudoscalar $A$ corresponds to the invariant mass of the $\ell\ell b\bar{b}$ or $\ell\ell WW$ final state. An analytical parameterization of the signal mass shape is used to interpolate between simulated mass points, enabling a continuous scan in $m_A$ and $m_H$. The parameterized and simulated mass distributions are compared in Fig.~\ref{fig:ATLAS_llbb_mass}. The measured $m_{bb}$ distributions before the $m_{bb}$ window requirements are shown in Fig.~\ref{fig:ATLAS_llbb_mass2}.

\begin{figure}[H]
\centering
\includegraphics[width=0.49\linewidth]{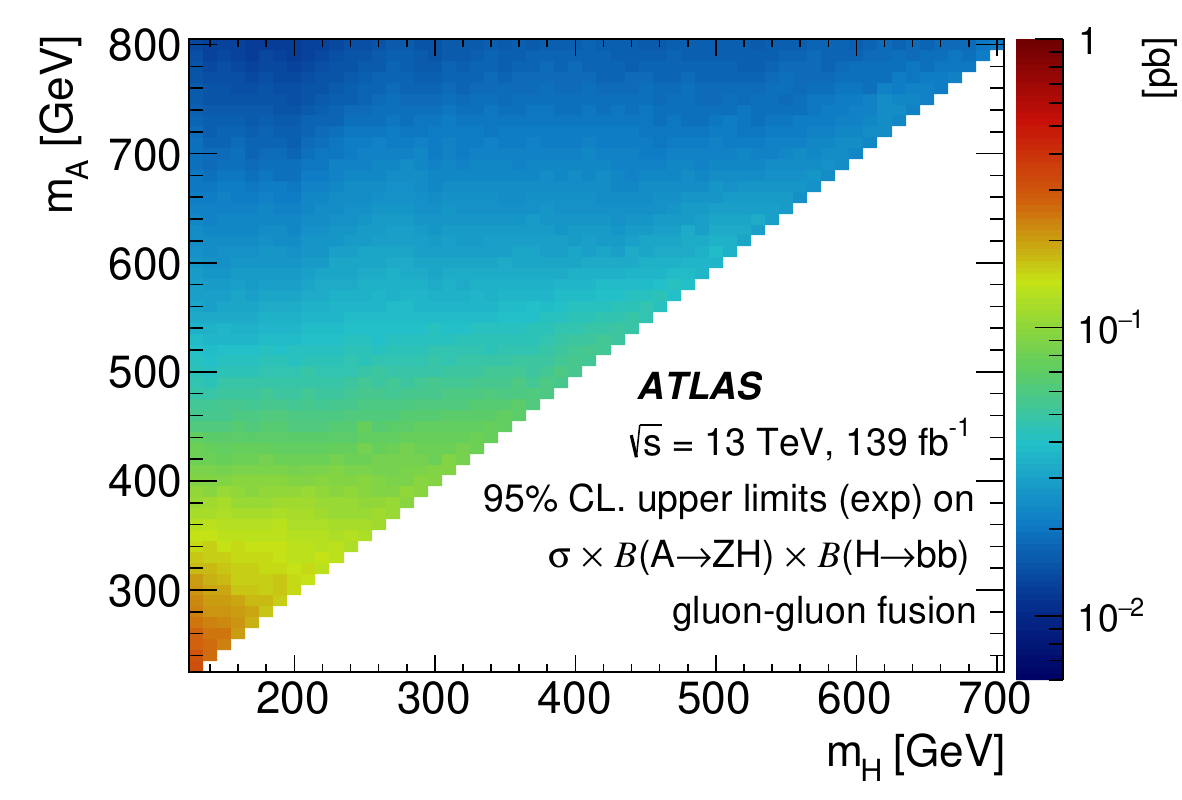}
\includegraphics[width=0.49\linewidth]{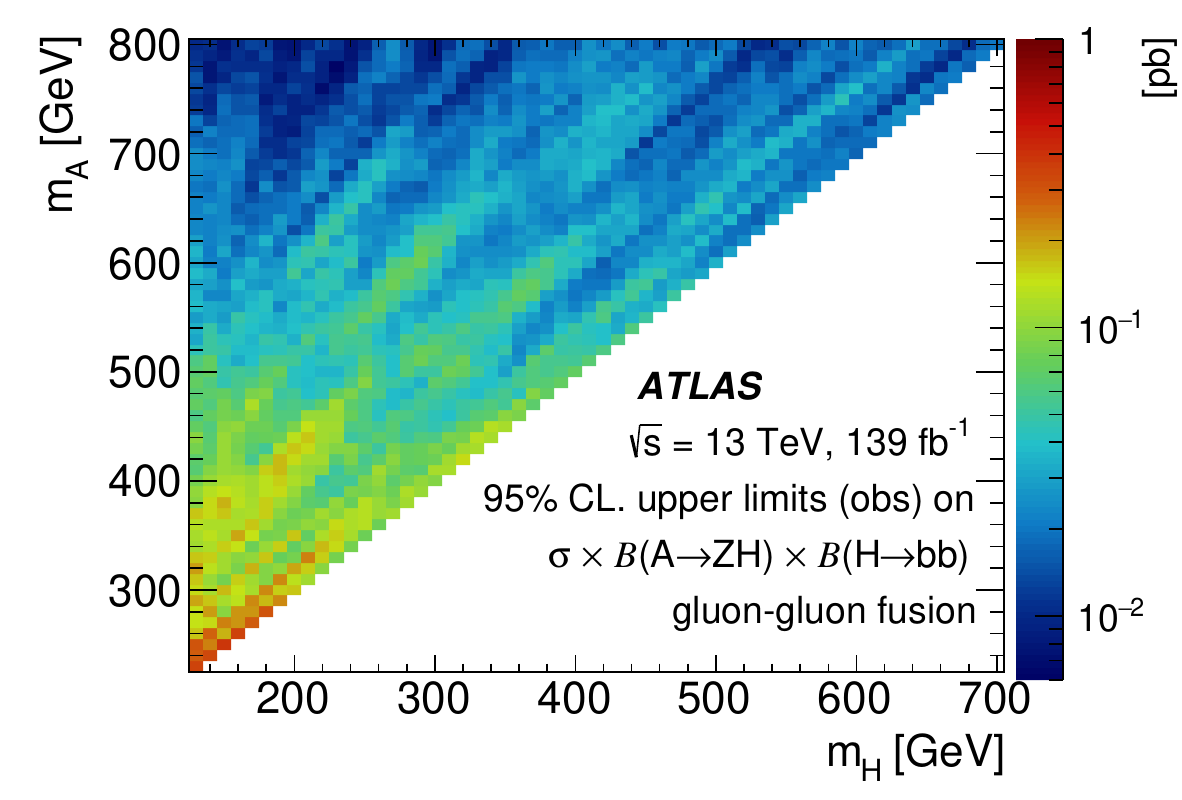}
\caption{
Expected (left) and observed (right) 95\% CL upper limits on $\sigma\mathcal{B}(A \to ZH)\mathcal{B}(H \to b\bar{b})$ (in pb) for gluon--gluon fusion production (Run-2, 13\,TeV).
From Ref.~\cite{ATLAS:2020gxx}.
}
\label{fig:ATLAS_llbb_limits}
\end{figure}

\begin{figure}[H]
\centering
\includegraphics[width=0.49\linewidth]{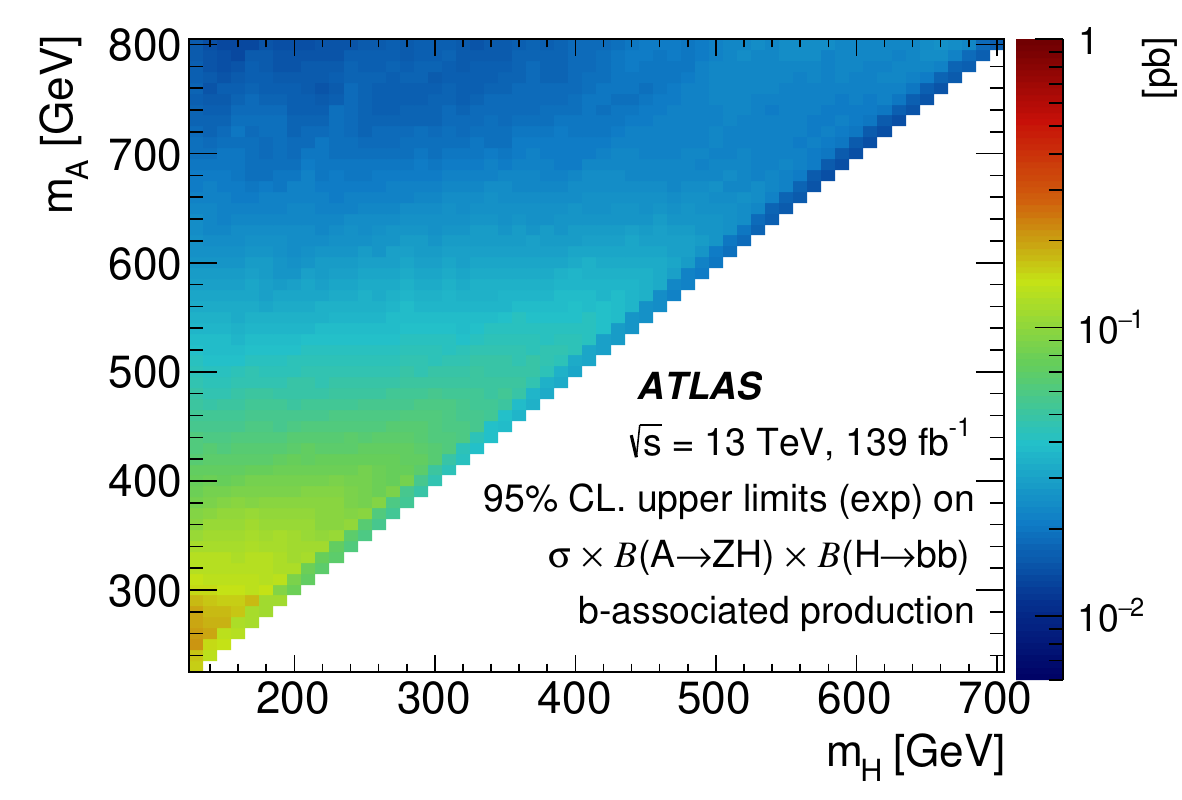}
\includegraphics[width=0.49\linewidth]{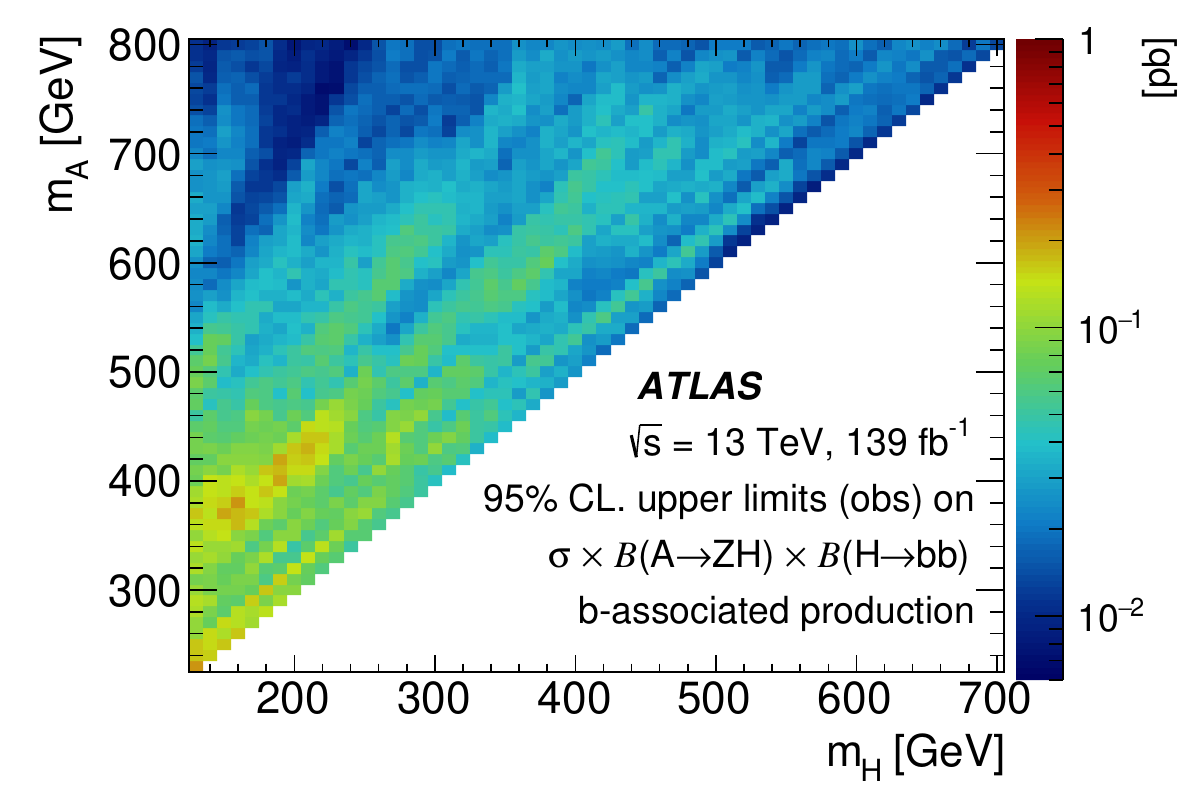}
\caption{
Expected (left) and observed (right) 95\% CL upper limits on $\sigma\mathcal{B}(A \to ZH)\mathcal{B}(H \to b\bar{b})$ (in pb) for $b$-associated production (Run-2, 13\,TeV).
From Ref.~\cite{ATLAS:2020gxx}.
}
\label{fig:ATLAS_llbb_limits2}
\end{figure}

\begin{figure}[H]
\centering
\includegraphics[width=0.49\linewidth]{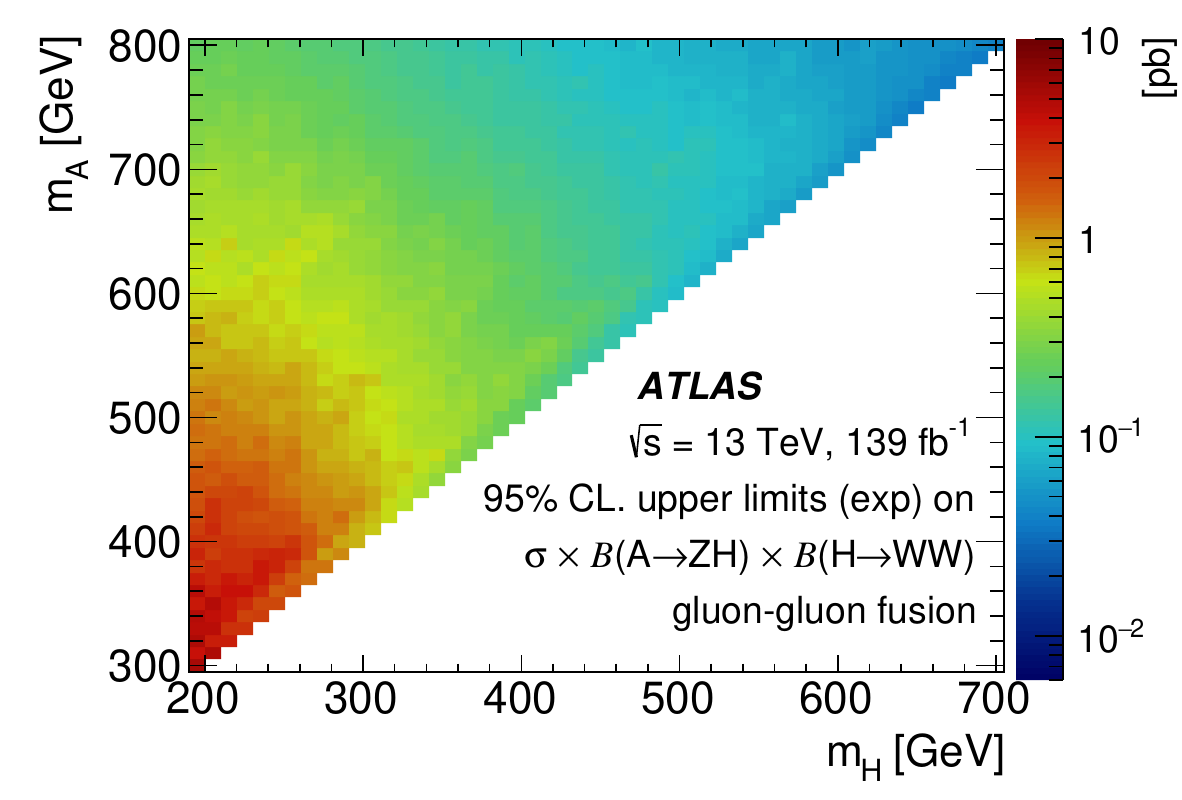}
\includegraphics[width=0.49\linewidth]{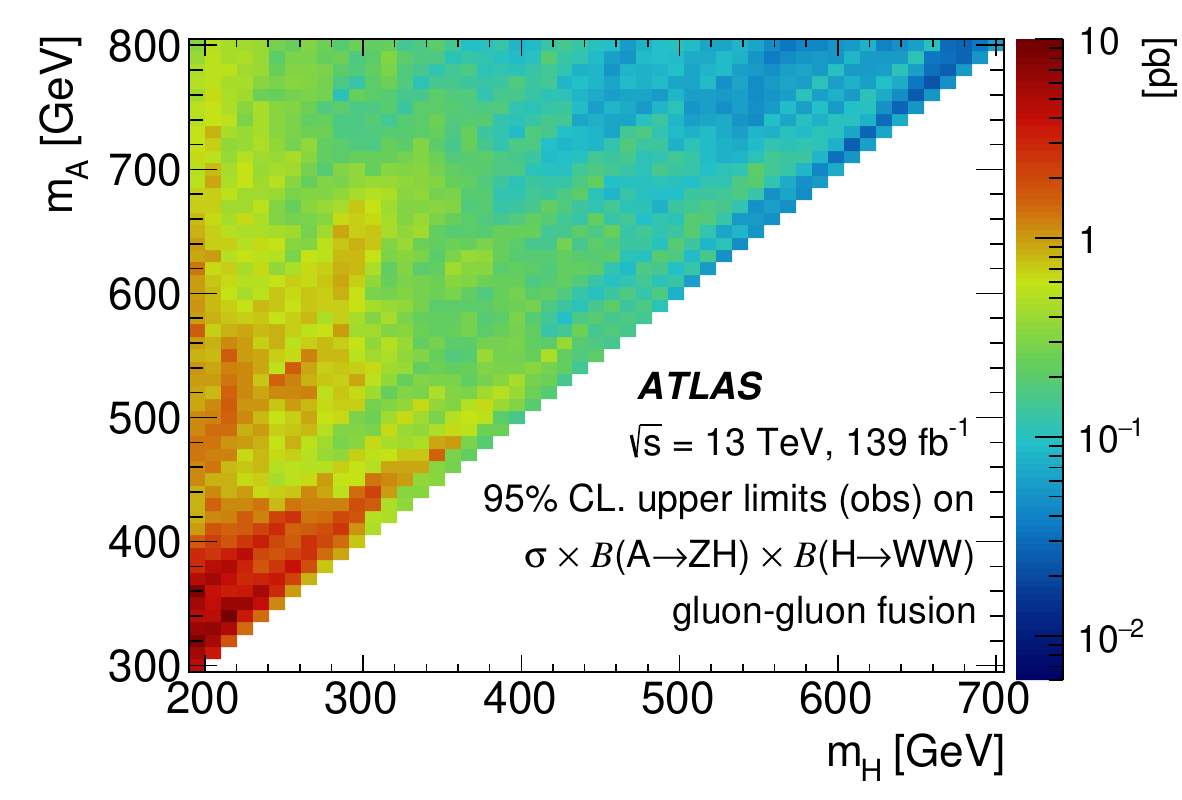}
\caption{
Expected (left) and observed (right) 95\% CL upper limits on $\sigma\mathcal{B}(A \to ZH)\mathcal{B}(H \to WW)$ (in pb) for gluon--gluon fusion production (Run-2, 13\,TeV).
From Ref.~\cite{ATLAS:2020gxx}.
}
\label{fig:ATLAS_llbb_limits3}
\end{figure}

\begin{figure}[H]
\centering
\includegraphics[width=0.49\linewidth]{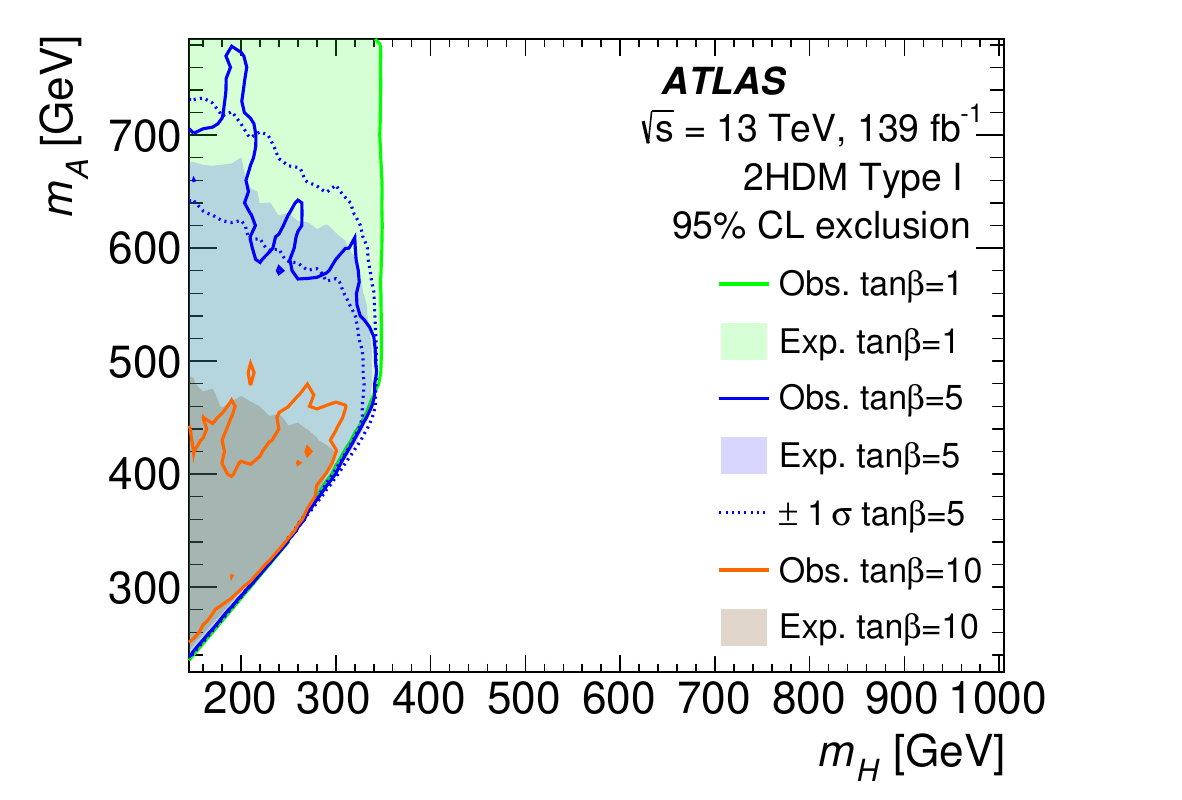}
\includegraphics[width=0.49\linewidth]{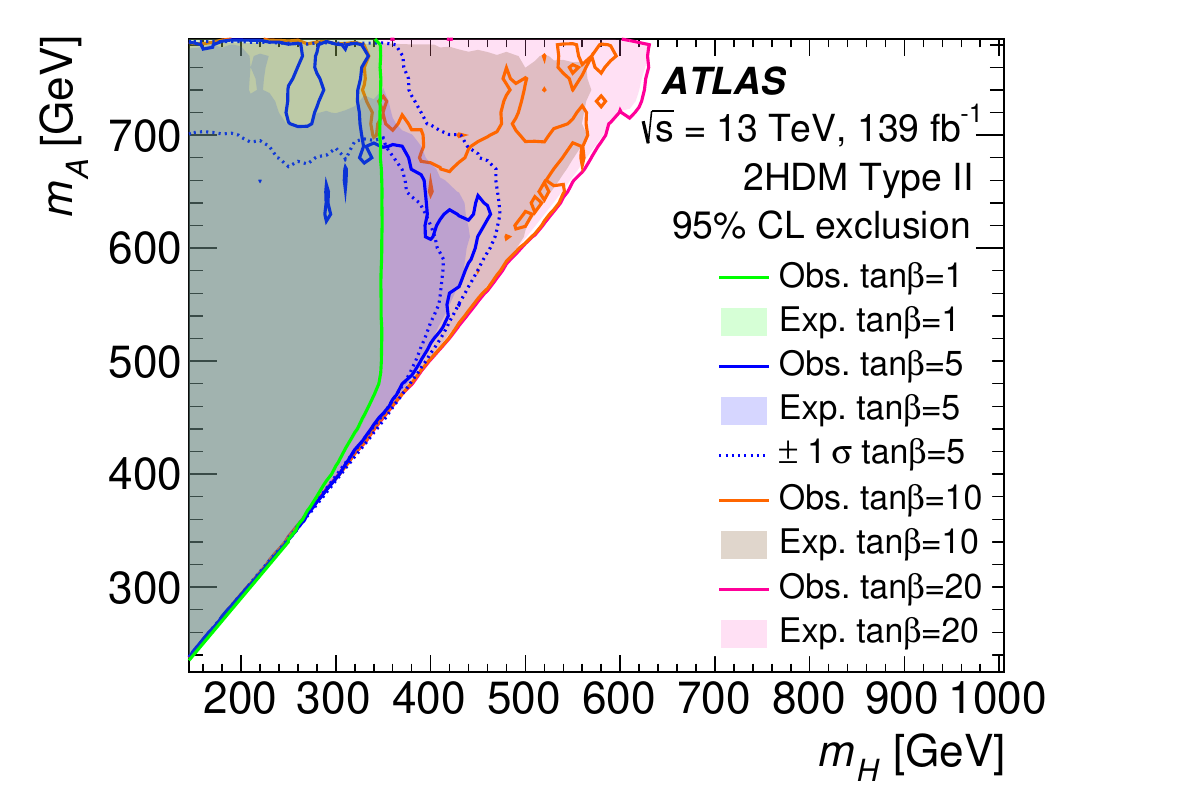}
\caption{Observed and expected 95\% CL exclusion regions for the
$\ell\ell b\bar{b}$ channel in the $(m_H, m_A)$ plane for various values of
$\tan\beta$ in
the type-I (left) and
type-II (right) two-Higgs-doublet model (2HDM),
assuming $\cos(\beta-\alpha) = 0$ (Run-2, 13\,TeV).
{Regions above the curves are excluded: larger $\tan\beta$ values are excluded at lower mass splittings $m_A - m_H$. The type-II model is more strongly constrained than type-I across most of the plane, owing to the enhanced $b$-quark Yukawa couplings in type-II that increase both production and decay rates. The unexcluded region at large $m_H$ corresponds to the kinematic boundary where $m_A < m_H + m_Z$.}
From Ref.~\cite{ATLAS:2020gxx}.
}
\label{fig:ATLAS_llbb_limits4}
\end{figure}

Expected and observed limits are shown for three production scenarios: gluon--gluon fusion with $H\rightarrow b\bar{b}$ (Fig.~\ref{fig:ATLAS_llbb_limits}),
$b$-associated production with $H\rightarrow b\bar{b}$ (Fig.~\ref{fig:ATLAS_llbb_limits2}), and
gluon--gluon fusion with $H\rightarrow WW$ (Fig.~\ref{fig:ATLAS_llbb_limits3}).
In addition, Figure~\ref{fig:ATLAS_llbb_limits4} shows expected and observed exclusion regions for the $\ell\ell b\bar{b}$ channel in the $(m_H, m_A)$ plane for various values of $\tan\beta$ in the type-I and type-II two-Higgs-doublet model (2HDM), assuming $\cos(\beta-\alpha) = 0$.

\subsection{$A\rightarrow V(\rightarrow \nu\nu,\ell\ell,\ell\nu)h(\rightarrow b\bar{b})$, SM h}
\begin{figure}[H]
\vspace{-5mm}
\centering
\includegraphics[width=0.32\linewidth]{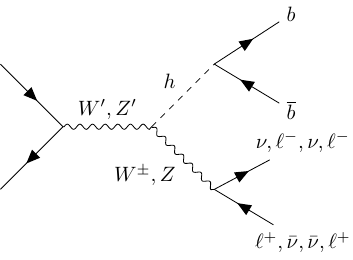}
\includegraphics[width=0.32\linewidth]{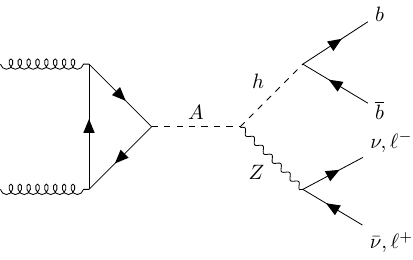}
\includegraphics[width=0.32\linewidth]{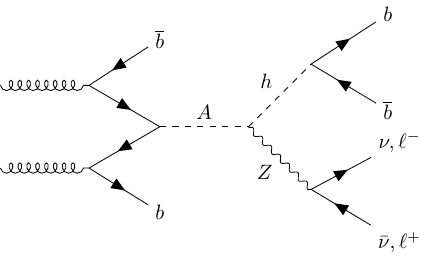}
\caption{Representative lowest-order Feynman diagrams for resonant $Vh$ production (Run-2, 13\,TeV) via
(left) quark--antiquark annihilation,
(middle) gluon--gluon fusion, and
(right) $b$-associated production,
with subsequent decays into the $\nu\nu b\bar{b}$, $\ell\nu bb$, and $\ell\ell b\bar{b}$
final states ($\ell = e, \mu, \tau$).
From Ref.~\cite{ATLAS:2022enb}.
}
\label{fig:ATLAS_VV_graph}
\vspace{-5mm}
\end{figure}

In this search the following channels are considered~\cite{ATLAS:2022enb}, based on the full ATLAS Run-2 dataset at 13\,TeV:
\begin{itemize}
\item
$A \to Zh \to \nu\nu b\bar{b}$, \quad
$A \to Zh \to \ell\ell b\bar{b}$,
\item
$W' \to Wh \to \ell\nu b\bar{b}$, \quad
$Z' \to Zh \to \ell\ell b\bar{b}$, and
\item
$bbA(\to Zh \to \nu\nu b\bar{b})$, \quad
$bbA(\to Zh \to \ell\ell b\bar{b})$.
\end{itemize}
Here $h$ denotes the SM Higgs boson at 125\,GeV. Alongside the 2HDM pseudoscalar $A$, this analysis also considers heavy vector resonances $W'$ and $Z'$ from the Heavy Vector Triplet (HVT) model. Both resolved (individual small-radius jets) and merged (single large-radius jet) topologies are reconstructed for the $h\to b\bar{b}$ decay. The corresponding Feynman diagrams are shown in Fig.~\ref{fig:ATLAS_VV_graph}.

\begin{figure}[H]
\centering
\includegraphics[width=0.49\linewidth]{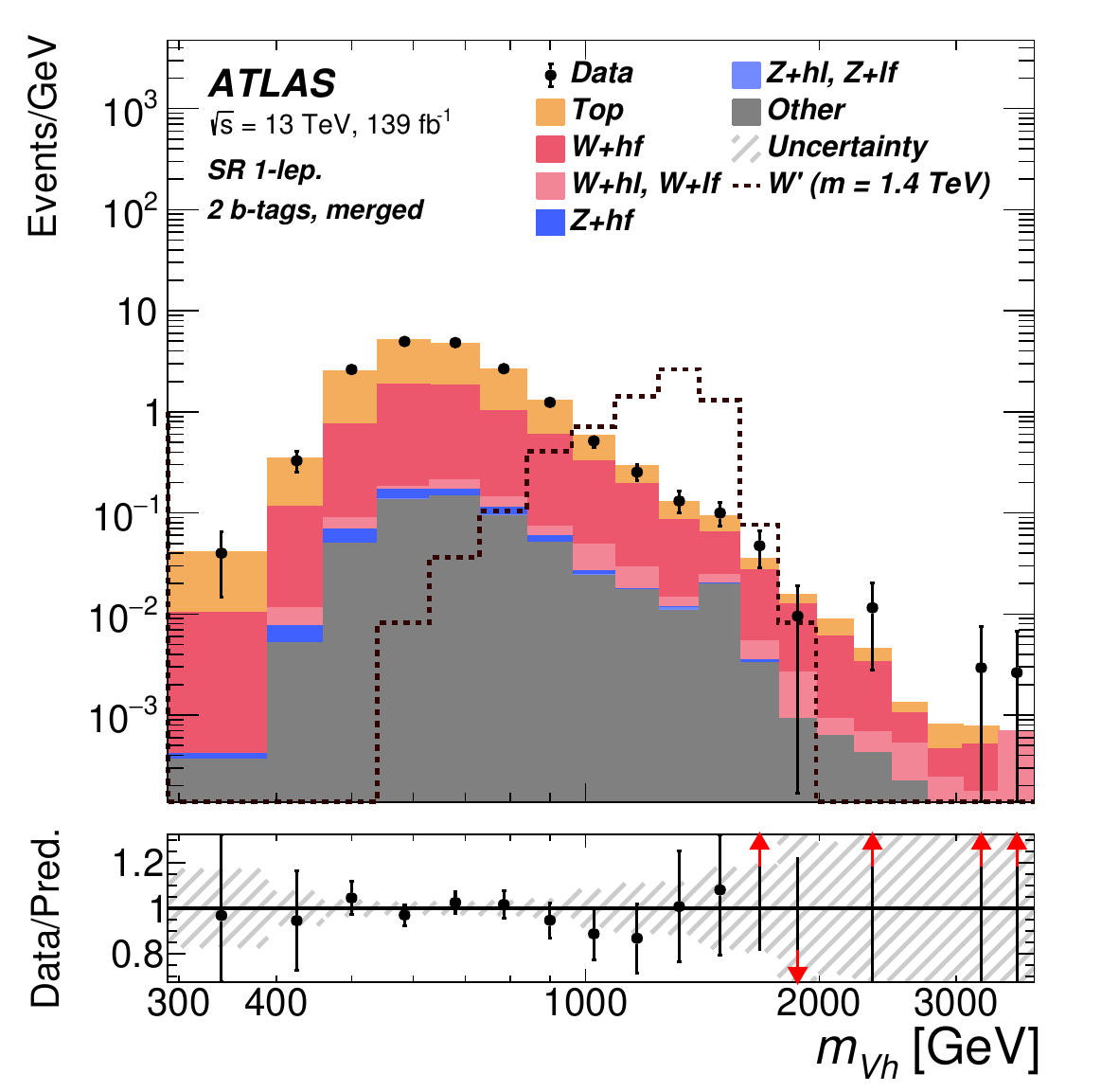}
\includegraphics[width=0.49\linewidth]{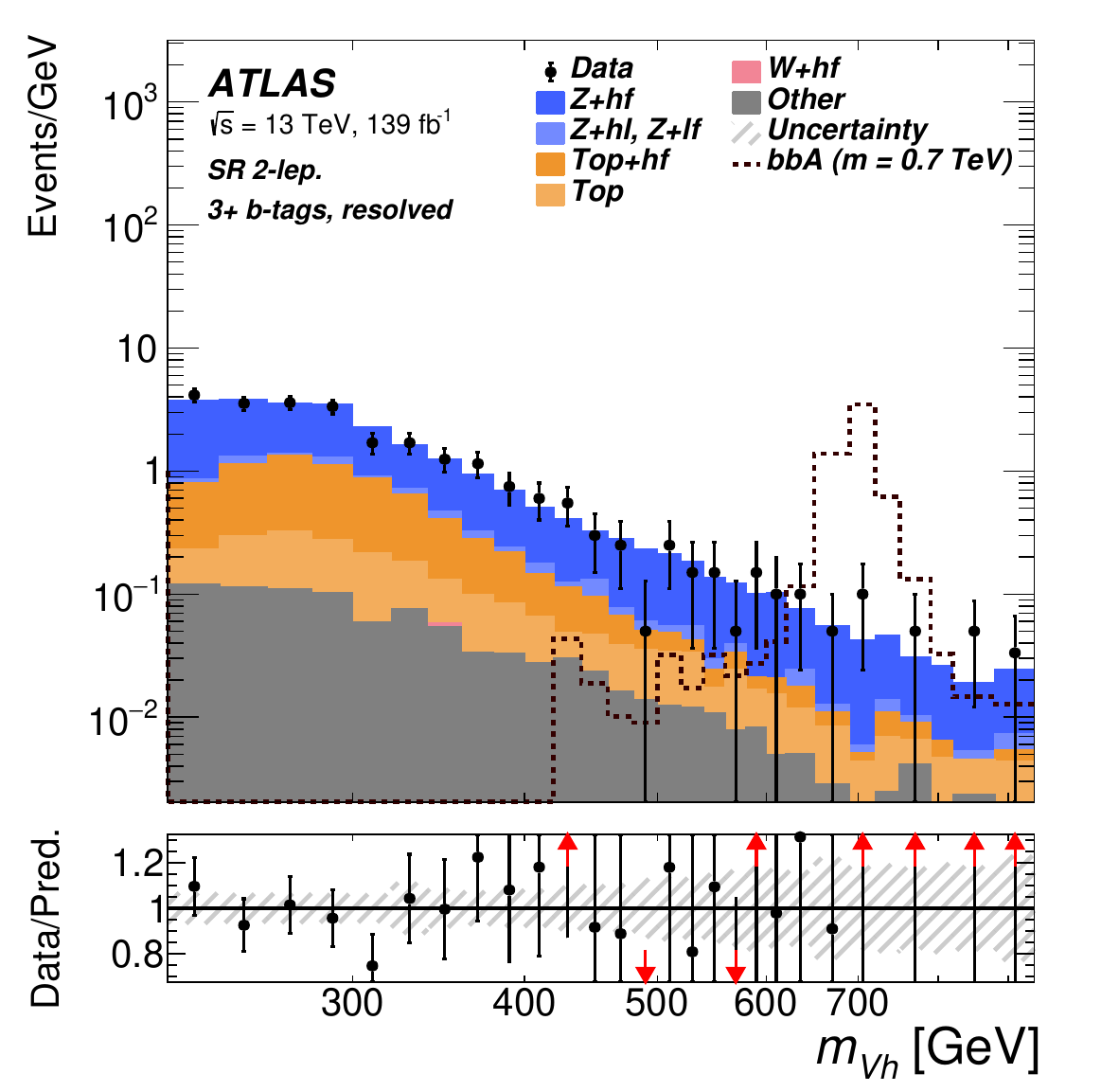}
\caption{
Post-fit distributions of $m_{Vh}$ (Run-2, 13\,TeV).
Left: one-lepton merged signal region of the $W'$ fit. The dashed line shows the HVT Model~A benchmark with $m_{W'} = 1.4\,\text{TeV}$, normalized to 0.1\,pb.
Right: two-lepton resolved three-or-more-tag signal-region categories of the $b\bar{b}A$ fit. The dashed line shows the 2HDM benchmark with $m_A = 0.7\,\text{TeV}$, normalized to $\sigma \times \mathcal{B}(Z h) \times \mathcal{B}(h \to b\bar{b}) = 0.1\,\text{pb}$.
In both panels, ``Top'' aggregates contributions from $t\bar{t}$, single top-quark, $t\bar{t}+h$, and $t\bar{t}+V$. The lower panels show the ratio of data to the SM background prediction.
From Ref.~\cite{ATLAS:2022enb}.
}
\label{fig:ATLAS_VV_mass}
\end{figure}

The reconstructed invariant mass of the $Vh$ system, $m_{Vh}$, is the primary discriminating variable. Figure~\ref{fig:ATLAS_VV_mass} shows representative post-fit $m_{Vh}$ distributions for the $W'$ search (one-lepton channel) and the $b\bar{b}A$ search (two-lepton channel).

\begin{figure}[H]
\vspace{-8mm}
\centering
\includegraphics[width=0.49\linewidth]{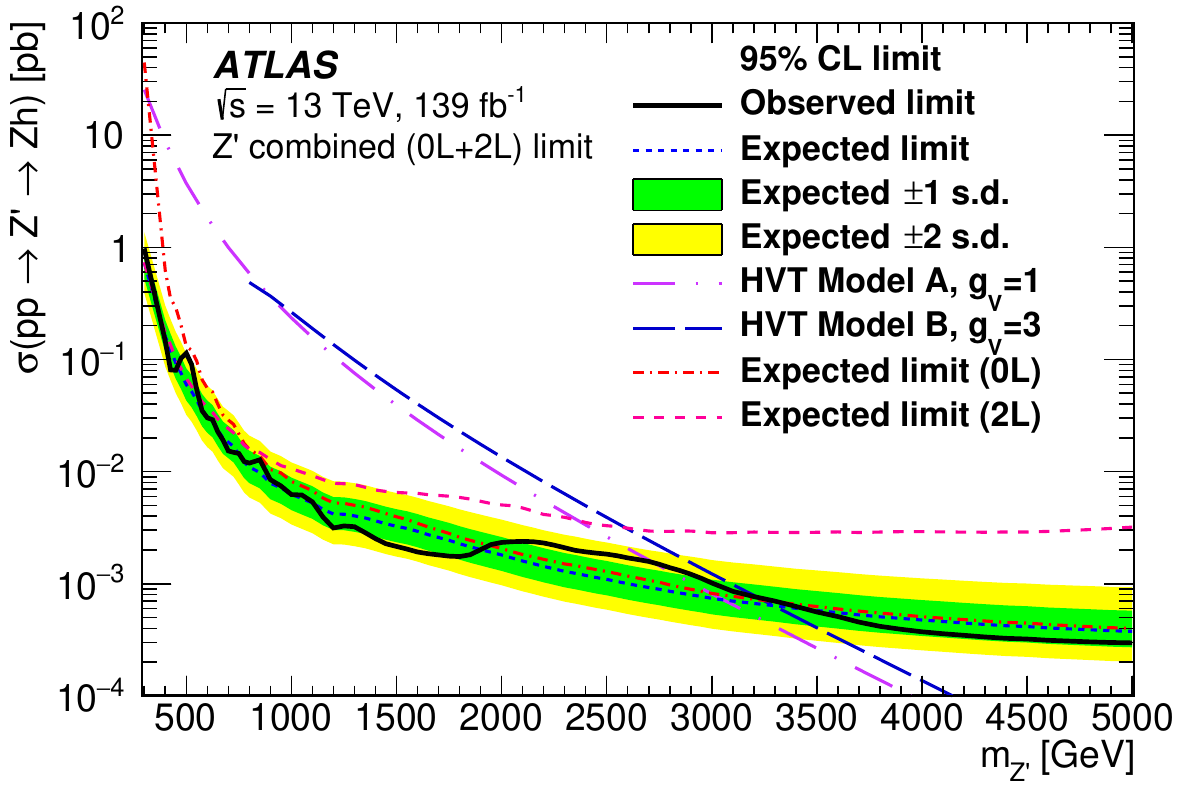}
\includegraphics[width=0.49\linewidth]{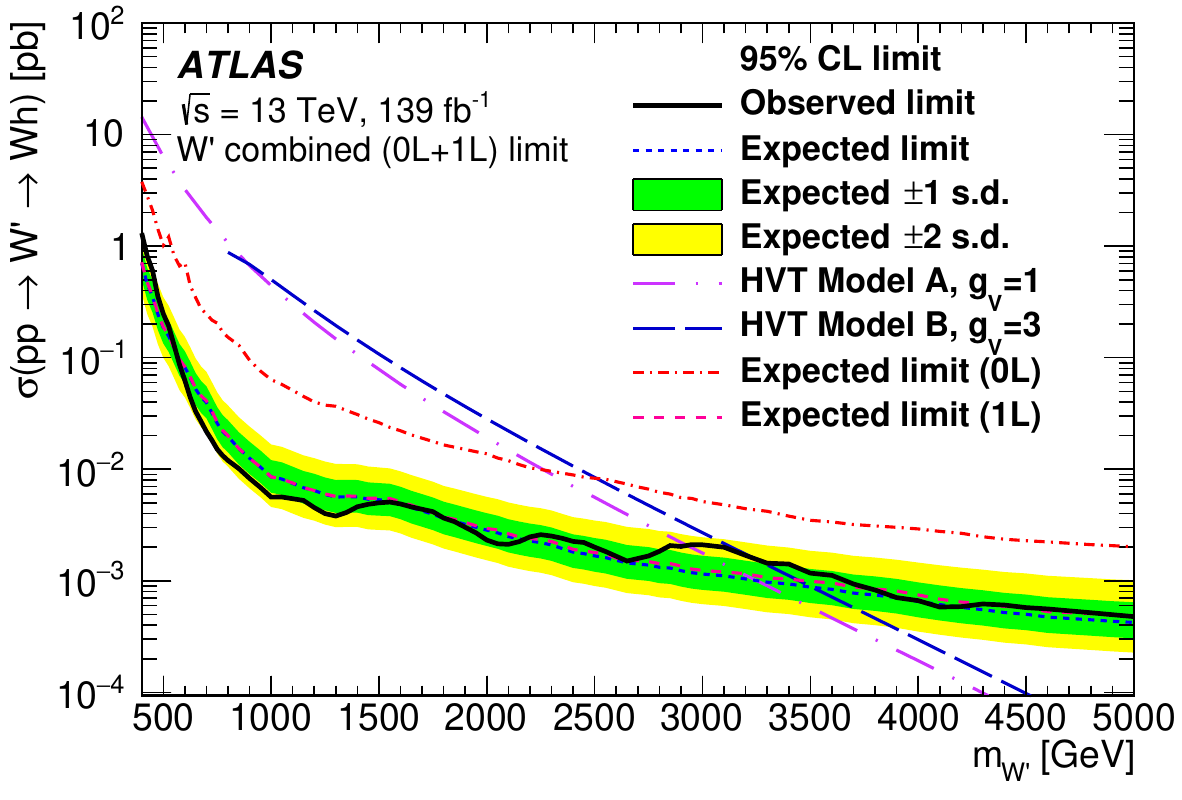}
\caption{
Upper limits at 95\% CL on $\sigma(pp \to Z')\mathcal{B}(Z'\to Zh)$ combining the 0-lepton and 2-lepton channels (left), and on $\sigma(pp \to W')\mathcal{B}(W'\to Wh)$ combining the 0-lepton and 1-lepton channels (right) (Run-2, 13\,TeV).
For both searches, $\mathcal{B}(h \to b\bar{b}/c\bar{c}) = 0.598$ is assumed.
From Ref.~\cite{ATLAS:2022enb}.
}
\label{fig:ATLAS_VV_limit}
\end{figure}

\subsection{$A\rightarrow Z(\rightarrow\ell\ell)h(\rightarrow \tau\tau)$, SM h}
A search has been performed in the $A\rightarrow Z(\rightarrow\ell\ell)h(\rightarrow \tau\tau)$ channel~\cite{CMS2025lltautau}, using the full CMS Run-2 dataset at 13\,TeV.
The pseudoscalar $A$ boson is produced via gluon-gluon fusion ($gg \to A$) and via associated production with a bottom quark-antiquark pair ($b\bar{b}A$). In each case, the $A$ boson decays to an SM-like $h$ boson and a $Z$ boson. The $\tau\tau$ final state is sensitive to $h$ decays because $\mathcal{B}(h \to \tau\tau) = 0.062$ is precisely predicted in the SM. Events are split into $b$-tag and no-$b$-tag categories to separately target the $gg \to A$ and $b\bar{b}A$ production modes.

\begin{figure}[H]
\centering
\includegraphics[width=0.49\linewidth]{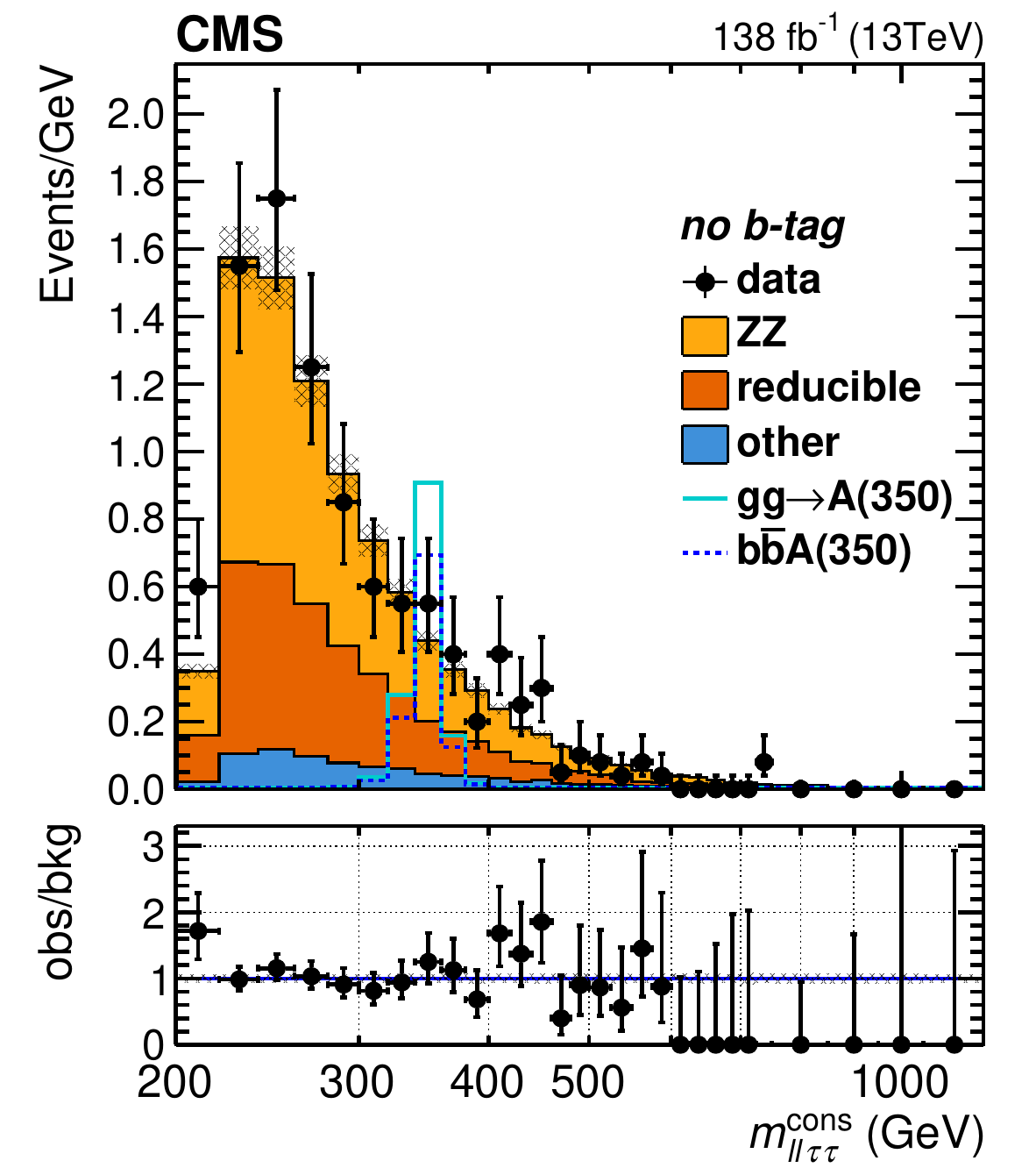}
\includegraphics[width=0.49\linewidth]{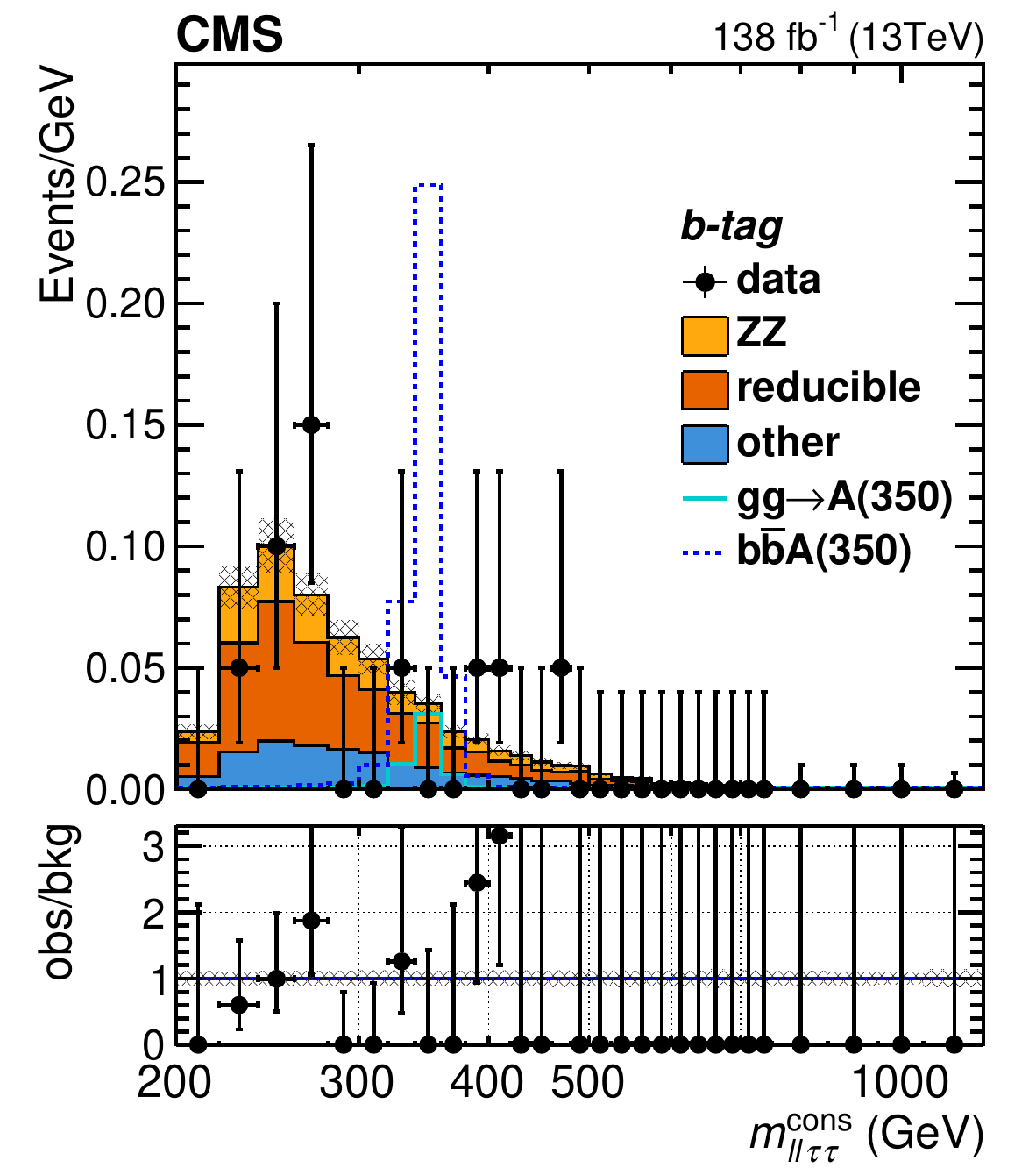}
\caption{
The reconstructed four-lepton mass $m_{\ell\ell\tau\tau}^{\text{cons}}$
in the no $b$-tag (left) and $b$-tag (right) categories (Run-2, 13\,TeV). Background
distributions are shown after a background-only maximum-likelihood fit to the data. Signal samples for $m_A = 350~\text{GeV}$ are overlaid with $\sigma\mathcal{B}(A \to Zh) = 1~\text{pb}$ for both $gg \to A$ and $b\bar{b}A$. Hatched bands indicate total background uncertainties; bin contents are divided by the bin width.
From Ref.~\cite{CMS2025lltautau}.
}
\label{fig:CMS_lltautau_mass}
\vspace{-5mm}
\end{figure}

The reconstructed invariant mass of the $\ell\ell\tau\tau$ system is shown in Fig.~\ref{fig:CMS_lltautau_mass}.

\begin{figure}[H]
\centering
\includegraphics[width=0.49\linewidth]{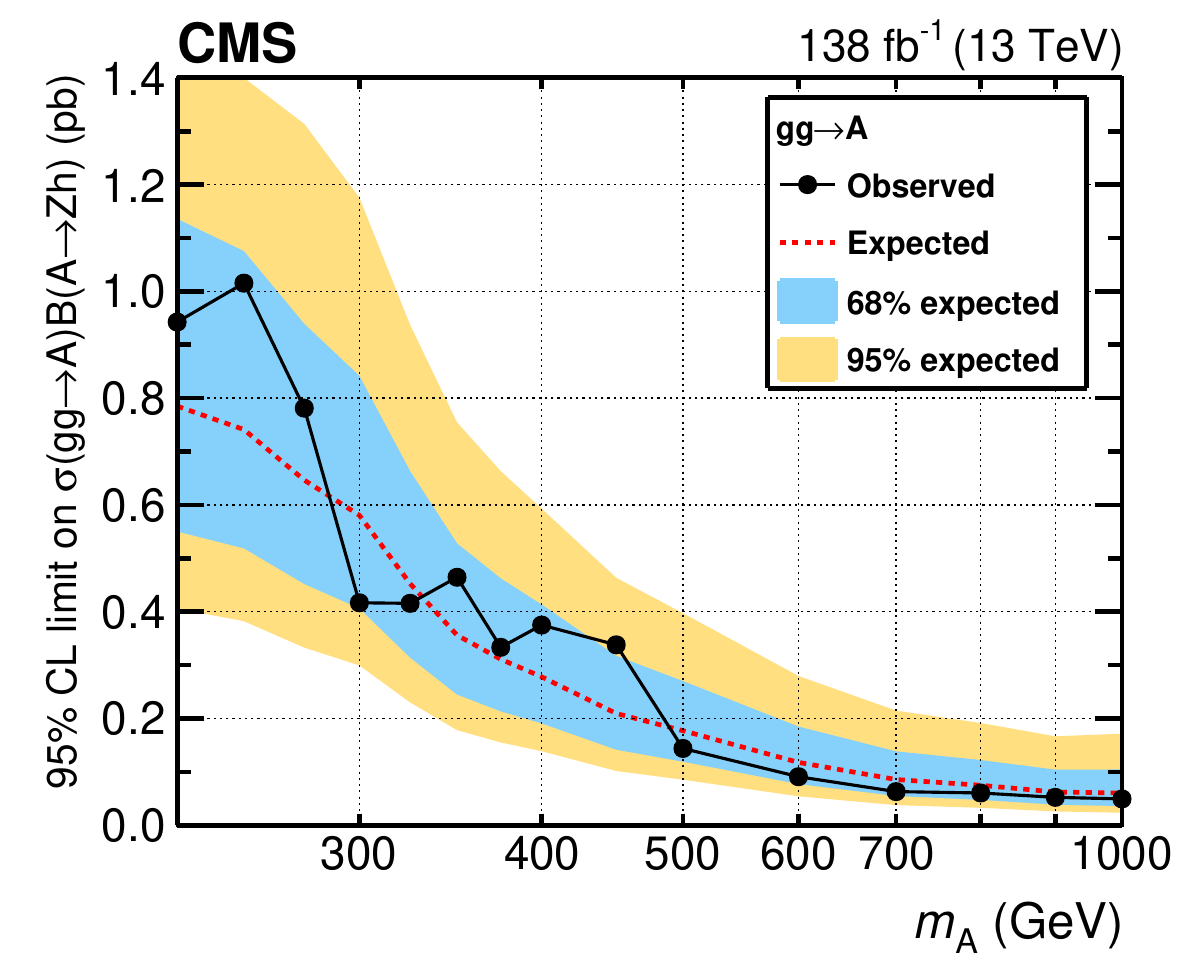}
\includegraphics[width=0.49\linewidth]
{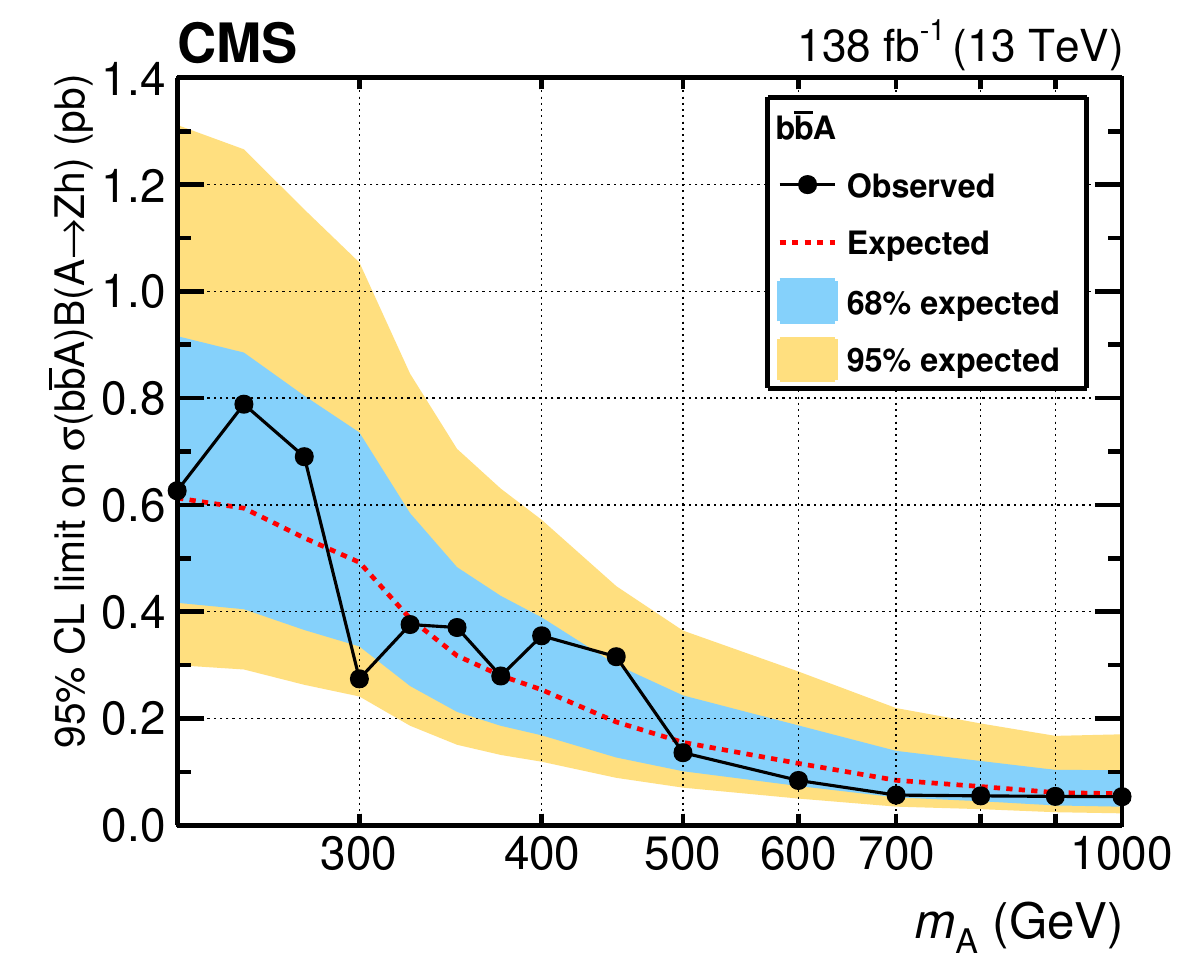}
\caption{
Expected (dashed) and observed (solid) 95\% CL upper limits on $\sigma\mathcal{B}(A \to Zh)$ for $gg \to A$ (left) and $b\bar{b}A$ (right) as functions of $m_A$ (Run-2, 13\,TeV). The rate of the other production mode is fixed to zero in each case. $\mathcal{B}(h \to \tau\tau) = 0.062$ (SM value) is assumed.
From Ref.~\cite{CMS2025lltautau}.
}
\label{fig:CMS_lltautau_limit}
\end{figure}

\begin{figure}[H]
\centering
\includegraphics[width=0.4\linewidth]{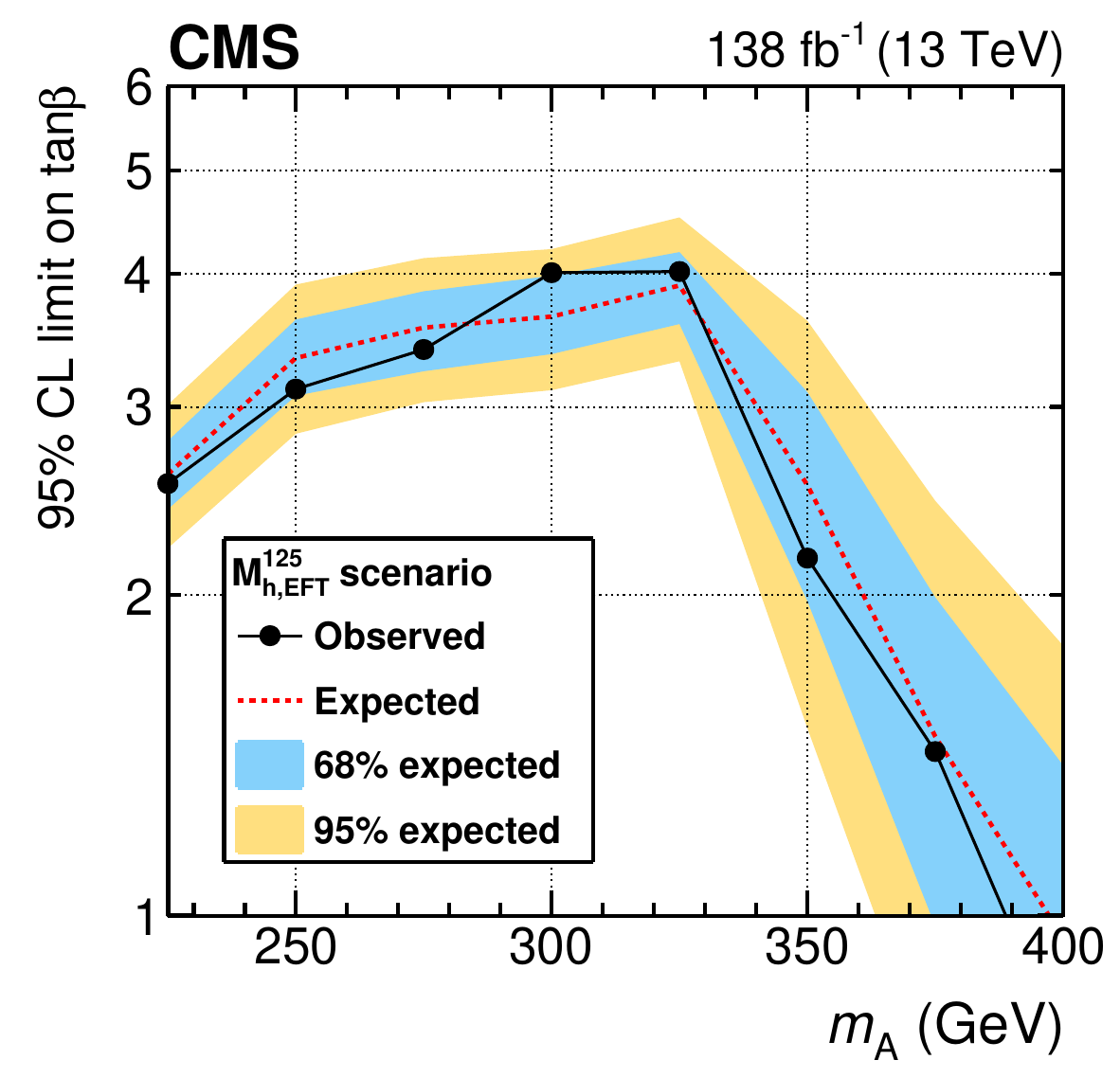}
\includegraphics[width=0.59\linewidth]
{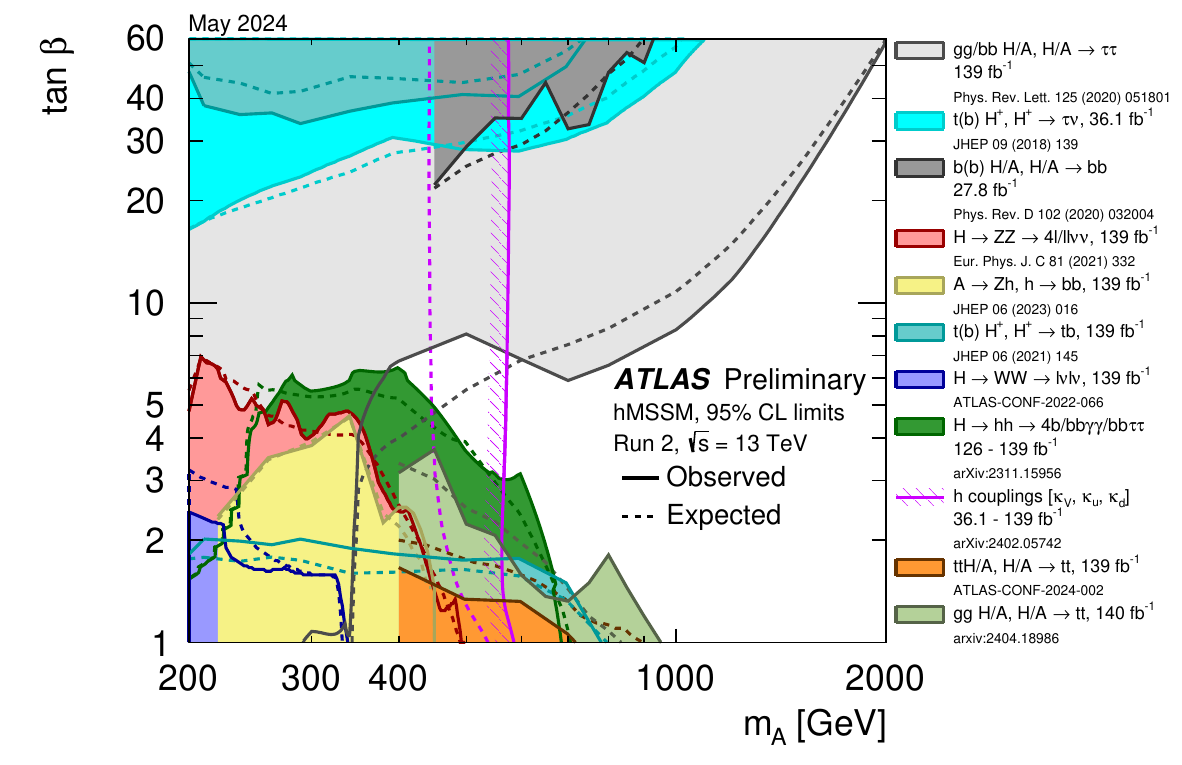}
\caption{
Left: Lower 95\% CL limit on $\tan\beta$ as a function of $m_A$ in the
$M_h^{125,\text{EFT}}$ MSSM scenario (Run-2, 13\,TeV). Values below the solid line are excluded at 95\% CL.
Right: Overview of exclusion regions in the $[m_A, \tan\beta]$ plane of the hMSSM from direct heavy Higgs boson searches and measurements of the 125\,GeV Higgs boson rates. Both observed (solid) and expected (dashed) limits are shown; shaded or hatched regions indicate observed exclusions. Cross-sections for $gg$-fusion and $b$-associated production are calculated at up to NNLO QCD using \textsc{SusHi}, and branching ratios using \textsc{HDECAY}.
From Ref.~\cite{CMS2025lltautau,ATL-PHYS-PUB-2025-042}.
}
\label{fig:CMS_lltautau_limit2}
\vspace{0mm}
\end{figure}

Expected and observed limits on the production cross-section for $A\rightarrow Z(\rightarrow\ell\ell)h(\rightarrow \tau\tau)$ are shown in Fig.~\ref{fig:CMS_lltautau_limit} separately for the $gg \to A$ and $b\bar{b}A$ production modes.
Figure~\ref{fig:CMS_lltautau_limit2} shows corresponding limits in the $(m_A, \tan\beta)$ plane of the $M_h^{125,\text{EFT}}$ MSSM scenario, and an overview plot including
$A\rightarrow Z(\rightarrow\ell\ell)h(\rightarrow b\bar{b})$.

\subsection{$A\rightarrow Z(\rightarrow\ell\ell,\nu\nu)H(\rightarrow t\bar{t},b\bar{b})$, heavy H}

A search for $A\rightarrow Z(\rightarrow\ell\ell,\nu\nu)H(\rightarrow t\bar{t},b\bar{b})$ production via gluon-gluon fusion and $b\bar{b}A$ production was conducted in the final state with two $t$-quarks and two leptons, or two or three $b$-quarks with missing transverse energy, using the full \mbox{ATLAS} Run-2 dataset at 13\,TeV~\cite{ATLAS:2023zkt}. Here $H$ denotes a heavy (non-SM) Higgs boson. The ATLAS Collaboration observed
the largest excess over the SM background prediction with a local significance of 2.85
standard deviations, corresponding to
$(m_A, m_H) = (650, 450)$\,GeV{. The excess arises in the $\ell\ell t\bar{t}$ channel, in events containing two same-flavour opposite-sign leptons consistent with a $Z$ boson decay and a pair of top quarks reconstructed from their decay products. The signal region is defined by requiring the invariant mass of the $\ell\ell t\bar{t}$ system to be consistent with $m_A \approx 650$\,GeV and the $t\bar{t}$ invariant mass to be consistent with $m_H \approx 450$\,GeV. The global significance, accounting for the look-elsewhere effect, is below $2\sigma$.}

\begin{figure}[H]
\centering
\includegraphics[width=0.49\linewidth]{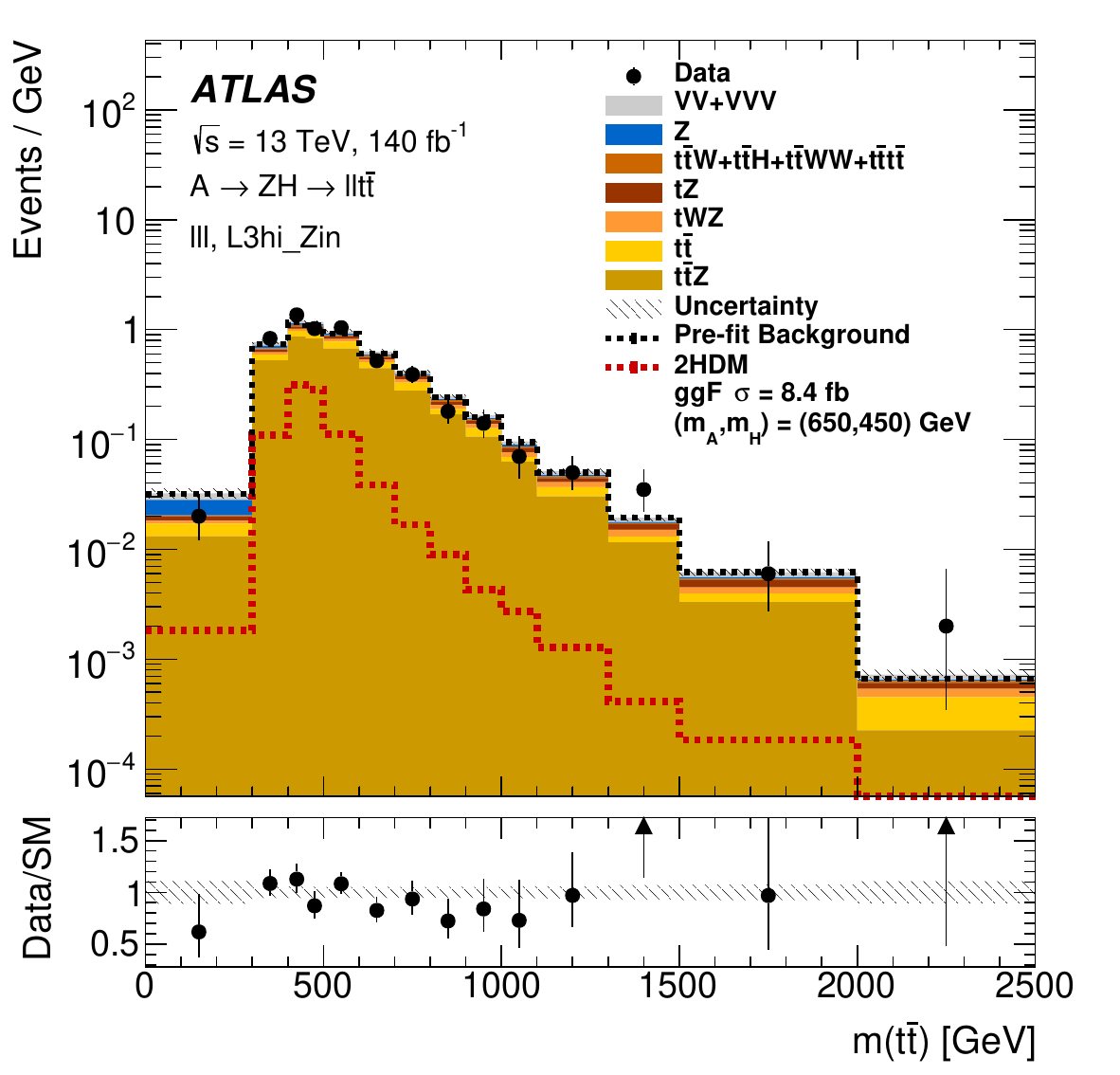}
\includegraphics[width=0.49\linewidth]{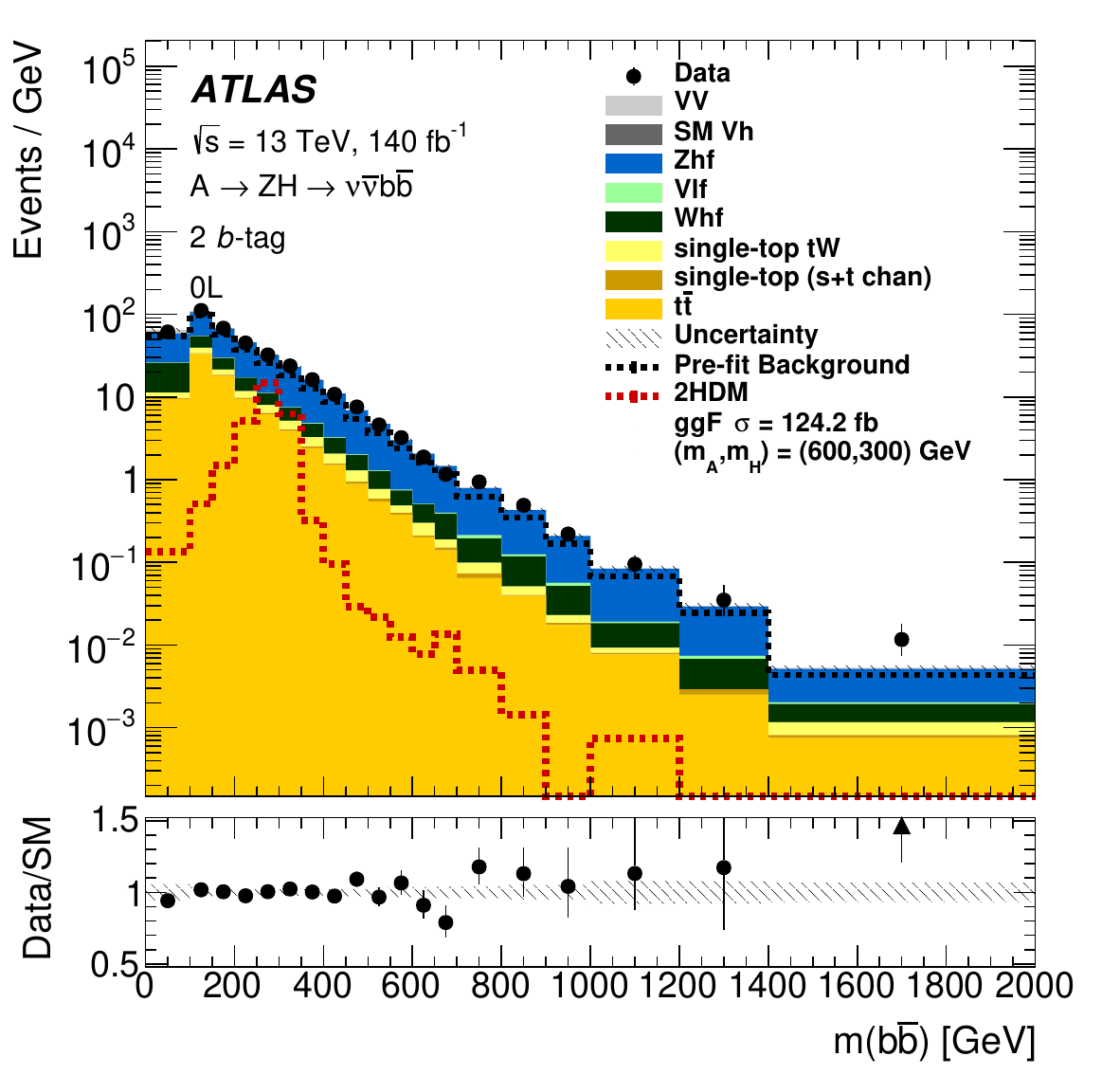}
\caption{
Post-fit distributions (Run-2, 13\,TeV): $m_{tt}$ in the $\ell^+\ell^- t\bar{t}$
channel (left); $m_{bb}$ in the $\nu\nu b\bar{b}$ channel (two-$b$-tag, right).
Signal distributions for $gg$-fusion or $b\bar{b}A$ production, normalized to theoretical cross-sections, are shown for comparison.
Data are shown as black points with statistical error bars. The hatched band indicates the combined statistical and systematic uncertainty in the total background.
From Ref.~\cite{ATLAS:2023zkt}.
}
\label{fig:ATLAS_lltautau_mass}
\vspace{-10mm}
\end{figure}

The reconstructed invariant mass of the $t\bar{t}$ system in the $\ell^+\ell^- t\bar{t}$ channel, and the invariant mass of the $b\bar{b}$ jets in the $\nu\nu b\bar{b}$ channel, are shown in Fig.~\ref{fig:ATLAS_lltautau_mass}. The data are well described by the background model except in the region of the reported excess.

\begin{figure}[H]
\vspace{-3mm}
\centering
\includegraphics[width=0.49\linewidth]{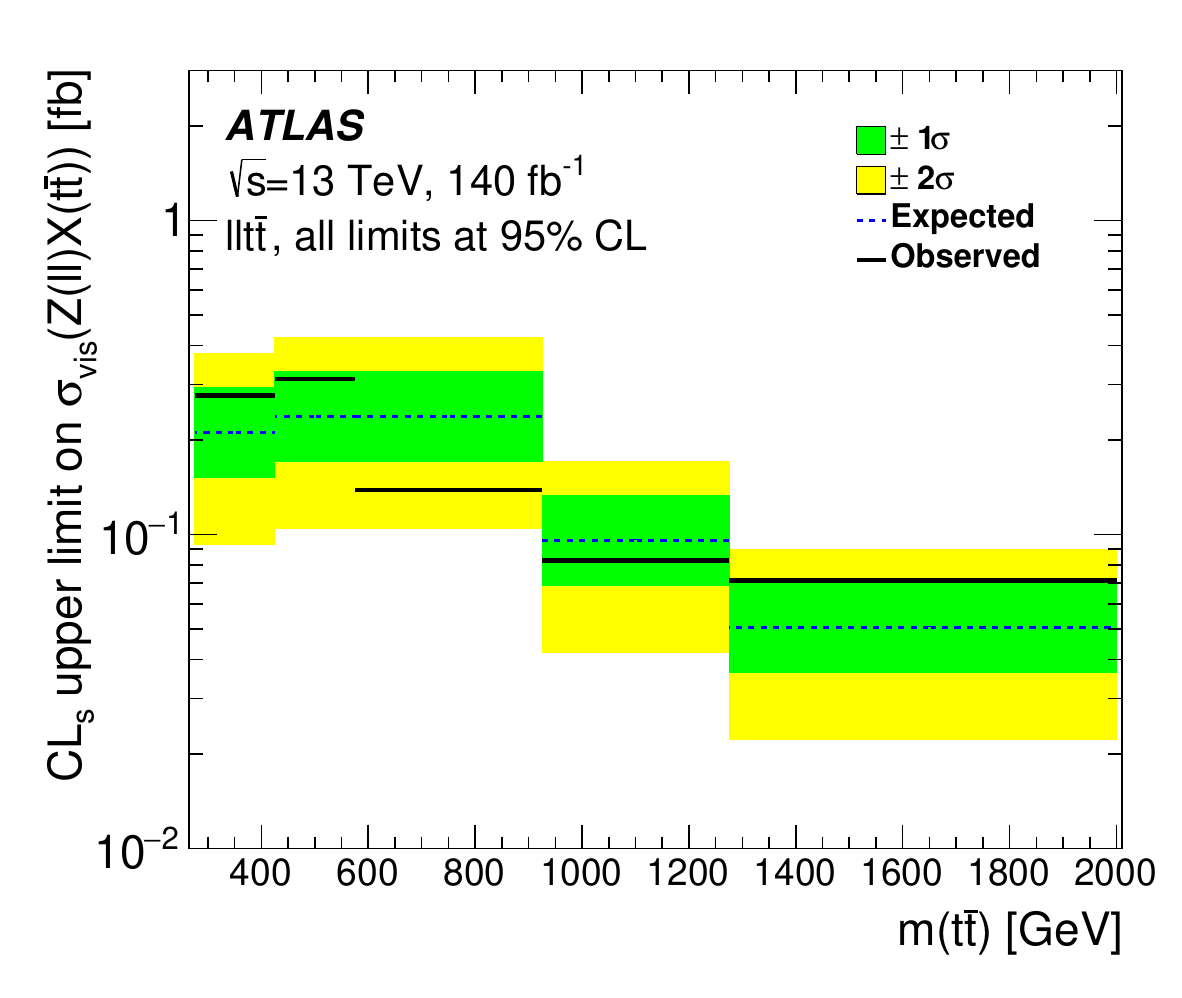}
\includegraphics[width=0.49\linewidth]{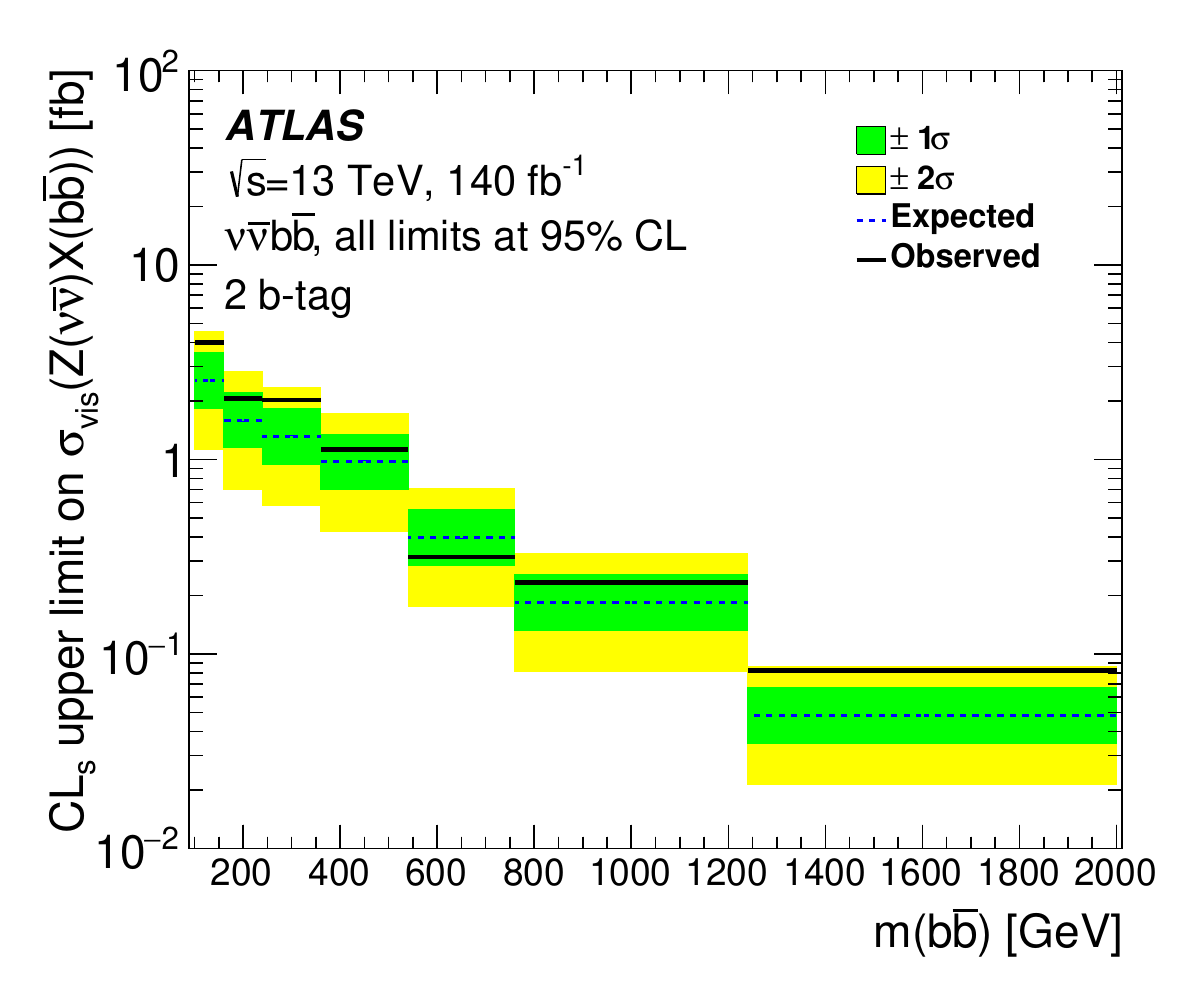}
\vspace{-3mm}
\caption{
Observed (solid) and expected (dashed) 95\% CL upper limits on $\sigma_{\text{vis}}(Z(\ell^+\ell^-)X(t\bar{t}))$ (left) and
$\sigma_{\text{vis}}(Z(\nu\nu)X(b\bar{b}))$ from the two-$b$-tag region (right) (Run-2, 13\,TeV).
The limits are derived in bins of the reconstructed $m_{tt}$ and $m_{bb}$ distributions.
The inner (green) and outer (yellow) bands indicate the $\pm1\sigma$ and $\pm2\sigma$ intervals around the expected limit.
From Ref.~\cite{ATLAS:2023zkt}.
}
\label{fig:ATLAS_lltt_limit}
\end{figure}

\begin{figure}[H]
\vspace{-5mm}
\centering
\includegraphics[width=0.49\linewidth]{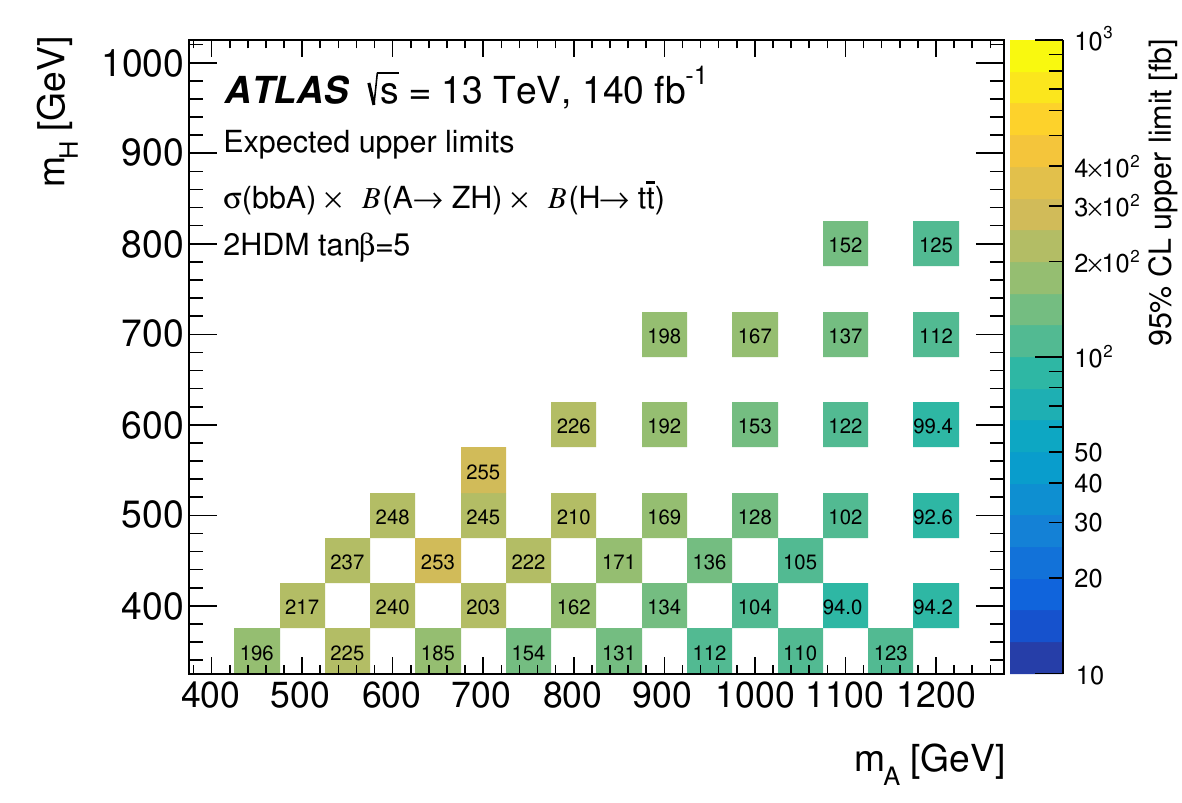}
\includegraphics[width=0.49\linewidth]{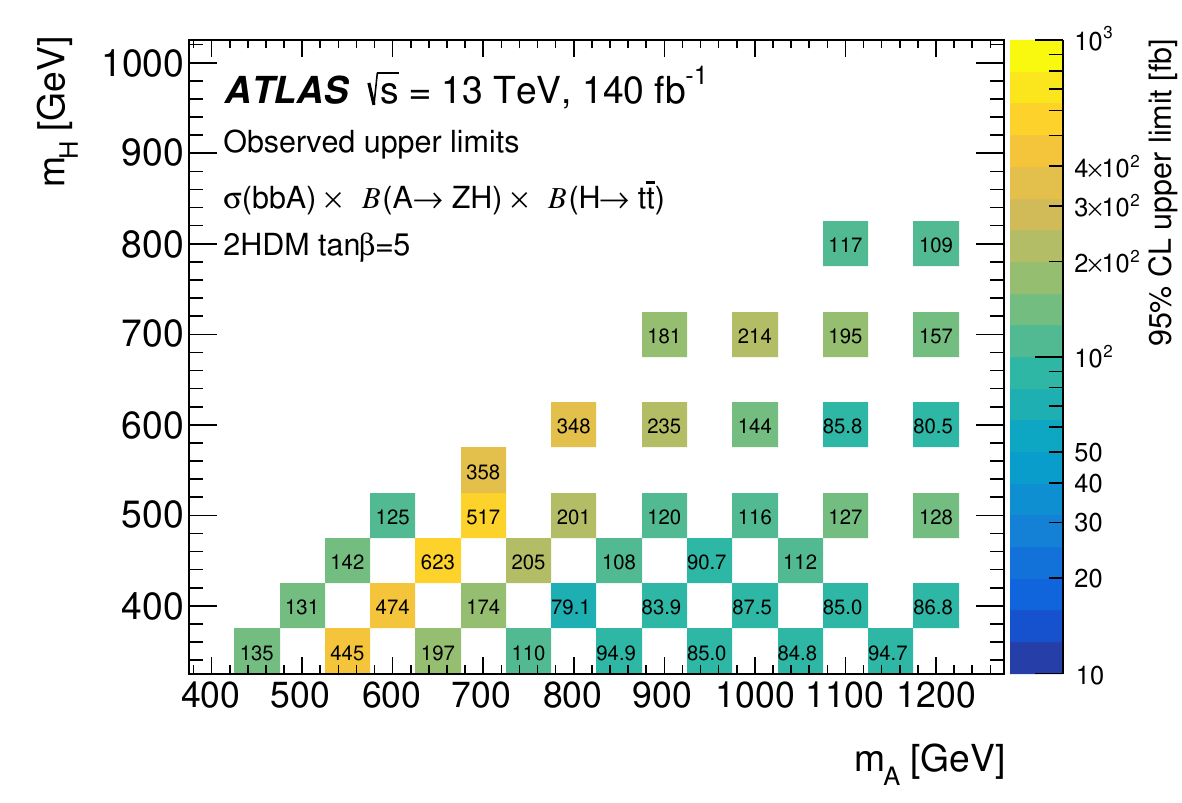}
\vspace{-2mm}
\caption{Expected (left) and observed (right) 95\% CL upper limits on
$\sigma(b\bar{b}A)\times\mathcal{B}(A \to ZH)\times\mathcal{B}(H \to t\bar{t})$
for $\tan\beta = 5$ (Run-2, 13\,TeV). The value of $\tan\beta$ affects only the choice of $A$ boson width. {The comparison of the expected and observed exclusion planes highlights the location of the excess reported by ATLAS. Around $(m_H,m_A)\approx(450,650)\,\mathrm{GeV}$, the observed exclusion is weaker than expected, corresponding to the local excess with a significance of $2.85\sigma$ discussed in the text.}
From Ref.~\cite{ATLAS:2023zkt}.
}
\label{fig:ATLAS_lltt_limit2}
\vspace{-1mm}
\end{figure}

Observed and expected 95\% CL upper limits on $\sigma_{\text{vis}}(Z(\ell^+\ell^-)X(t\bar{t}))$ and $\sigma_{\text{vis}}(Z(\nu\nu)X(b\bar{b}))$ are shown in Fig.~\ref{fig:ATLAS_lltt_limit}. Figure~\ref{fig:ATLAS_lltt_limit2} shows the cross-section limits in the $(m_H, m_A)$ plane.

\subsection{$A\rightarrow Z(\rightarrow\ell\ell)H(\rightarrow t\bar{t})$, $\Delta m \approx m_A-m_H$}
A search for the process $A\rightarrow Z(\rightarrow\ell\ell)H(\rightarrow t\bar{t})$
was also conducted by the CMS Collaboration, using the full CMS Run-2 dataset at 13\,TeV~\cite{CMS:2025ttZ}. This analysis specifically targets the phase-space region where $\Delta m = m_A - m_H$ is close to $m_Z$, which leads to a $Z$ boson produced nearly at rest in the $A$ rest frame. The key discriminating variable is $p_T^Z \times \Delta m$, which combines the $Z$ boson's transverse momentum with the mass difference.

\begin{figure}[H]
\vspace{-6mm}
\centering
\includegraphics[width=0.64\linewidth]{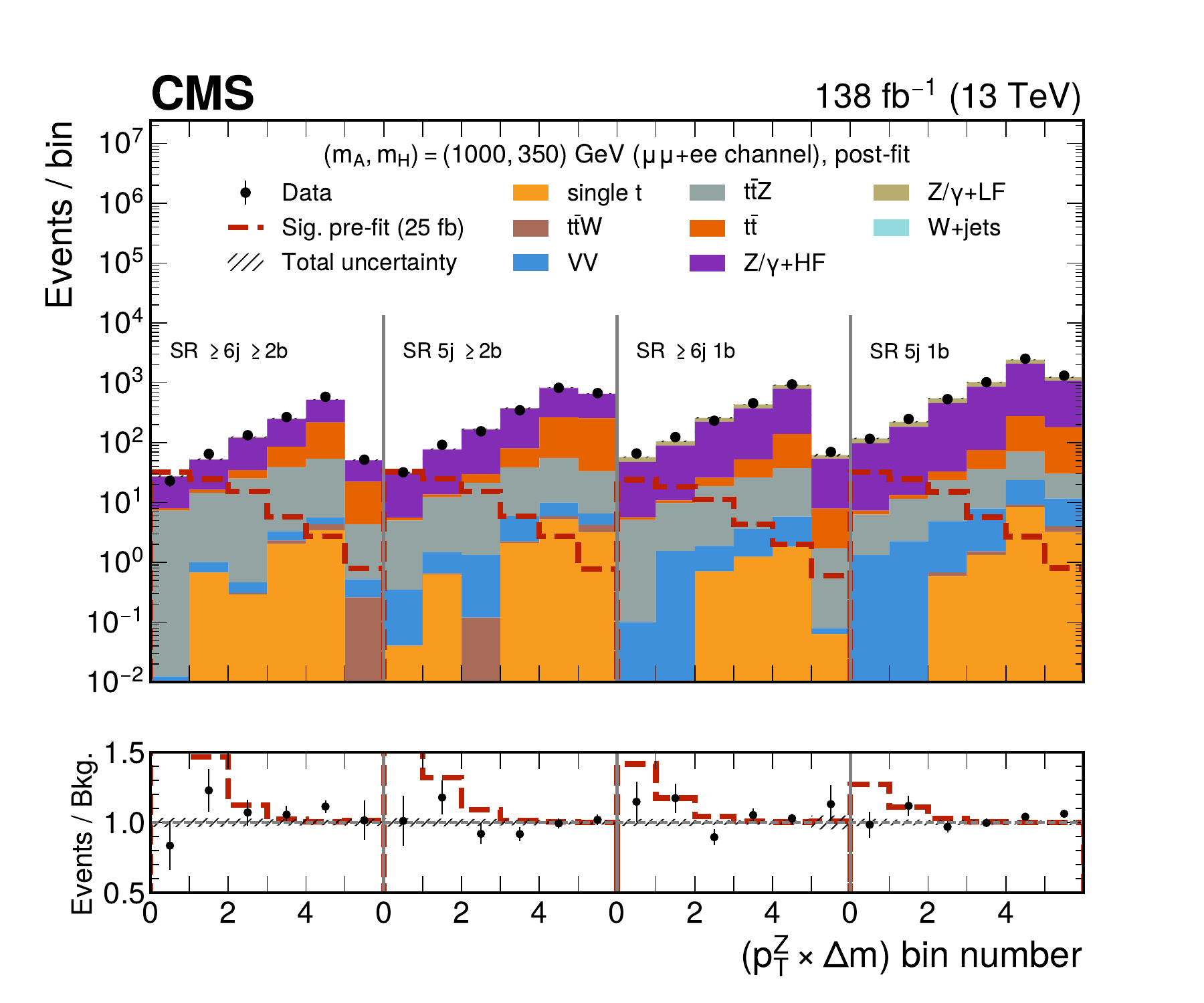}
\vspace{-2mm}
\caption{
Distributions of $p_{\text{T}}^{Z}\times\Delta m$ in the signal regions after the
fit to data (Run-2, 13\,TeV), for $m_A = 1000\,\text{GeV}$ and $m_H = 350\,\text{GeV}$.
Signal (solid red) and background (colored histograms) are shown with best-fit normalizations (post-fit).
The post-fit signal cross-section is $0.0 \pm 0.1\,\text{fb}$ in the left panel; accordingly the signal histogram is not displayed. The signal is also shown pre-fit, normalized to 25\,fb (dashed red).
Lower panels show the ratio of data to background (black points) and the expected signal-plus-background to background ratio (red lines). Hatched areas represent total uncertainties.
From Ref.~\cite{CMS:2025ttZ}.
\label{fig_CMS_lltt_mass}
}
\vspace{-10mm}
\end{figure}

Figure~\ref{fig_CMS_lltt_mass} shows the
distributions of $p_{\text{T}}^{Z}\times\Delta m$ in the signal regions.
The excess of events reported by the ATLAS Collaboration~\cite{ATLAS:2023zkt} at $(m_A, m_H) = (650, 450)$\,GeV
is not observed by CMS~\cite{CMS:2025ttZ}.

\begin{figure}[H]
\vspace{-4mm}
\centering
\includegraphics[width=0.49\linewidth]{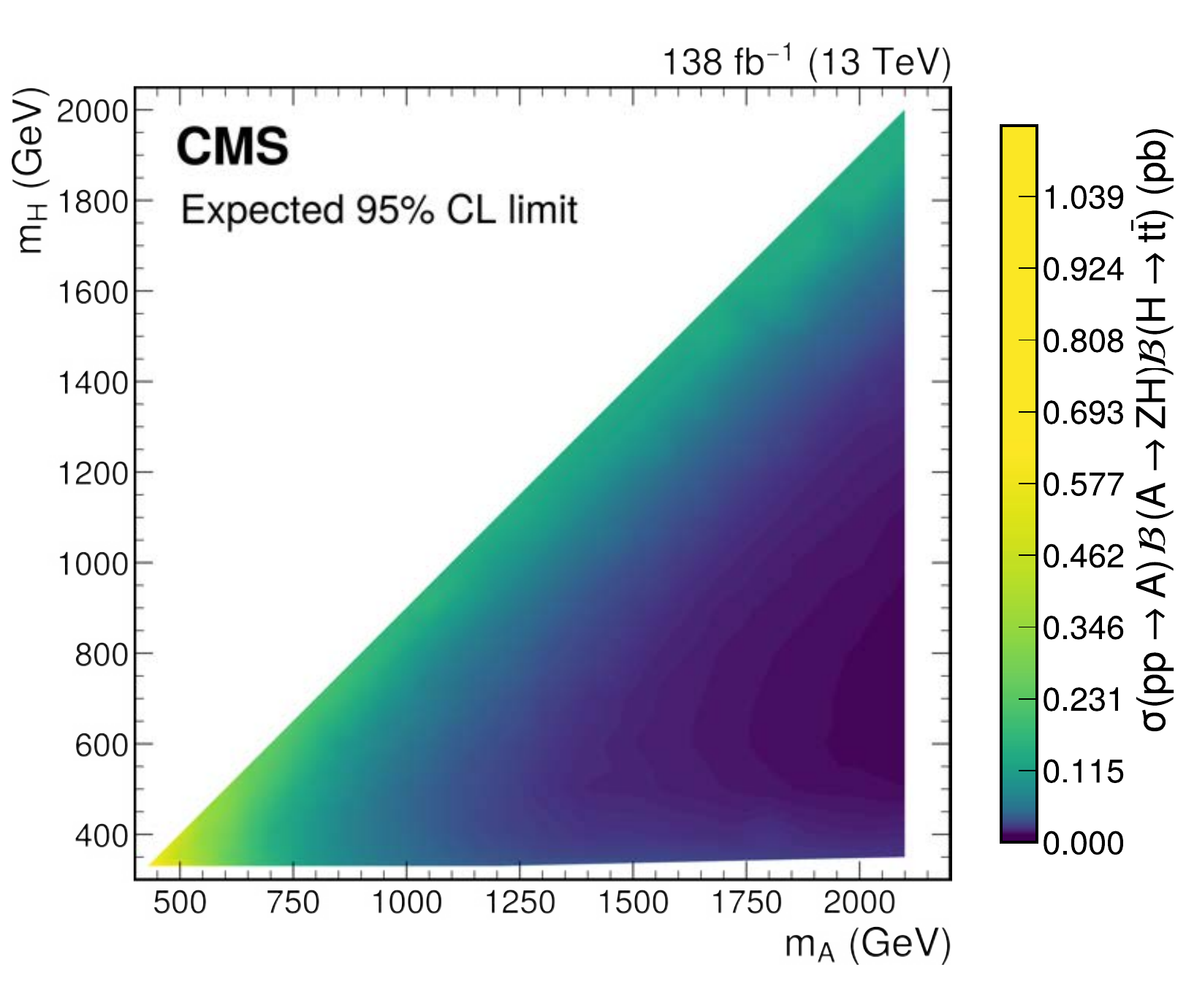}
\includegraphics[width=0.49\linewidth]{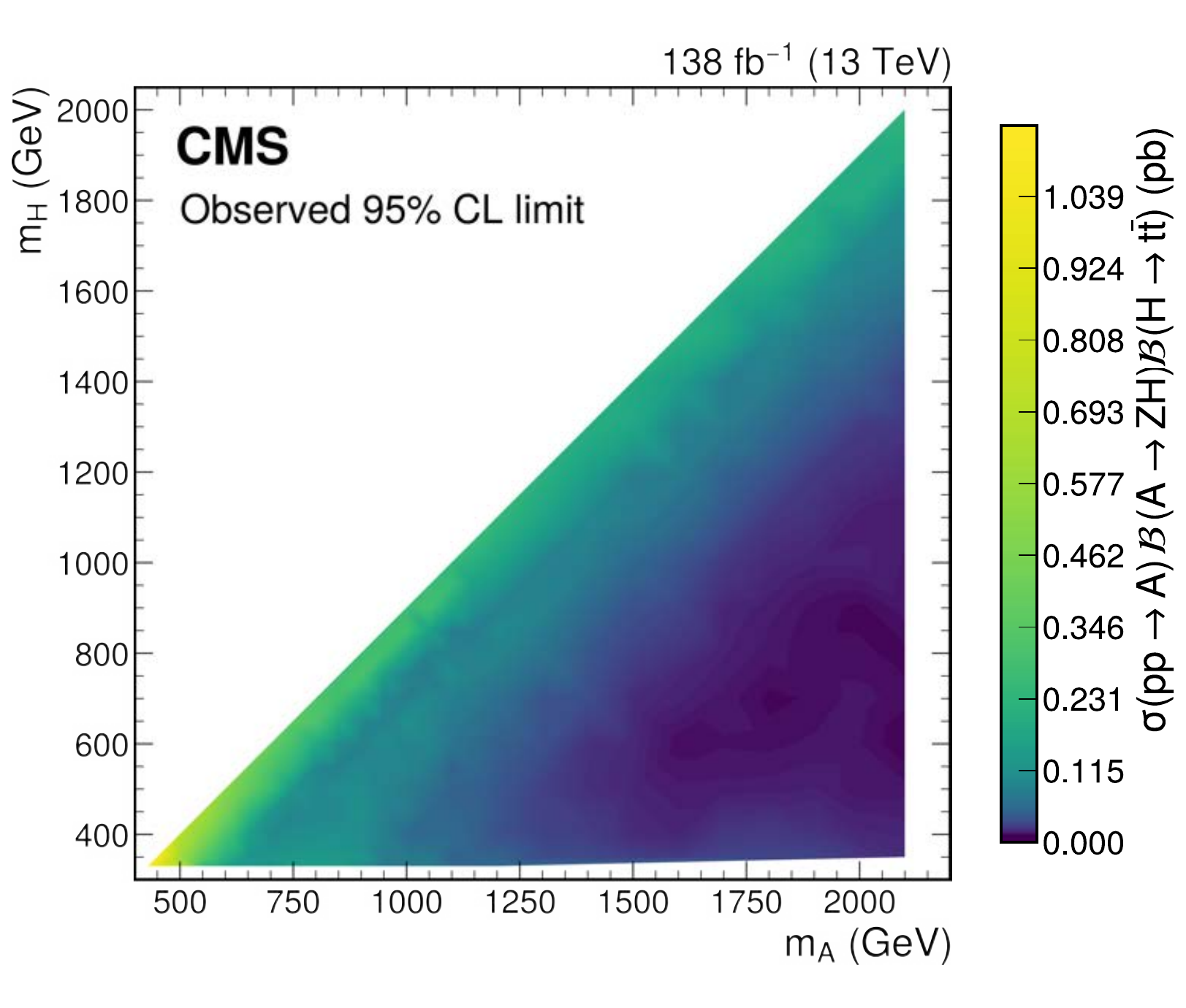}
\vspace{-4mm}
\caption{
Expected (left) and observed (right) 95\% CL upper limits on the
product of production cross-section and branching fractions for the
$A \to ZH \to Zt\bar{t}$ process in the $(m_A, m_H)$ plane (Run-2, 13\,TeV).
From Ref.~\cite{CMS:2025ttZ}.
}
\label{fig:CMS_lltt}
\end{figure}

Expected and observed limits are shown in Fig.~\ref{fig:CMS_lltt} in the
$(m_A, m_H)$ plane.
With no signal observed, the expected and observed limits agree within uncertainties.

\section{Heavy Charged Higgs Boson Decaying into Another Neutral Higgs Boson and a $W$ Boson
}

The existence of charged Higgs bosons is a generic prediction of extended Higgs sectors. Because the SM Higgs mechanism requires only a single $\text{SU}(2)_L$ doublet to give masses to the $W^\pm$ and $Z$ bosons, the physical spectrum contains no electrically charged scalar. Any observation of a charged Higgs boson would therefore constitute \emph{unambiguous} evidence of physics beyond the SM. Charged Higgs bosons arise in all models with two or more Higgs doublets (e.g.\ the 2HDM and the MSSM), as well as in models with triplet representations. At the LHC, heavy charged Higgs bosons ($m_{H^\pm} \gtrsim m_t + m_b$) are predominantly produced in association with a top-bottom quark pair ($pp \to tbH^\pm$). The decay channel $H^+ \to W^+ H$ (or $H^+ \to W^+ h$) is particularly interesting because it is the charged-Higgs analog of the $A \to ZH$ decay discussed in the previous section, and is also predicted in the 2HDM. Two searches for this topology are reviewed below, using Run-2 data at 13\,TeV.

\subsection{$H^+\rightarrow h(\rightarrow b\bar{b})W(\rightarrow \ell\nu)$, SM h}

\begin{figure}[H]
\centering
\includegraphics[width=0.49\linewidth]{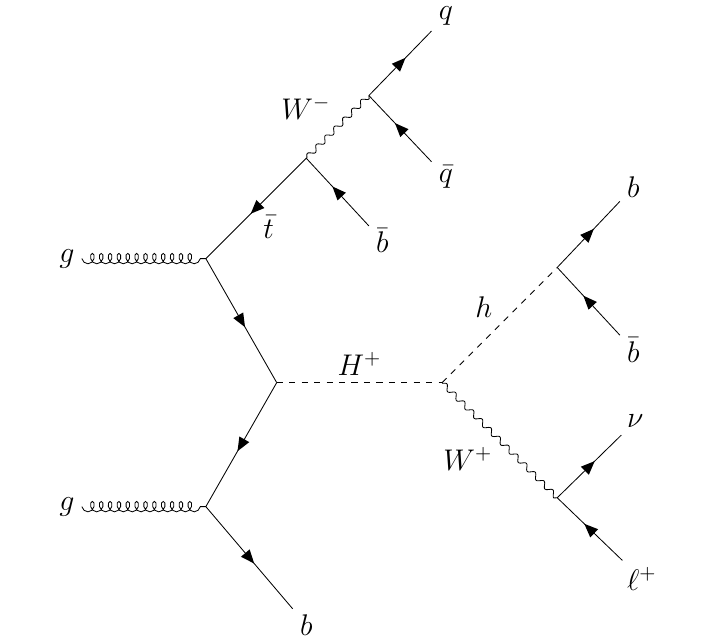}
\includegraphics[width=0.49\linewidth]{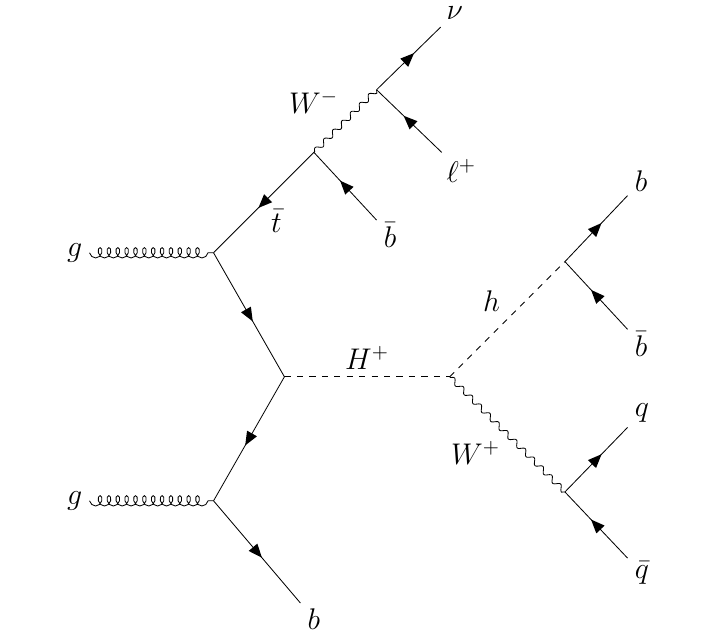}
\caption{
Representative lowest-order Feynman diagrams for $pp \to tbH^{+}$ production
and subsequent decay via
(left) $H^{+} \to W^{+}h \to \ell^{+}\nu b\bar{b}$ and
(right) $H^{+} \to W^{+}h \to q\bar{q}b\bar{b}$ (Run-2, 13\,TeV).
From Ref.~\cite{ATLAS:2024rcu}.
}
\label{fig:hplus_bb}
\end{figure}

A search for $H^+\rightarrow h(\rightarrow b\bar{b})W(\rightarrow \ell\nu)$ production was conducted in the final state with two or three $b$-quark jets and one light lepton, using the full ATLAS Run-2 dataset at 13\,TeV~\cite{ATLAS:2024rcu}. Here $h$ is the SM Higgs boson at 125\,GeV. The search covers charged Higgs boson masses from 300\,GeV to 2\,TeV, using two complementary reconstruction strategies: a \emph{resolved} analysis for $m_{H^\pm} \lesssim 900$\,GeV (where the $h\to b\bar{b}$ decay products form individual small-radius jets) and a \emph{merged} analysis at higher masses (where the boosted $h$ is reconstructed as a single large-radius jet). The corresponding Feynman diagrams are shown in Fig.~\ref{fig:hplus_bb}.

\begin{figure}[H]
\centering
\includegraphics[width=0.49\linewidth]
{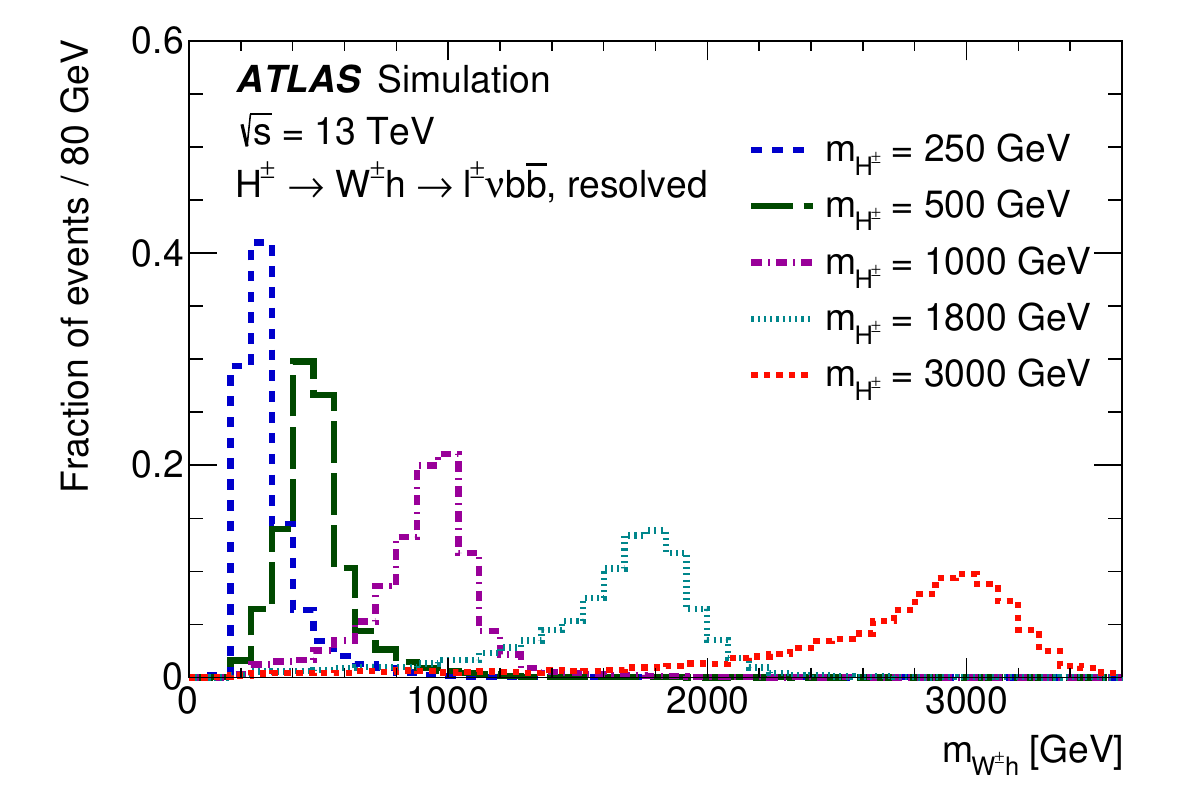}
\includegraphics[width=0.49\linewidth]
{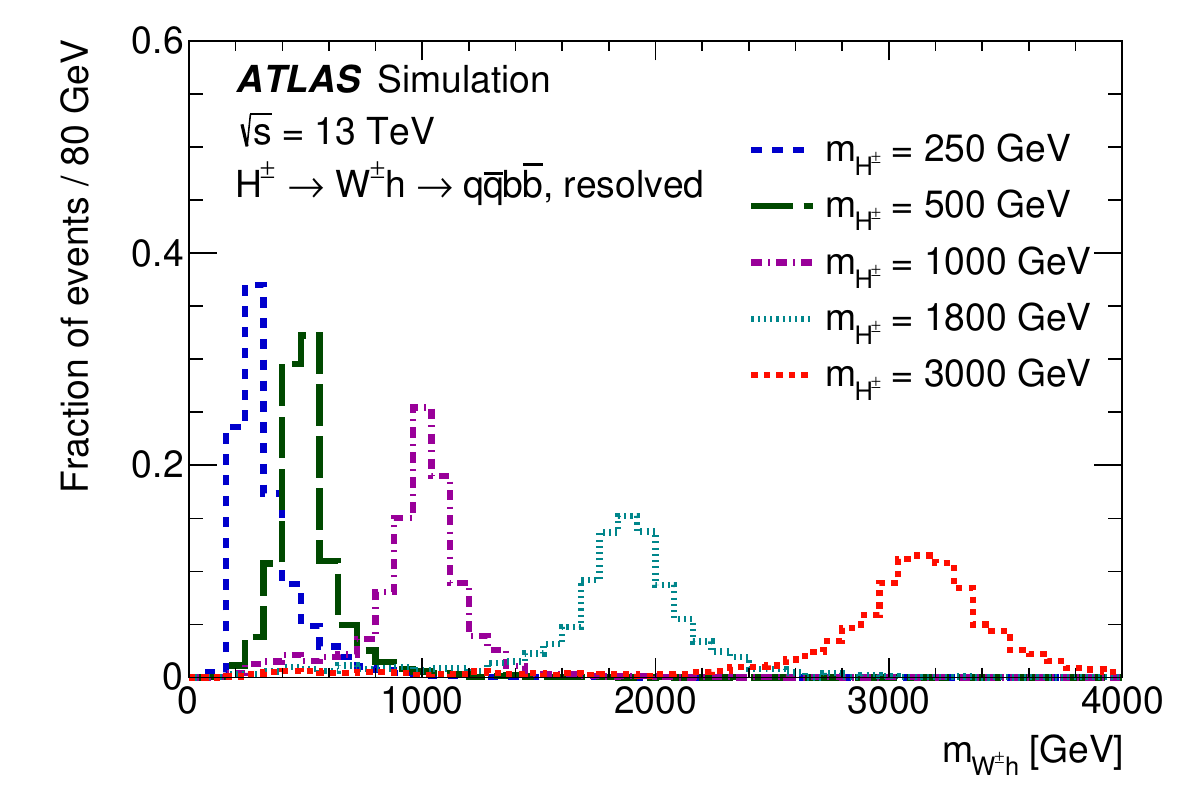}
\caption{
Invariant mass distributions for
(left) $H^{\pm} \to W^{\pm}h \to \ell^{\pm}\nu b\bar{b}$ and
(right) $H^{\pm} \to W^{\pm}h \to q\bar{q}b\bar{b}$
for selected charged Higgs boson pole masses (Run-2, 13\,TeV), reconstructed using BDTs in the resolved topology. All distributions are normalized to unit area.
From Ref.~\cite{ATLAS:2024rcu}.
}
\label{fig:ATLAS_hplus_bb}
\end{figure}

\begin{figure}[H]
\centering
\includegraphics[width=0.49\linewidth]{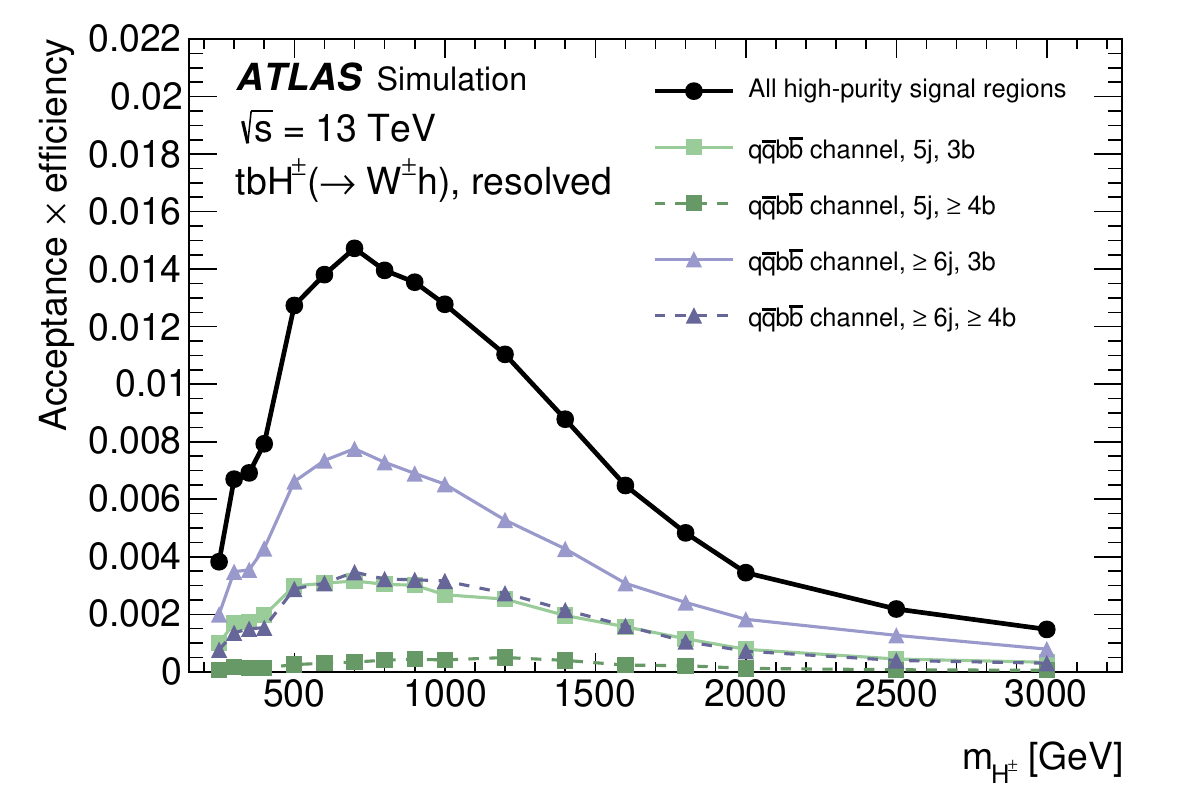}
\includegraphics[width=0.49\linewidth]{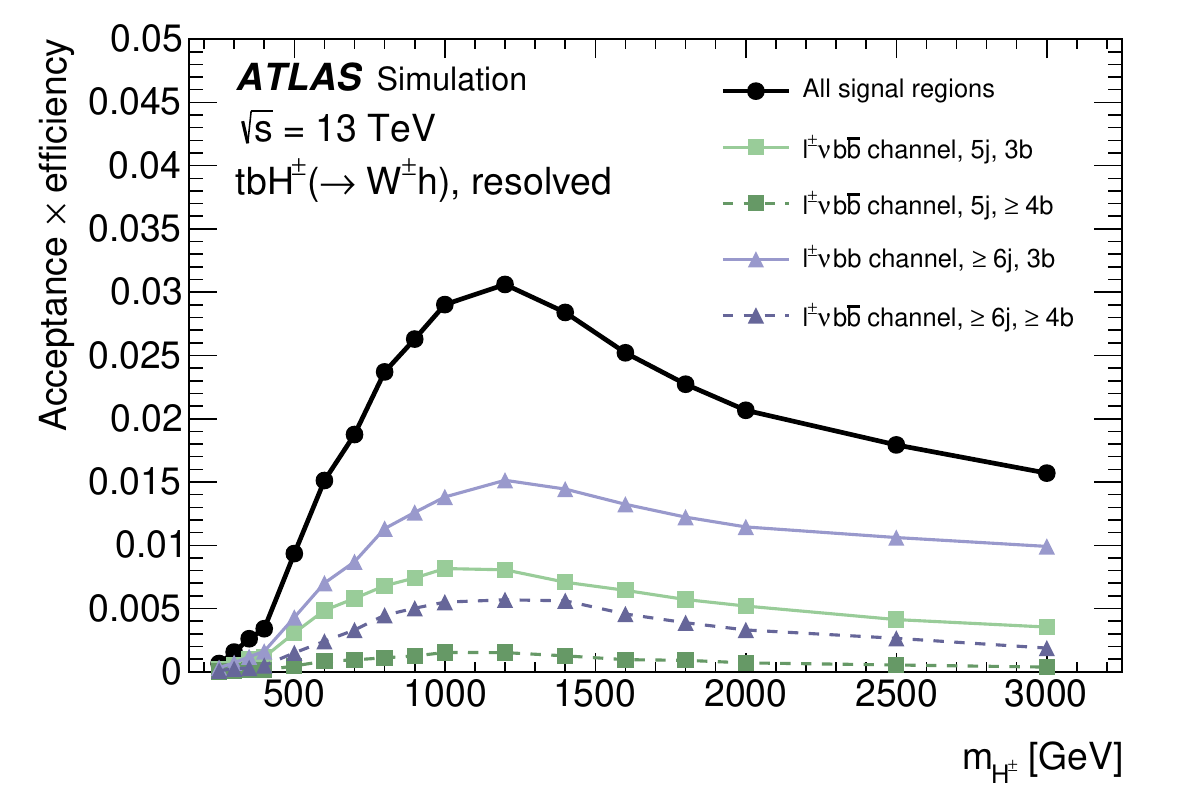}
\caption{
Product of acceptance and efficiency for $pp \to tbH^{\pm}(\to W^{\pm}h)$ as a function of $m_{H^\pm}$ for
(left) the resolved $q\bar{q}b\bar{b}$ high-purity signal regions and
(right) the resolved $\ell^{\pm}\nu b\bar{b}$ signal regions (Run-2, 13\,TeV).
From Ref.~\cite{ATLAS:2024rcu}.
}
\label{fig:ATLAS_hplus_bb_acc}
\end{figure}

The invariant mass of the $Wh$ system, $m_{Wh}$, is reconstructed using a BDT applied to events passing the resolved preselection. Figure~\ref{fig:ATLAS_hplus_bb} shows the signal $m_{Wh}$ distributions for several charged Higgs boson mass hypotheses, illustrating that the distribution peak shifts with $m_{H^\pm}$ and that the $\ell^\pm\nu b\bar{b}$ and $q\bar{q}b\bar{b}$ final states provide complementary sensitivities. Figure~\ref{fig:ATLAS_hplus_bb_acc} shows the product of acceptance and efficiency as a function of the charged Higgs boson mass for both signal regions.

\begin{figure}[H]
\centering
\includegraphics[width=0.49\linewidth]{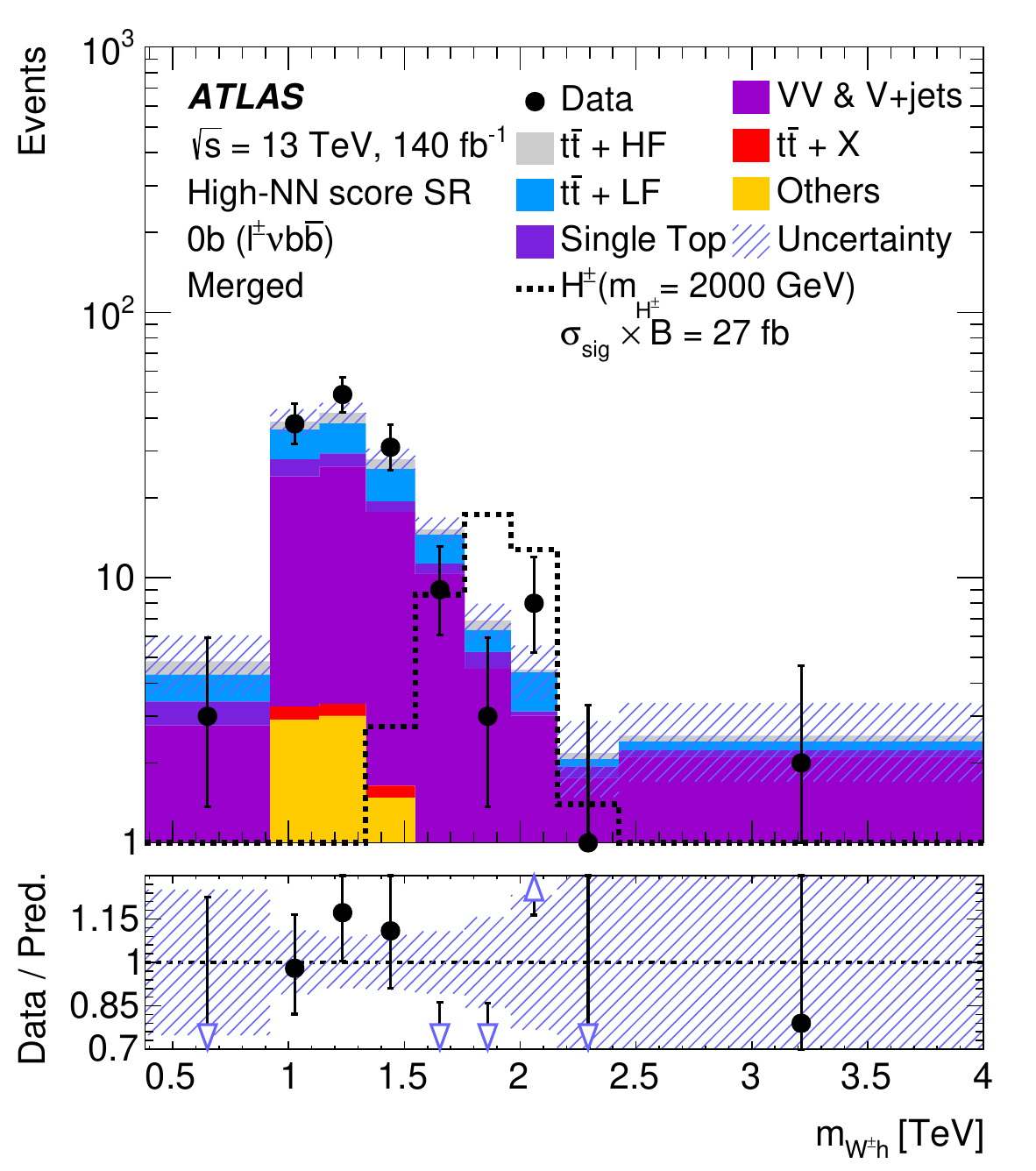}
\includegraphics[width=0.49\linewidth]{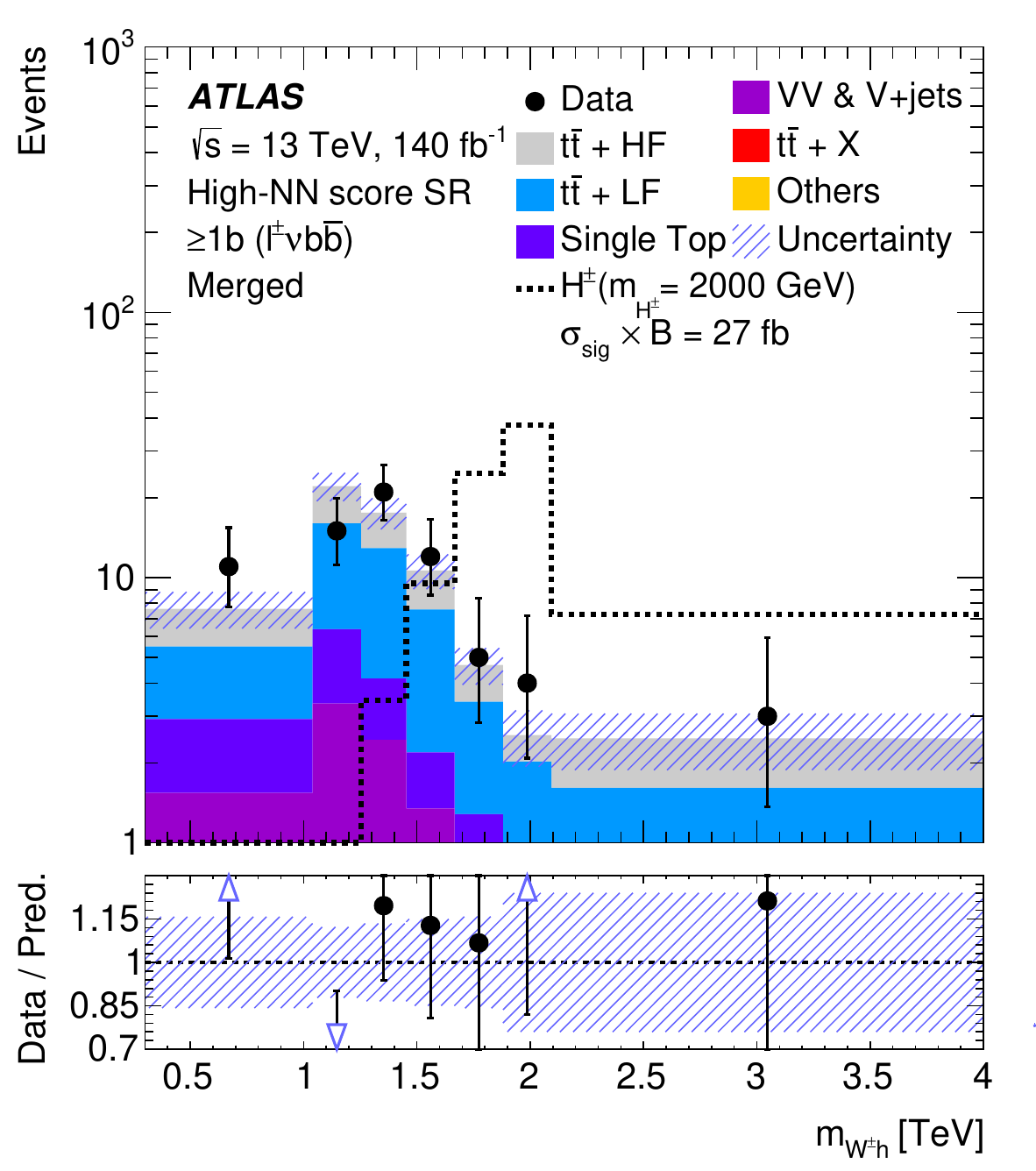}
\caption{
Post-fit distributions of $m_{Wh}$ in the high neural-network score signal regions of the merged $\ell^{\pm}\nu b\bar{b}$ event categories (Run-2, 13\,TeV).
``Others'' summarizes contributions from $tHjb$, $tWh$, $t\bar{t}t\bar{t}$, and SM $Vh$.
The shaded bands represent the total post-fit uncertainty in the background prediction.
The lower panels show the ratio of observed data to the estimated SM background.
The expected signal contribution for $m_{H^{\pm}} = 2000\,\text{GeV}$ is shown as a dashed histogram, normalized to $\sigma_{\text{sig}} \times \mathcal{B} = 27\,\text{fb}$.
From Ref.~\cite{ATLAS:2024rcu}.
}
\label{fig:ATLAS_hplus_bb2}
\end{figure}

\begin{figure}[H]
\centering
\includegraphics[width=0.7\linewidth]{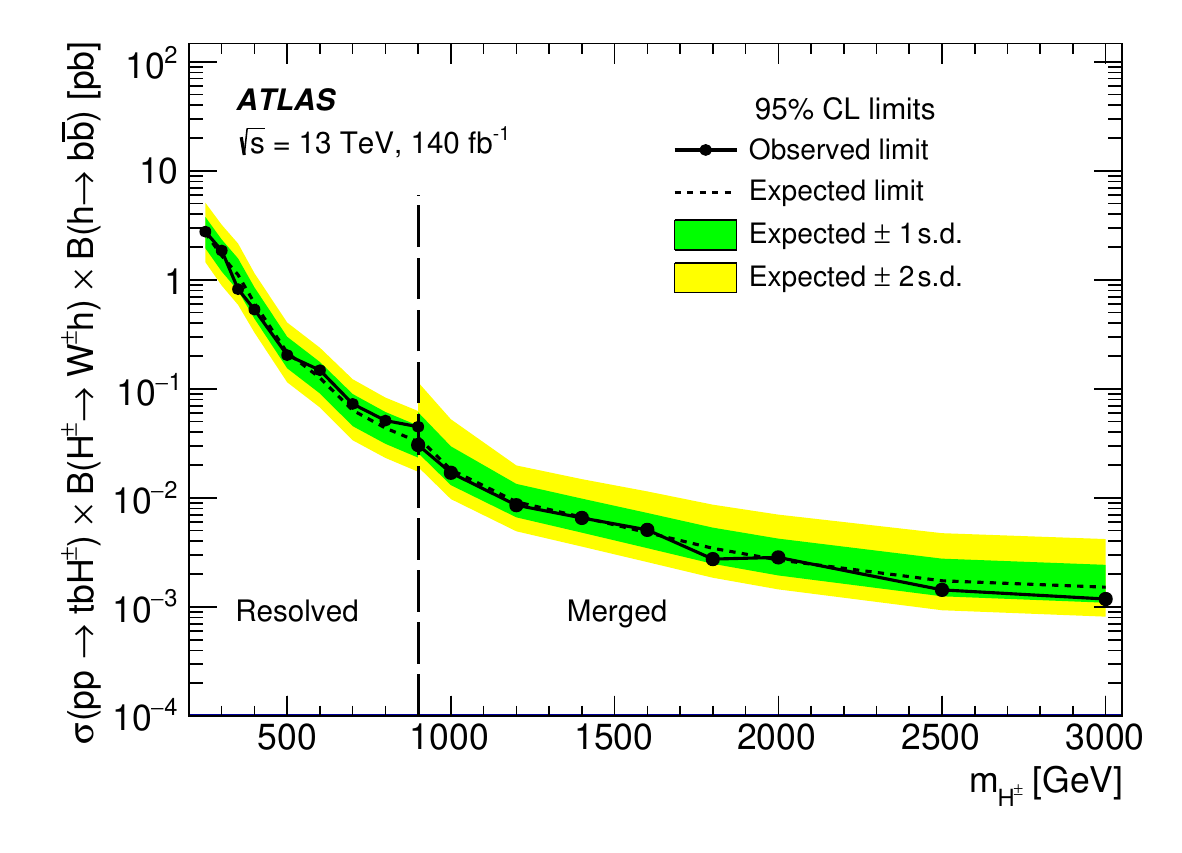}
\caption{
Upper limits at 95\% CL on $\sigma(pp \to tb H^\pm) \times \mathcal{B}(W^\pm h) \times \mathcal{B}(h \to b\bar{b})$ from the combined fit to all signal and control regions of the resolved and merged analyses (Run-2, 13\,TeV). The $\pm1\sigma$ and $\pm2\sigma$ intervals around the expected limit are shown. The resolved analysis is used up to 900\,GeV; the merged analysis is used at higher masses.
From Ref.~\cite{ATLAS:2024rcu}.
}
\label{fig:ATLAS_hplus_limit}
\end{figure}

Expected and observed limits from the combined resolved and merged analyses are shown in Fig.~\ref{fig:ATLAS_hplus_limit} as a function of $m_{H^\pm}$.

\subsection{$H^+\rightarrow H(\rightarrow\tau\tau)W(\rightarrow qq,\ell\nu)$, heavy H}

\begin{figure}[H]
\centering
\includegraphics[width=0.49\linewidth]{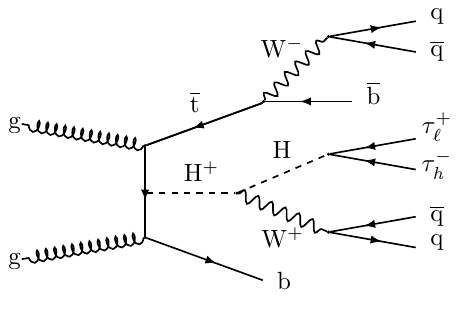}
\includegraphics[width=0.49\linewidth]{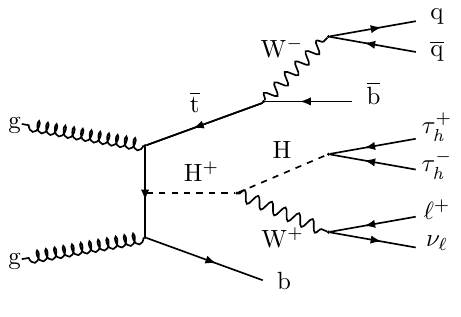}
\caption{
Feynman diagrams for the signal processes targeted by this analysis (Run-2, 13\,TeV): production of $H^+$ followed by $H^+ \to H W^+$ and $H \to \tau\tau$, yielding $\ell\tau_h$ (left) and $\ell\tau_h\tau_h$ (right) final states. The $\ell\tau_h$ final state can also arise from the right diagram when one $\tau_h$ from $H \to \tau\tau$ is not reconstructed.
From Ref.~\cite{CMS:2022jqc}.
}
\label{fig:CMS_hplus}
\end{figure}

A search for $H^+\rightarrow H(\rightarrow \tau\tau)W(\rightarrow qq,\ell\nu)$ production was conducted in the final state with one hadronically-decaying tau lepton and multi-leptons or hadronic jets, using the full CMS Run-2 dataset at 13\,TeV~\cite{CMS:2022jqc}. Here $H$ is a \emph{heavy} (non-SM) Higgs boson. The presence of two tau leptons from the heavy $H\to\tau\tau$ decay, together with the large mass splitting between $H^\pm$ and $H$, produces a distinctive kinematic signature that is exploited using a BDT with gradient boosting (BDTG). The corresponding Feynman diagrams are shown in Fig.~\ref{fig:CMS_hplus}.

\begin{figure}[H]
\centering
\includegraphics[width=0.32\linewidth]{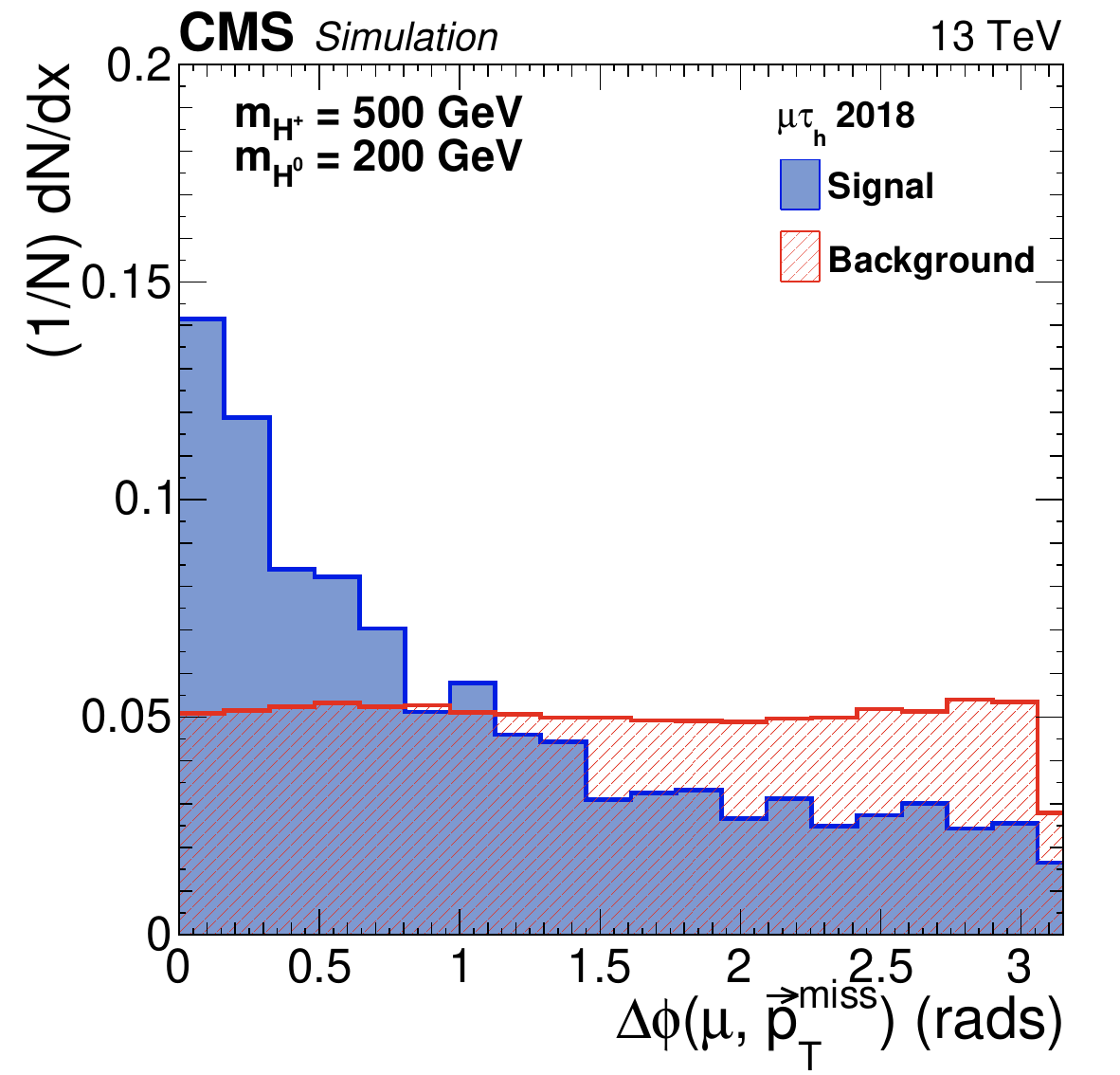}
\includegraphics[width=0.32\linewidth]{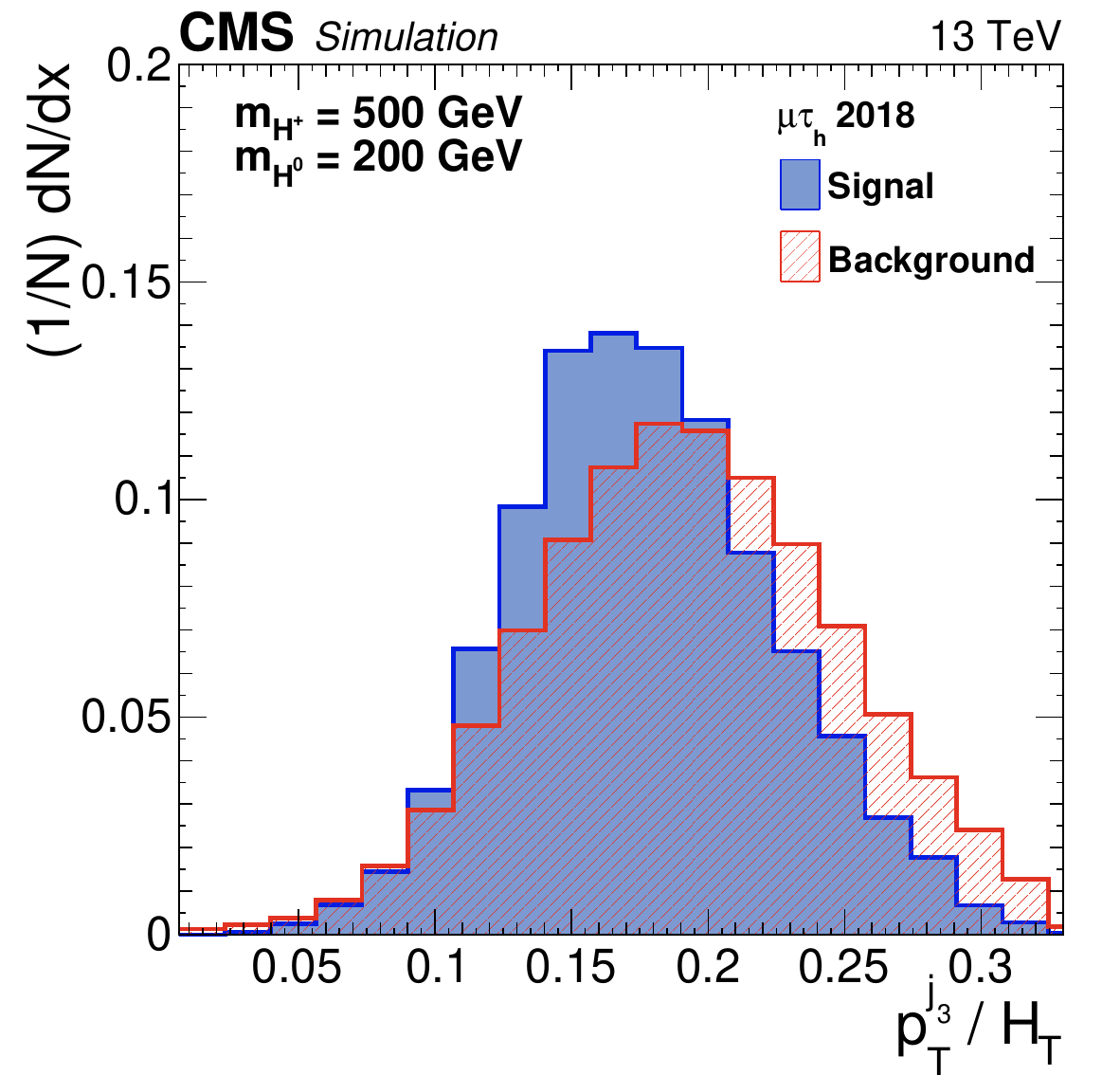}
\includegraphics[width=0.32\linewidth]{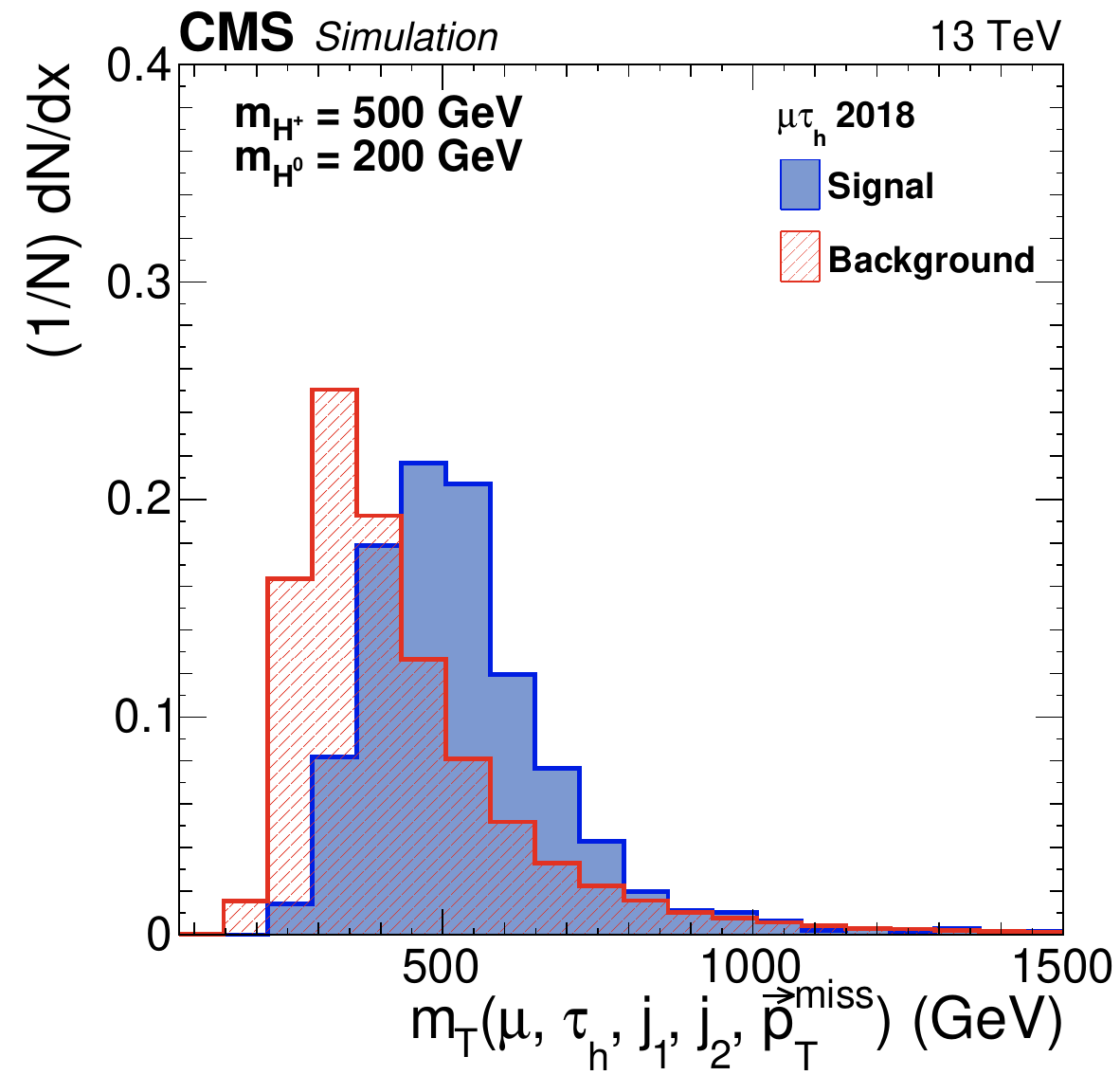}
\caption{
Three BDTG input variables for the $\mu\tau_h$ final state, for signal with $m_{H^\pm}=500~\text{GeV}$ and 2018 data-taking conditions (Run-2, 13\,TeV): the azimuthal angle between $\mu$ and $\vec{p}_T^{\text{miss}}$ (left), the ratio of the third leading jet $p_T$ to $H_T$ (middle), and the transverse mass from $\mu$, $\tau_h$, $j_1$, $j_2$, and $\vec{p}_T^{\text{miss}}$ (right). All distributions are normalized to unit area.
From Ref.~\cite{CMS:2022jqc}.
}
\label{fig:CMS_hplus_features}
\end{figure}

Three representative BDTG input variables for the $\mu\tau_h$ final state are shown in Fig.~\ref{fig:CMS_hplus_features} for a signal mass hypothesis of $m_{H^\pm} = 500$\,GeV: the azimuthal angle between the muon and $\vec{p}_T^{\text{miss}}$, the ratio of the third leading jet $p_T$ to the total hadronic transverse energy $H_T$, and the transverse mass $m_T$ reconstructed from all visible and invisible decay products. These variables probe different aspects of the event kinematics and together provide good discrimination against the dominant $t\bar{t}$ background.

\begin{figure}[H]
\centering
\includegraphics[width=0.49\linewidth]{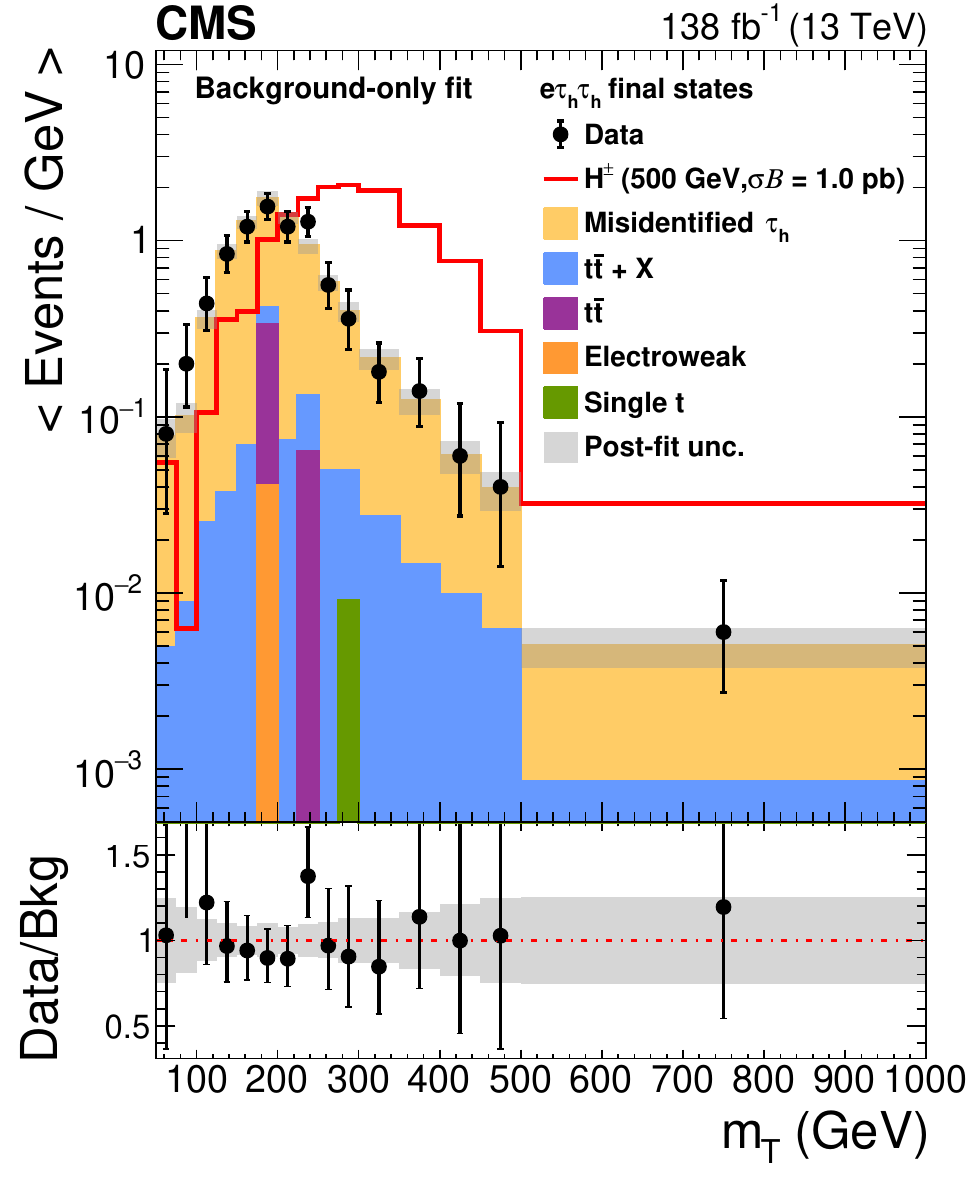}
\includegraphics[width=0.49\linewidth]{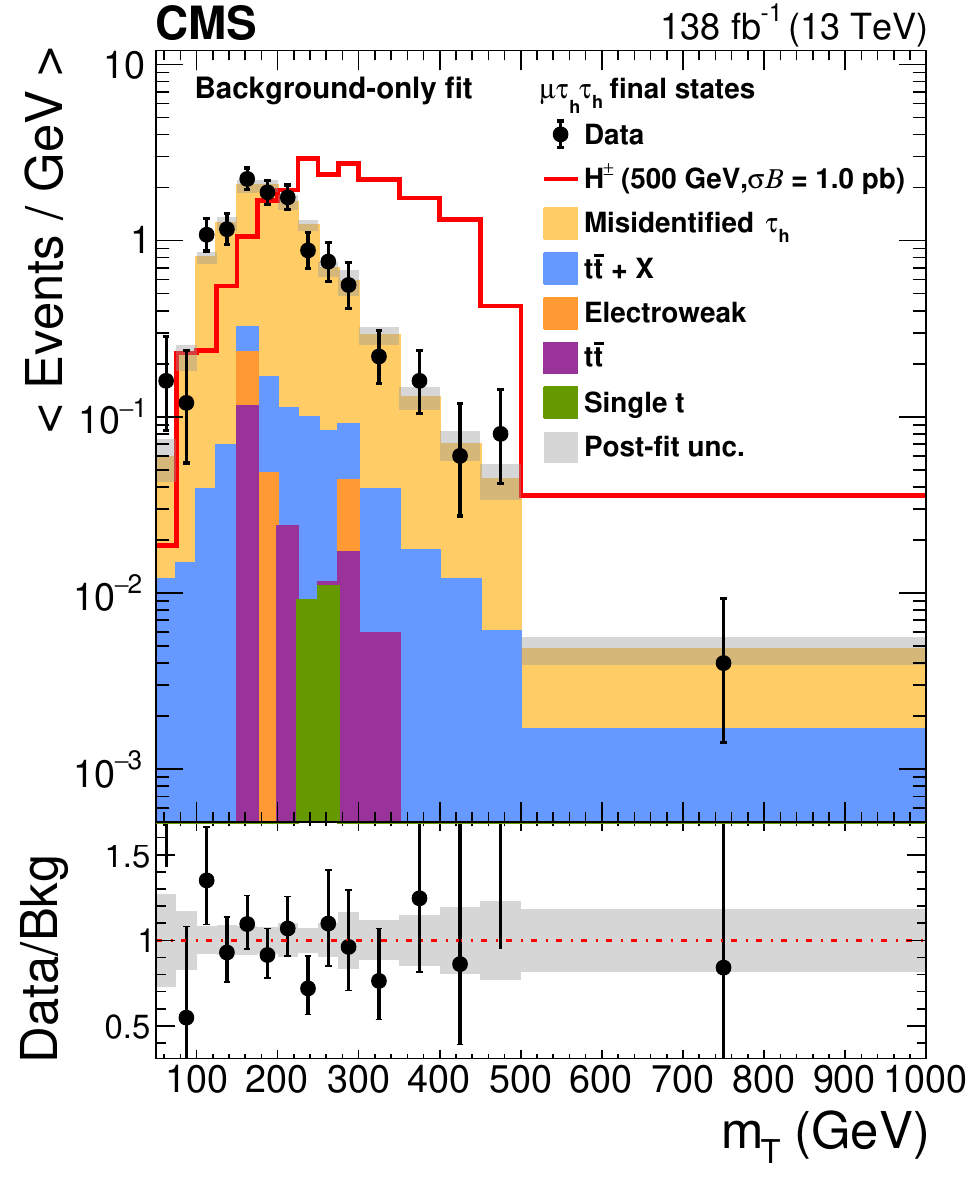}
\caption{
Post-fit $m_T$ distributions for the $e\tau_h\tau_h$ (left) and $\mu\tau_h\tau_h$ (right) final states (Run-2, 13\,TeV). Pre-fit signal contribution from $H^{\pm}\to HW^{\pm}$ with $m_{H^{\pm}}=500~\text{GeV}$, $m_H=200~\text{GeV}$, and $\sigma\mathcal{B}=1~\text{pb}$ is overlaid. The brackets $\langle Events / GeV\rangle$ denote averaging over an interval where the event frequency may have varied considerably.
From Ref.~\cite{CMS:2022jqc}.
}
\label{fig:CMS_hplus_mass}
\end{figure}

The transverse mass distributions after a background-only fit to the data are shown for the $e\tau_h\tau_h$ and $\mu\tau_h\tau_h$ final states in Fig.~\ref{fig:CMS_hplus_mass}.

\begin{figure}[H]
\vspace{-2mm}
\centering
\includegraphics[width=0.46\linewidth]{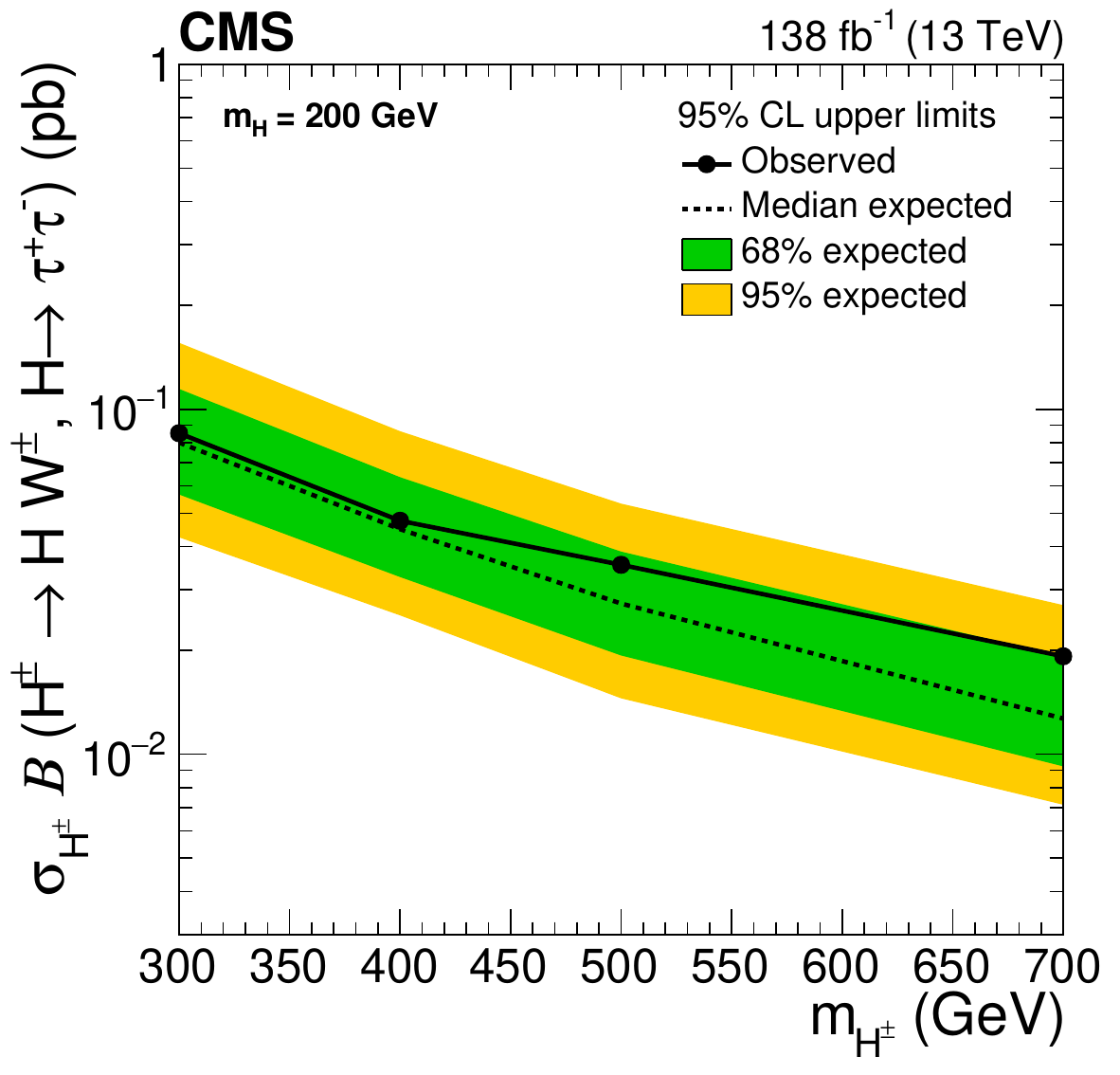}
\vspace{-2mm}
\caption{
Expected (dashed) and observed (solid) 95\% CL upper limits on $\sigma_{H^{\pm}}\mathcal{B}(H^{\pm}\to HW^{\pm},\,H\to\tau\tau)$ as a function of $m_{H^{\pm}}$, assuming $m_H=200~\text{GeV}$, for the combination of all final states (Run-2, 13\,TeV). The inner (green) and outer (yellow) bands indicate the 68\% and 95\% intervals around the expected limit.
From Ref.~\cite{CMS:2022jqc}.
}
\label{fig:CMS_hplus_limit}
\end{figure}

Limits on the production cross-section are set, as shown in Fig.~\ref{fig:CMS_hplus_limit}.
The observed limits are consistent with the expected limits within uncertainties.

\section{Combined Limits}
\begin{figure}[H]
\centering
\vspace{-2mm}
\includegraphics[width=0.51\linewidth]{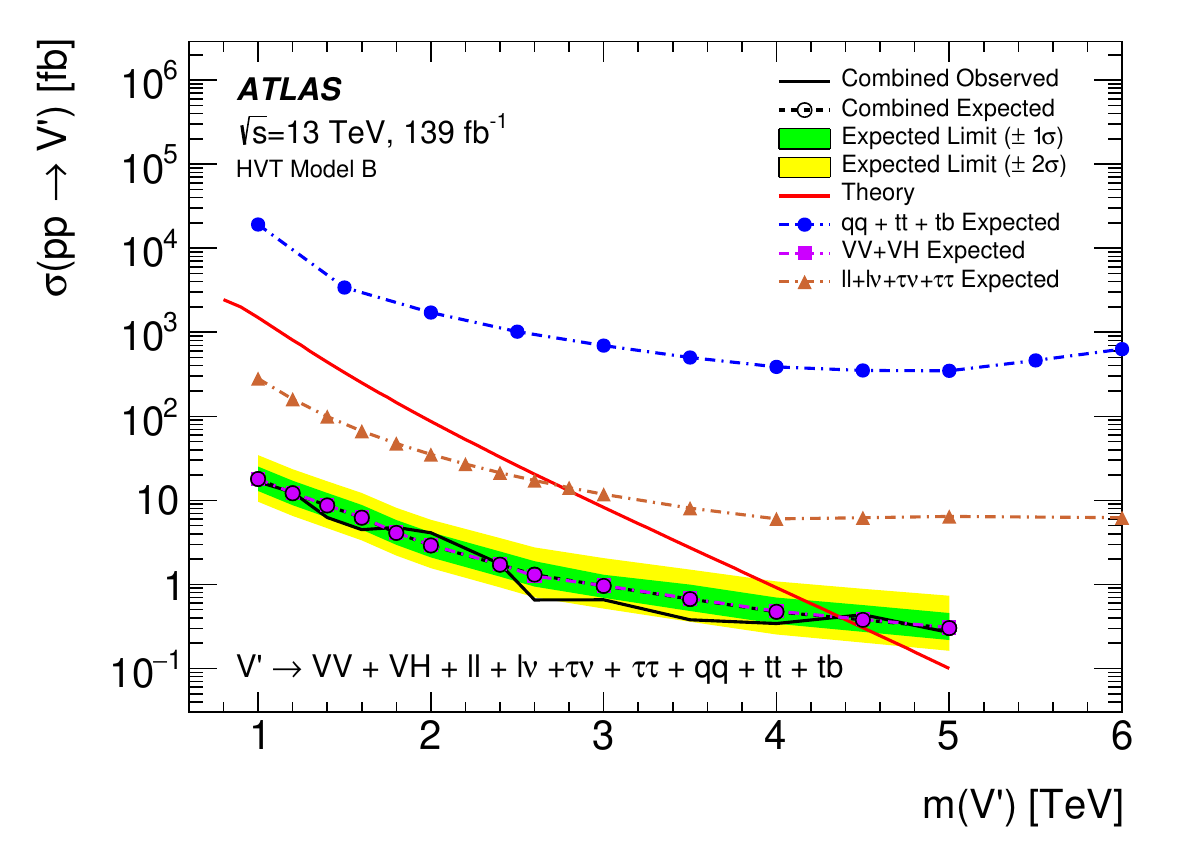}
\includegraphics[width=0.48\linewidth]{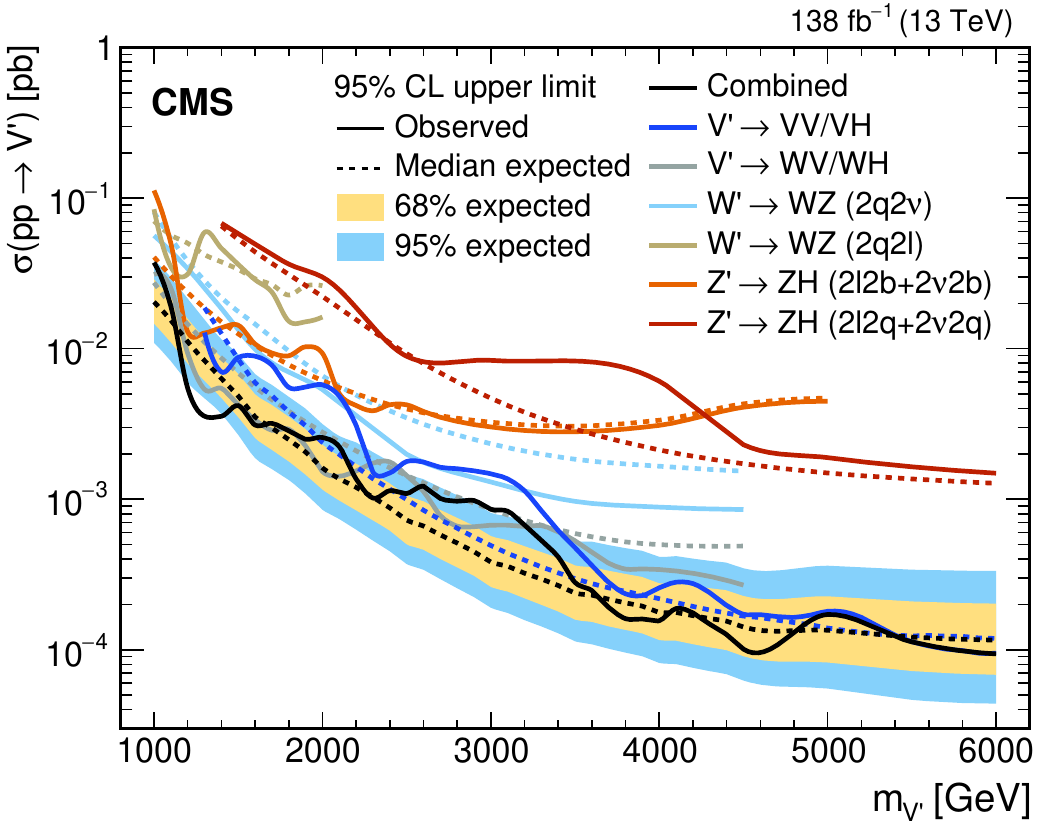}
\vspace{-5mm}
\caption{
Left: Observed and expected 95\% CL upper limits on the $V'$ production cross-section as a function of pole mass, for the subcombinations and the full combination, assuming benchmark points and the $q\bar{q}$ production mode.
Right: Expected and observed 95\% CL upper limits on $\sigma(V')$ as a function of $m_{V'}$, shown separately for the bosonic final states.
From Ref.~\cite{ATLAS:2024qvg,CMS:2026fjw}.
}
\label{fig:combined}
\vspace{-2mm}
\end{figure}

The ATLAS and CMS Collaborations released also combined limits~\cite{ATLAS:2024qvg,CMS:2026fjw}.
Observed and expected 95\% CL upper limits on the $V'$ production cross-section as a function of the resonance mass $m_{V'}$ for the individual analyses targeting
quarkonic, leptonic and bosonic final states, as well as for their combination, are shown in
Fig.~\ref{fig:combined}.

\section{Conclusions}
This review summarizes the results of present searches by the ATLAS and CMS collaborations for the resonant production of a heavy scalar $X$ decaying into a Standard Model Higgs boson and a lighter scalar $S$, as well as searches for a heavy neutral Higgs boson decaying into another neutral Higgs boson and a $Z$ boson. Searches for a heavy charged Higgs boson decaying into another neutral Higgs boson and a $W$ boson are also discussed.

The searches are motivated by well-motivated BSM frameworks. Cascade decays $X \to SH$ are characteristic signatures of extended scalar sectors such as the NMSSM. Decays $A \to ZH$ are particularly sensitive probes of the 2HDM in the alignment limit ($\cos(\beta-\alpha) \approx 0$). The observation of a charged Higgs boson via $H^\pm \to WH$ would provide unambiguous evidence of physics beyond the SM, since charged scalars are absent in the SM.

Currently, no evidence for new physics has been found based on the LHC \mbox{Run-2} dataset at 13\,TeV, corresponding to an integrated luminosity of about $140\,\text{fb}^{-1}$. A notable exception is the 2.85$\sigma$ local excess observed by ATLAS in the $A \to ZH \to Zt\bar{t}$ channel at $(m_A, m_H) = (650, 450)$\,GeV, which was not confirmed by CMS. In summer 2026, Run-3 of the LHC is expected to complete data-taking at 13.6\,TeV, giving a combined total integrated luminosity of approximately $500\,\text{fb}^{-1}$ for Runs~1--3. The analysis of this substantially increased dataset is expected to lead to significantly improved sensitivities in all investigated channels.

\section*{Acknowledgment}
The research is supported by the Ministry of
Education, Youth and Sports of the Czech
Republic under the project number LM 2023040.

\bibliographystyle{ws-ijmpa}
\bibliography{main}
\end{document}